\documentclass[useAMS,usenatbib]{mn2e}

\usepackage{natbib}

\usepackage{latexsym,graphicx}
\usepackage{color}
\usepackage{fixltx2e}
\usepackage{verbatim} \usepackage{float}
\usepackage{amsmath,amssymb}
\usepackage{times}

\usepackage{setspace}
\usepackage{rotating}
\usepackage{pdflscape}
\usepackage{subfig}

\usepackage[squaren, Gray, cdot]{SIunits}

\newcommand\hii{H\,{\sc ii} \,}

\def\apgt{\ {\raise-.5ex\hbox{$\buildrel>\over\sim$}}\ }
\def\aplt{\ {\raise-.5ex\hbox{$\buildrel<\over\sim$}}\ }

\let\oldhat\hat
\renewcommand{\hat}[1]{\oldhat{\mathbf{#1}}}

\title[The burst mode of accretion in massive star formation]{Parameter study for the burst mode of accretion in massive star formation}

\author[D. M.-A.~Meyer et al.]
       {D. M.-A.~Meyer$^{1}$\thanks{E-mail: dmameyer.astro@gmail.com}, E.~I.~Vorobyov$^{2,3}$, V.~G.~Elbakyan$^{3}$, 
       J.~Eisl\" offel$^{4}$, A.~M.~Sobolev$^{5}$  \newauthor and M. St\" ohr$^{6,7}$ \\ 
       $^{1}$Institut f\" ur Physik und Astronomie, Universit\" at Potsdam, Karl-Liebknecht-Strasse 24/25, 14476 Potsdam, Germany \\
       $^{2}$Department of Astrophysics, The University of Vienna, Vienna, A-1180, Austria \\
       $^{3}$Institute of Astronomy, Russian Academy of Sciences, Pyatnitskaya str. 48, Moscow 119017, Russia \\
       $^{4}$Th\" uringer Landessternwarte Tautenburg, Sternwarte 5, D-07778 Tautenburg, Germany \\ 
       $^{5}$Ural Federal University, 51 Lenin Str., 620051 Ekaterinburg, Russia \\       
       $^{6}$VSC Research Center, TU Wien, Operngasse 11, A-1040 Vienna  \\
       $^{7}$BOKU-IT, University of Natural Resources and Life Sciences, Peter-Jordan-Strasse 82, A-1190 Vienna \\
       }

\begin{document}

\date{Received; accepted}

\maketitle

\label{firstpage}

\begin{abstract} 
\textcolor{black}{
It is now a widely held view that, in their formation and early evolution, stars build up mass in bursts.
The burst mode of star formation} scenario proposes that the stars grow in mass 
via episodic accretion of fragments migrating from their gravitationally-unstable circumstellar discs and 
it naturally explains the existence of observed pre-main-sequence bursts from high-mass protostars. 
We present a parameter study of hydrodynamical models of massive young stellar objects (MYSOs) that explores 
the initial masses of the collapsing clouds ($M_{\rm c}=60$--$200\, \rm M_{\odot}$) and ratio of 
rotational-to-gravitational energies \textcolor{black}{($\beta=0.005$--$0.33$)}. 
An increase in $M_{\rm c}$ and/or $\beta$ produces protostellar accretion discs that are more 
prone to develop gravitational instability and to experience bursts. We find that all MYSOs have 
bursts even if their pre-stellar core is such that \textcolor{black}{$\beta\le0.01$}. Within our assumptions, the 
lack of stable discs is therefore a major difference between low- and high-mass star formation 
mechanisms. 
All our disc masses and disk-to-star mass ratios $M_{\rm d}/M_{\star}>1$ scale as a power-law 
with the stellar mass. 
Our results confirm that massive protostars accrete about $40$$-$$60\%$ of their mass in the 
burst mode. 
\textcolor{black}{
The distribution of time periods between two consecutive bursts is bimodal: there is a short duration ($\sim 1$$-$$10~\rm yr$) 
peak corresponding to the short, faintest bursts and a long-duration peak (at $\sim 10^{3}$$-$$10^{4} \rm yr$) 
corresponding to the long, FU-Orionis-type bursts appearing in later disc evolution, i.e., 
around $30\, \rm kyr$ after disc formation.
}
\textcolor{black}{
We discuss this bimodality in the context of the structure of massive protostellar jets as 
potential signatures of accretion burst history.
}
\end{abstract}



\section{Introduction}
\label{sect:intro}

\textcolor{black}{Stars are born in collapsing  pre-stellar cores}, made of cold molecular material. 
Although the early, classical picture for star formation concluded that young stellar objects gain 
their mass by constant mass accretion via spherical accretion~\citep{larson_mnras_145_1969,shu_apj_214_1977}, 
the free-falling gas \textcolor{black}{landing onto an accretion disc rather than interacting with the 
protostellar surface}. This continual mass-loading sustains the disk in a gravitationally unstable state, 
which is characterized by highly variable accretion rates, in agreement with those monitored by observations 
of low-mass star-forming regions~\citep{vorobyov_apj_704_2009}. 
Amongst many disc-based models developed to describe the way stars gain their mass, the burst mode 
of accretion is a picture developed \textcolor{black}{in the context of the formation of low-mass stars}~\citep{voroboyov_apj_650_2006, vorobyov_apj_719_2010,vorobyov_apj_805_2015}. 
This depiction of star formation processes includes the \textcolor{black}{gravitational collapse of a parent cloud, 
followed by the establishment} and fragmentation of a gravitationally unstable circumstellar accretion disc, and the inward migration of gas clumps towards the star. The inward-migrating clumps trigger an increase of the accretion rate and 
generate accretion-driven luminosity outbursts as they are tidally destroyed in the vicinity of the star.   It was successfully  applied to solve the so-called "luminosity 
problem", stating that young protostars are on average less luminous than expected from simple spherical collapse calculations~\textcolor{black}{\citep{2011ApJ...736...53O,dunham_apj_747_2012,2014ApJ...797...32P}}, showed consistencies with observations of FU-Orionis  flares~\citep{vorobyov_apj_805_2015} and demonstrated agreement with the knot spacing in \textcolor{black}{protostellar jets} (Vorobyov et al. 2018).
The clump-infall-triggered mechanism of accretion bursts in low-mass stars was confirmed and further elaborated in three-dimensional (magneto)-hydrodynamical simulations~\citep{zhao_mnras_473_2018} and in semi-analytic studies (Nayakshin \& Lodato 2012).
%
These results and the observational discovery of a luminous flare from the massive young stellar object (MYSO) S255IR-NIRS3, 
triggered by a sudden increase of its accretion rate, raised the question of the existence of a scaling relationship 
\textcolor{black}{between the forming mechanisms of low- and high-mass stellar objects, respectively}. 
%

Observations of the circumstellar medium of proto-OB stars have accumulated, increasing our knowledge of the 
formation of massive stellar objects. In particular, the works of~\citet{fuente_aa_366_2001,testi_2003,cesaroni_natur_444_2006} 
\textcolor{black}{revealed that the mechanisms involved in the formation of massive stars are characterized by the presence of certain features such as} converging accretion flows~\citep{keto_apj_637_2006} 
and jets~\citep{Cunningham_apj_692_2009,caratti_aa_573_2015,burns_mnras_467_2017,
burns_2018IAUS,reiter_mnras_470_2017,purser_mnras_475_2018,samal_mnras_477_2018,2019arXiv191208510B,2019arXiv191111447Z}. 
\textcolor{black}{
Differences lie in the fact that young massive stars exhibit \textcolor{black}{lobed} \textcolor{black}{ bubbles of 
ionized gas}~\citep{cesaroni_aa_509_2010,purser_mnras_460_2016}. 
}
At the same time, a growing number of (Keplerian) disc-like structures has been reported in interferometric 
observations~\citep{johnston_apj_813_2015,ilee_mnras_462_2016,
forgan_mnras_463_2016,2018arXiv180410622G,maud_aa_620_2018,2018arXiv181110245B,ahmadi_aa_618_2018,sanna_aa_623_2019}, some 
of them revealing the presence of substructures in it such as MM1-Main~\citep{maud_467_mnras_2017}, 
the massive double-core proto-system G350.69-0.49~\citep{chen_apj_835_2017}, 
the protomassive object G11.920.61 MM 1~\citep{2018ApJ...869L..24I},
the AFGL 4176 mm1~\citep{2019arXiv191109692J} and the O-type (proto-)star G17.64+0.16~\citep{maud_aa_627_2019}, 
G353.273+0.641~\citep{motogi_apj_877_2019}, suggesting similar qualitative formation mechanisms to those 
in the low-mass regime of star formation~\citep{bosco_aa_629_2019}, see also~\citet{wurster_mnras_486_2019,2019arXiv190612276W}. 
\textcolor{black}{
Most recent high-angular {\sc alma} observations in the region S255IR-SMA1 show a clear consistency between the 
predictions of the burst mode of accretion in high-mass star formation and the properties of the accretion flow of the 
circumstellar medium of S255IR-NIRS3~\citep{2020arXiv201009199L}. 
}

These observations have been supported by \textcolor{black}{3D hydrodynamics and radiative transfer calculations}, predicting \textcolor{black}{how 
accretion discs surrounding young high-mass stars form}~\citep{1998MNRAS.298...93B,2002ApJ...569..846Y,krumholz_apj_656_2007,
peters_apj_711_2010,seifried_mnras_417_2011,harries_mnras_448_2015,klassen_apj_823_2016,harries_2017,rosen_apj_887_2019,ahmadi_aa_632_2019,anezlopez_apj_888_2020}. 
Multiplicity, as an indissociable characteristic of massive star formation, suggests that disc
fragmentation can play a crucial role in the formation of the (spectroscopic) companions observed in most systems involving
OB stars~\citep{2013A&A...550A..27M,2014ApJS..213...34K,chini_424_mnras_2012,kraus_apj_835_2017}.
\textcolor{black}{
Young massive stars are also sites of strongly variable maser emission, see in particular strong maser 
flares of the MYSOs NGC~6334~I, S255IRNIRS3 and G358.93-0.03 associated with accretion 
bursts~\citep{szymczak_aa_617_2018,macleod_mnras_478_2018,burns_natas_2020}. 
It is now established that the methanol emission traces accretion disks around 
MYSOs~\citep[][ and references therein]{sanna_aa_603_2017} while water maser emission 
traces well outflows from these objects~\citep[][and references therein]{brogan_apj_866_2018}. 
New maser species and a growing number of Class II CH3OH maser lines are discovered 
from massive star-forming regions~\citep{brogan_apj_866_2018,macLeod_mnras_489_2019,chen_apj_890_2020,breen_apj_876_2019}. 
Lastly, it is worth mentioning the evidence of non-thermal synchrotron emission from the outflows 
reported in a number of MYSOs~\citep{carrasco_sci_330_2010,obonyo_486_MNRAS_2019}  
and probable detection of the synchrotron emission from accretion disk~\citep{Shchekinov_aa_418_2004}. 
}
It is now established that water maser emission trace well protostellar outflows~\citep{2019asrc_confE_91H} 
and a growing number of Class II $\rm CH_{3}OH$ maser lines are discovered from massive star-forming regions. 
Lastly, it is worth mentioning that the evidence for non-thermal \textcolor{black}{synchrotron radiation from an outflow 
originating from a MYSO has been reported} in~\citet{obonyo_486_MNRAS_2019}.

The radiation-hydrodynamics simulations of~\citet{meyer_mnras_464_2017} discovered the burst mode 
of accretion in \textcolor{black}{the formation of massive stars}. The bursts are 
triggered by the accretion of fast-moving circumstellar gaseous clumps, which migrate inwards from the gravitationally 
fragmenting spiral arms towards the star. Moreover, we showed that some fragments have internal thermodynamical 
properties (e.g., temperature $>2000$$-$$3000\,\rm K$) consistent with the onset of molecular hydrogen dissociation 
and run-away collapse, \textcolor{black}{showing the disc fragmentation channel to be a viable route for 
making high-mass spectroscopic protobinaries}~\citep{meyer_mnras_473_2018}. 
\textcolor{black}{
The setups developed for the burst mode in accretion by~\citet{meyer_mnras_464_2017} 
and~\citet{meyer_mnras_473_2018} have been further, elegantly used in~\citet{2020arXiv200813653A}. 
}
We then calculated that \textcolor{black}{MYSOs} spend \textcolor{black}{only} ($\le 2\%$) in the bursting 
phase, while they can therethrough accrete up to $50\%$ of their \textcolor{black}{final mass}~\citep{meyer_mnras_482_2019}. 
The episodic increase of \textcolor{black}{the mass transfers onto} the \textcolor{black}{surface of the} protostar 
induces bloating of \textcolor{black}{its} radius, provoking quick excursions towards \textcolor{black}{redder} 
region of the \textcolor{black}{temperature-luminosity} diagram. 
This process is accompanied by intermittency of the photon fluxes, which fill and irradiate the bipolar outflow 
as an \hii region~\citep{2019MNRAS.484.2482M}. 
Last, we have performed synthetic images of the accretion discs around our 
\textcolor{black}{
massive protostars and predicted their {\sc alma} signature~\citep{meyer_487_MNRAS_2019,jankovic_mnras_482_2019}. 
}
However, given to the computationally-expensive aspect of massive star formation calculations, such results 
were so far obtained on a limited number of star-disc models, which raises the question of the effects of 
the pre-stellar core properties used as initial conditions in numerical simulations.

This paper performs a parameter study for the burst mode of accretion \textcolor{black}{in the context of forming high-mass stars}. 
Using methods developed in~\citet{meyer_mnras_464_2017}, we investigate here the effects of the \textcolor{black}{mass of the core, 
together with its rotational-to-gravitational energy ratio}, on the accretion history and 
protostellar mass evolution. For each model, we analyse (i) the disc properties developing around the protostars 
and (ii) the accretion-driven burst properties, using the method presented in~\citet{meyer_mnras_482_2019}. 
If such a parameter study is original \textcolor{black}{for} high-mass stars, similar works \textcolor{black}{exist 
for low-mass stars}~\citet{vorobyov_apj_728_2011}. \textcolor{black}{Our results show that}, in opposite to low-mass star formation, 
all models exhibit highly-variable accretion rate histories and that their associated lightcurves are interspersed 
with episodic bursts, i.e. no young massive stars appear to be burstsless. 
Particularly, we discuss our findings \textcolor{black}{within observations of massive protostars} which exhibited 
accretion variability and/or (probable signs of) disc fragmentation, such as S255IR-NIRS3 and NGC 6334I-MM1. 
We further consider our results in connection with the morphology and temporal domain of protostellar jets 
\textcolor{black}{of some massive protostars}.

\textcolor{black}{
In Section~\ref{sect:method} we introduce our numerical methods and specify \textcolor{black}{which} 
parameter space is explored} in this paper. 
We detail the properties of our simulated accretion discs in Section~\ref{sect:discs} and analyse the burst 
properties for our whole sample of MYSOs in Section~\ref{sect:bursts}. \textcolor{black}{Our outcomes are discussed in 
Section~\ref{sect:discussion} and we conclude in Section~\ref{sect:conclusion}.}


\section{Method}
\label{sect:method}

\textcolor{black}{We hereby present our} numerical methods and initial conditions used to perform our 
gravito-radiation-hydrodynamics disc models, from which we extract accretion discs masses and 
time-dependent protostellar accretion rate histories.

\subsection{Governing equations}
\label{sect:eq}

The hydrodynamics of the gas \textcolor{black}{obeys} the conservation of mass,
\begin{equation}
	   \frac{\partial \rho}{\partial t}  + 
	   \bmath{\nabla}  \cdot (\rho\bmath{v}) =   0,
\label{eq:euler1}
\end{equation}
\textcolor{black}{the conservation of} momentum
\begin{equation}
	   \frac{\partial \rho \bmath{v} }{\partial t}  + 
           \bmath{\nabla} \cdot ( \rho  \bmath{v} \otimes \bmath{v})  + 
           \bmath{\nabla}p 			      =   \bmath{f},
\label{eq:euler2}
\end{equation}
and \textcolor{black}{the conservation of} energy
\begin{equation}
	  \frac{\partial E }{\partial t}   + 
	  \bmath{\nabla} \cdot ((E )+p) \bmath{v})  =	   
	  \bmath{v} \cdot \bmath{f} ,
\label{eq:euler3}
\end{equation}
\textcolor{black}{
with the fluid density $\rho$, velocity $\bmath{v}$, and thermal pressure $p$. The latter is defined as  
\begin{equation}
	p=(\gamma-1)E_{\rm int},
	\label{dddd}
\end{equation}
with the adiabatic index $\gamma=5/3$. In Eq.~(\ref{dddd}), $E_{\rm int}$ stands for the gas internal energy,
and the total energy is written as}
\begin{equation}
	  E  = E_{\rm int} + \rho \frac{1}{2} \bmath{v}^{2}. 
\label{eq:etot}
\end{equation}
Our model considers the total gravitational potential 
\begin{equation}
     \Phi_{\rm tot}  = \Phi_{\star} + \Phi_{\rm sg},  
   \label{eq:potential}    
\end{equation}
where the stellar contribution reads
\begin{equation}
      \Phi_{\star}=-G \frac{M_{\star}}{r},  
\end{equation}
with $M_{\star}$ being the \textcolor{black}{protostellar mass and $G$ the universal constant of gravity. 
Self-gravity is found by numerically solving for the Poisson equation} 
\begin{equation}
	  \bmath{\Delta} \Phi_{\rm sg}   =  4\pi G\rho.
\label{eq:poisson}
\end{equation}
Our setup does not include artificial shear viscosity~\citep{hosokawa_2015}.

\begin{table}
	\centering
	\caption{
	Initial characteristics of the solid-body-rotating pre-stellar cores in our grid of simulations. 
	The table gives the mass of the molecular pre-stellar core $M_{\rm c}$, the ratio of rotational-to-gravitational 
	energy $\beta$ (in $\%$), the final simulation time $t_{\rm end}$ and the final stellar mass $M_{\star}(t_{\rm end})$ 
	of each model, respectively.   
	}
	\begin{tabular}{lcccr}
	\hline
	$\rm Models$               & $M_{\rm c}$ ($\rm M_{\odot}$)   &  $\beta$ ($\%$)  &  $t_{\rm end}$ ($\rm kyr$)  &  $M_{\star}$ ($t_{\rm end}$   )  \\ 
	\hline    
  	$\rm Run-60-4$$\%^{(a)}$       & 60                              &  4             &  65.2       &  20.0      \\  
	$\rm Run-80-4$$\%$            & 80                              &  4             &  53.6       &  26.6      \\  
	$\rm Run-100-4$$\%^{(a)}$      & 100                             &  4             &  47.6       &  33.3      \\  	
	$\rm Run-120-4$$\%$           & 120                             &  4             &  44.3       &  40.0      \\
	$\rm Run-140-4$$\%$           & 140                             &  4             &  41.0       &  46.6      \\	
	$\rm Run-160-4$$\%$           & 160                             &  4             &  39.0       &  53.3      \\
	$\rm Run-180-4$$\%$           & 180                             &  4             &  36.5       &  60.0      \\
	$\rm Run-200-4$$\%$           & 200                             &  4             &  33.7       &  66.6      \\
	$\rm Run-60-0.1$$\%$          & 60                              &  0.1           &  60.0       &  41.6      \\  		
	$\rm Run-60-0.3$$\%$          & 60                              &  0.3           &  60.0       &  31.6      \\  		
	$\rm Run-60-0.5$$\%$          & 60                              &  0.5           &  60.0       &  29.9      \\  
	$\rm Run-60-1$$\%$            & 60                              &  1             &  60.0       &  13.7      \\  	
	$\rm Run-100-2$$\%$           & 100                             &  2             &  60.0       &  51.6      \\  
	$\rm Run-100-5$$\%^{(b)}$      & 100                             &  5             &  60.0       &  41.5      \\  	
	$\rm Run-100-6$$\%$           & 100                             &  6             &  60.0       &  39.3      \\
	$\rm Run-100-8$$\%$           & 100                             &  8             &  60.0       &  34.0      \\	
	$\rm Run-100-10$$\%^{(b)}$     & 100                             &  10            &  60.0       &  34.1      \\
	$\rm Run-100-12$$\%$          & 100                             &  12            &  60.0       &  33.8      \\
	$\rm Run-100-14$$\%$          & 100                             &  14            &  60.0       &  29.5      \\
	$\rm Run-100-16$$\%$          & 100                             &  16            &  60.0       &  29.6      \\
	$\rm Run-100-18$$\%$          & 100                             &  18            &  60.0       &  22.2      \\	
	$\rm Run-100-20$$\%$          & 100                             &  20            &  60.0       &  25.0      \\
	$\rm Run-100-25$$\%$          & 100                             &  25            &  60.0       &  19.8      \\
	$\rm Run-100-33$$\%$          & 100                             &  33            &  60.0       &  27.4      \\		
        \hline
	\end{tabular}
\label{tab:models}\\
\footnotesize{ ${(a)}$~\citet{meyer_mnras_473_2018,2019MNRAS.484.2482M}, ${(b)}$~\citet{meyer_mnras_482_2019} }\\
\end{table}

The source term function $\bmath{ f }$ in Eqs.~(\ref{eq:euler2}) and~(\ref{eq:euler3}) is the force density vector. It reads 
\begin{equation}
	  \bmath{ f } = -\rho \bmath{\nabla} \Phi_{\rm tot} 
			- \lambda \bmath{\nabla} E_{\rm R} 
			- \bmath{\nabla} \cdot \Big( \frac{ \bmath{F_{\star}} }{ c } \Big) \bmath{e}_{\rm r},
\label{eq:f}
\end{equation}
where $\lambda$ represents the flux limiter, $E_{\rm R}$ the thermal \textcolor{black}{radiation} energy 
density, $\bmath{e}_{\rm r}$ the radial unit vector, $\bmath{F_{\star}}$ the stellar radiation flux and $c$ the speed 
of light. 
The equation of radiation transport, 
\begin{equation}
	  \frac{\partial }{\partial t} \Big( \frac{ E_{\rm R} }{ f_{\rm c}  }  \Big)  + 
	  \bmath{\nabla} \cdot \bmath{F}  =	   
	  -\bmath{\nabla} \cdot \bmath{F_{\star}},
\label{eq:rad1}
\end{equation}
governs  the thermal radiation energy density $E_{\rm R}$ with
\begin{equation}
      f_{\rm c} = \frac{ 1 }{ \frac{ c_{\rm v} \rho}{4 a T^{3} } + 1 },
\end{equation}
where $a$ is the radiation constant and $c_{\rm v}$ the specific heat capacity. 
We solve it within the so-called flux-limited diffusion formalism, i.e., 
\begin{equation}
    \bmath{ F } = -D \bmath{\nabla} E_{\rm R},
\end{equation}
stands for the radiation flux with the diffusion constant,
\begin{equation}
      D = \frac{\lambda c}{ \rho \kappa_{\rm R}},
\end{equation}
with the average Rosseland opacity $\kappa_{\rm R}$. Therefore,
\begin{equation}
	  \frac{ \bmath{F_{\star}}(r)  }{ \bmath{F_{\star}}( R_{\star})  } = \Big( \frac{ R_{\star} }{ r } \Big)^{2} e^{-\tau(r)},
\label{eq:rad2}
\end{equation}
accounts for diminishing the incident stellar radiation flux as it penetrates through the circumstellar medium. 
The quantity $R_{\star}$ denotes the radius of the MYSO and the optical depth of the medium is,
\begin{equation}
    \tau(r) = \int_{r_{\rm in}}^{r} G(r') dr',  
\end{equation} 
while the total opacity includes radiation attenuation by dust and gas, with 
$\rm in$ the inner boundary of the grid in the radial direction (see below). The function $G(r)$ reads, 
\begin{equation}
    G(r) = \kappa_{\rm g}  \rho_{\rm g}(r)  + \kappa_{\rm d}(r)  \rho_{\rm d}(r), 
\end{equation}    
where \textcolor{black}{the quantities $\kappa_{\rm d}$ and $\kappa_{\rm g}$ are respectively 
the opacities of the dust and gas components of the disc material, respectively. 
The gas-to-dust mass ratio is initially set to 100, the gas opacity is taken to 
a constant value $\kappa_{\rm g}=0.01\, \rm cm^{2}\, \rm g^{-1}$ while the opacity of the 
dust comes from~\citet{laor_apj_402_1993}. 
Therefore, }
\begin{equation}
	  a T^{4}  = E_{\rm R} + \frac{\kappa(r)}{\kappa_{\rm P}(T)} \frac{ |\bmath{F_{\star}}| }{ c }, 
\label{eq:radfield}
\end{equation}
with \textcolor{black}{
\begin{equation}
    \kappa(r)=\kappa_{\rm g}(r)  + \kappa_{\rm d}(r),
\end{equation}
using $\kappa_{\rm P}$,} the Planck opacity. 
\textcolor{black}{
Last, the stellar flux $\bmath{F_{\star}}( R_{\star})$ through the sink cell is the total irradiation, constituted by 
$L_{\star}$, the photospheric luminosity, and the accretion luminosity of the MYSO. 
The values of the effective temperature $T_{\rm eff}$ and stellar radius $R_{\star}$ are taken from the stellar 
evolutionary tracks of~\citet{hosokawa_apj_691_2009}. 
}

\begin{figure*}
        \centering
        \begin{minipage}[b]{ 1.0\textwidth} 
                \includegraphics[width=1.0\textwidth]{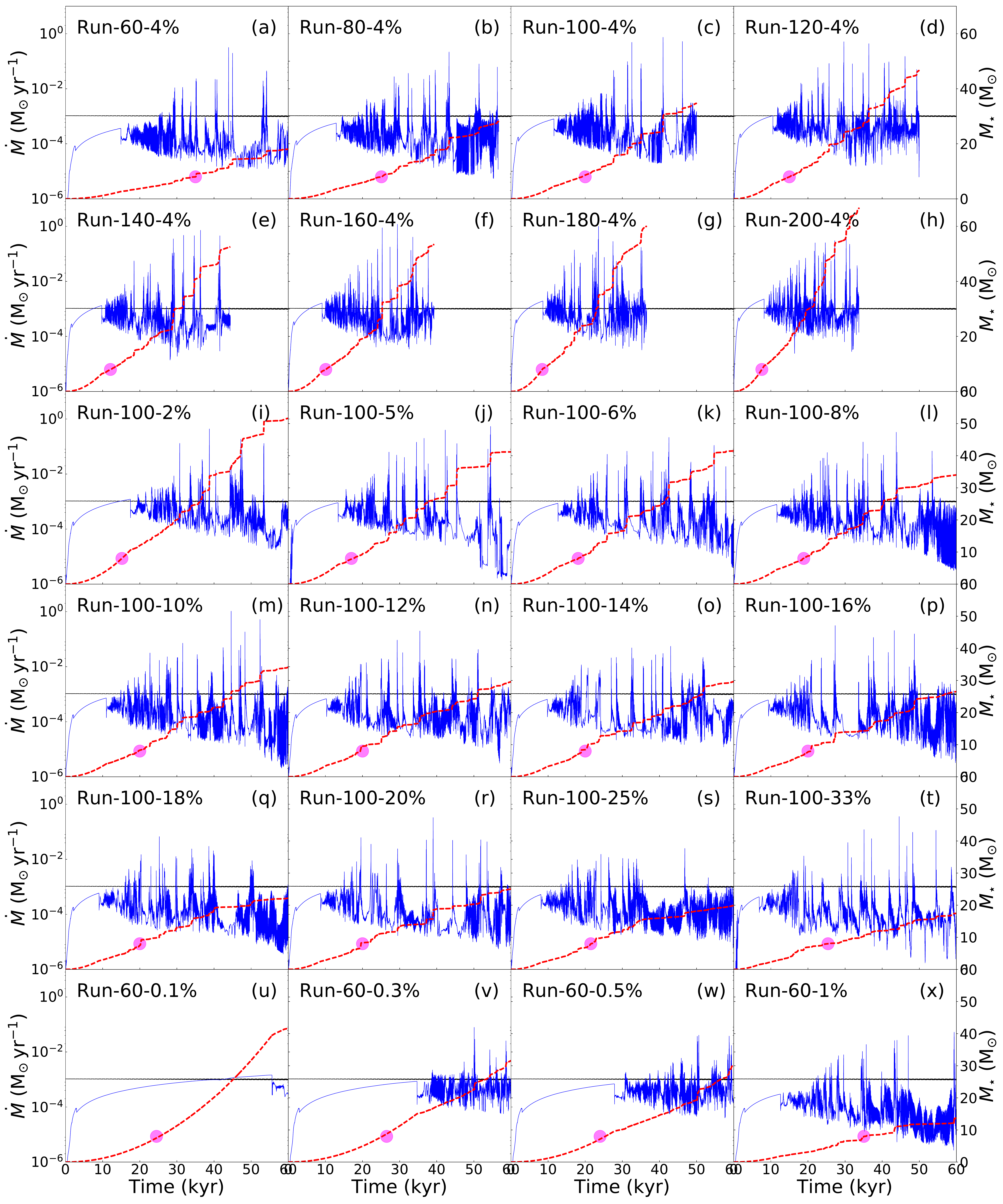}
        \end{minipage}     
        \caption{ 
                 Accretion rate histories $\dot{M}(t)$ in our parameter study. For each 20 models, the figure shows the 
                 accretion rate (in $\rm M_{\odot}\, \rm yr^{-1}$, thin blue line) and the time evolution of the 
                 stellar mass $M_{\star}(t)$ (thick dotted red line, in $\rm M_{\odot}$).  
        	     The thin horizontal black line indicates a rate of $\dot{M}=10^{-3}\, \rm M_{\odot}\, \rm yr^{-1}$, 
        	     the magenta dot marks times where the protostar enters the high-mass regime ($M_{\star}=8\, \rm M_{\odot}$). 
                 }      
        \label{fig:rates}  
\end{figure*}

\subsection{Numerical scheme, initial conditions, parameter space}
\label{sect:dens}

The 3D models are carried out in spherical coordinates $(r,\theta,\phi)$ with a static grid. 
Under the simplifying assumption of the mid-plane symmetry, 
the size of the grid is $[r_{\rm in},r_{\rm max}]\times[0,\pi/2]\times[0,2\pi]$ 
along the different radial, polar and azimuthal directions. 
It is constructed of $128\times21\times128$ grid zones, while the mesh expands along 
$r$ as a logarithm, along $\theta$ as a cosine,  and is kept uniform along $\phi$. 
The inner and outer boundaries are $r_{\rm in}=10\, \rm AU$ and $r_{\rm max}=R_{\rm c}=0.1\, \rm pc$, 
where $R_{\rm c}$ stands for the core radius, respectively. 
Outflow conditions are assigned at both boundaries of the radial 
directions $r$ so that we can measure the accretion rate onto the protostar $\dot{M}$ as 
the mass of the gas crossing $r_{\rm in}$. 
The set of above described equations are solved using a $2^{\rm nd}$ order in space and time numerical scheme 
with the {\sc pluto} code~\citep{mignone_apj_170_2007,migmone_apjs_198_2012} including 
stellar evolution, radiation transport and self-gravity~\citep{meyer_mnras_464_2017,meyer_mnras_473_2018}. 
Our scheme treats the protostellar radiation, by which the photons propagate from the atmosphere of the MYSO 
to the accretion disc and their subsequent propagation into the disk by flux-limited diffusion performed 
in the gray approximation.  
Finally, our multidimensional scheme is solved making use of 
the Strang operator splitting 
available in the \textsc{pluto} code, which permits to calculate fluxes such as radiation 
fluxes as a series of independent one-dimensional problems.

We initialise our models with a spinning molecular core characterised by the density 
profile,
\begin{equation}
    \rho(r) = K_{\rho} r^{ \beta_{\rho} },
    \label{eq:density_profile}
\end{equation}
with $K_{\rho}$ being a constant and where $\beta_{\rho}$ is negative.
The core mass that is embedded inside a given radius $r$ is,
\begin{equation}
   M(r) = M_{\rm c} \Big( \frac{r}{R_{\rm c}} \Big)^{ \beta_{\rho} + 3 },
    \label{eq:mass2}   
\end{equation}
and it determines the quantity $K_{\rho}$. Hence, one can find the density profile, 
\begin{equation}
    \rho(r) = \frac{ ( \beta_{\rho} + 3 ) }{ 4\pi  } \frac{ M_{\rm c} }{  R_{\rm c}^{ \beta_{\rho} + 3 } } r^{ \beta_{\rho} },
    \label{eq:density_prof_tot}
\end{equation}
where $r$ is the radial coordinate. The angular momentum distribution  is,
\begin{equation}
    \Omega(R) = \Omega_{0} \Big( \frac{ R }{ r_{0} } \Big)^{ \beta_{ \Omega } },
    \label{eq:momentum_distribution1}    
\end{equation}
with 
\begin{equation}
    R = r \sin(\theta)
\end{equation}
the so-called cylindrical radius and $\Omega_{0}$ a normalization constant. It   
is a function of the ratio of kinetic-to-gravitational energy, 
\begin{equation}
\beta = \frac{ E_{\rm rot} }{ E_{\rm grav}},
\end{equation}
which fixes the initial rotation properties of the system. Finally, the cloud total gravitational 
energy reads
\begin{equation}
    E_{\rm grav} =  \frac{ \beta_{\rho} + 3 }{ 2\beta_{\rho} + 5 }  \frac{G M_{\rm c}^{2}}{R_{\rm c}}, 
   \label{eq:Egrav}    
\end{equation}
whereas its rotational kinetic energy is, 
\begin{equation}
    E_{\rm rot} = \frac{ ( \beta_{\rho} + 3 )  }{ 4 ( \beta_{\rho} + 2\beta_{\Omega} + 5 )  }
		  \frac{ \Omega_{0}^{2} M_{\rm c} r_{0}^{ -2\beta_{\Omega} }  }{  R_{\rm c}^{ -2(\beta_{\Omega} + 1 ) }  } 
                  \int_{ 0 }^{ \pi } d\theta \sin( \theta )^{ 3+2\beta_{\Omega} },
   \label{eq:Erot}                  
\end{equation}
which must be solved prior to the numerical simulations to find $\Omega_{0}$. 
We initially set the molecular core with $\beta=4\%$.
The radial profile for the distribution reads,  
\begin{equation}
    v_{\phi}(R)=R\Omega(R), 
\end{equation}
but $v_{\rm r}=0$ and $v_{\theta}=0$. 
The system's thermal pressure is 
\begin{equation}
     p=\frac{ R \rho T_{\rm c} }{\mu},
\end{equation}
where $\mu$ is the mean molecular weight, $R$ is the ideal gas constant and where $T_{\rm c}=10\, \rm K$ 
is the core temperature. We initialise the simulations by setting, 
\begin{equation}
  T_{\rm d}=T_{\rm g}=T_{\rm c}, 
\end{equation}  
and we do not distinguish between gas and dust temperature throughout the simulation.
The gas and dust temperatures are obtained by solving Eq.~\ref{eq:radfield} where $E_\mathrm{R}$ is 
calculated from Eq.~\ref{eq:rad1}.

We run a series of simulations exploring the effects of the mass $M_{\rm c}$ and 
the initial ratio $\beta$ of the pre-stellar core. 
Instead of running the simulations up to the complete collapse of all the core material, 
we implicitly account for stellar feedback and its role in stopping accretion. 
Estimating when a protostar reaches the ZAMS is complicated, however, our stellar evolution 
calculations in~\citet{2019MNRAS.484.2482M} concluded that a $100\, \rm M_{\odot}$ cloud 
with $\beta=4\%$ produces a protostar reaching the ZAMS $\approx 50\, \rm kyr$ after the 
beginning of the collapse, when $M_{\rm c}/3\approx 33.3\, \rm M_{\odot}$ of the core has 
been accreted. This is the criterion we applied as an educated guess to terminate the simulations  
for the line of increasing $M_{\rm c}$. 
Otherwise, the mass-loading from the infalling core would continue to replenish the disk with material 
during the entire collapse phase if the simulation had been allowed to continue. It would thus 
sustain strong gravitational instability and fragmentation and hence the production of bursts 
which qualitative properties would remain unchanged with respect to the present study. 
We summarise our models in Table~\ref{tab:models}.


\section{Discs properties}
\label{sect:discs}

This section investigates the \textcolor{black}{protostellar mass evolution, the mass of the accretion disc, 
and the ratio of the disc-to-star masses} in our simulations for the formation of young massive stellar 
objects. We discuss these quantities for the \textcolor{black}{different initial conditions of our models 
(masses and spins of parent pre-stellar cores). }

\begin{figure*}
        \centering
        \begin{minipage}[b]{ 0.95\textwidth} 
                \includegraphics[width=1.0\textwidth]{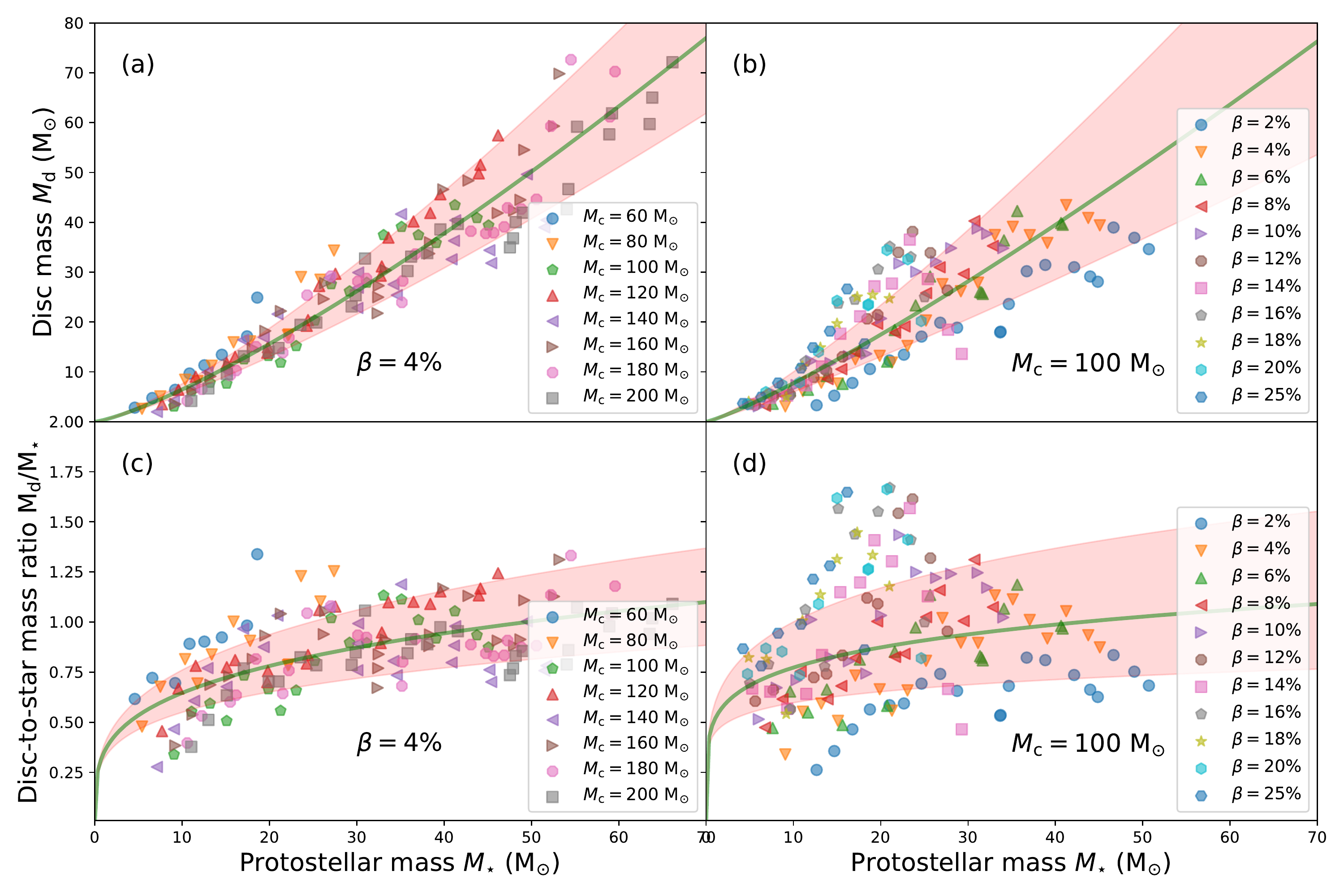}
        \end{minipage}     
        \caption{ 
                 Disc masses $M_{\star}$ (in $\rm M_{\odot}$, top panels) and \textcolor{black}{ratio of the disc-to-star masses}  
                 $M_{\rm d}/M_{\star}$ (bottom panels) \textcolor{black}{shown according to the} protostellar mass $M_{\star}$ 
                 (in $\rm M_{\odot}$), \textcolor{black}{for all models with changing core mass} $M_{\rm c}$ (left panels),  
                 \textcolor{black}{and for all models with changing} $\beta$-ratio (right panels). The green lines show the 
                 power-law fits, respectively, and the red zone represents the  
                 errors to the fits. 
                 }      
        \label{fig:masses}  
\end{figure*}

\subsection{Gravitational collapse and disc fragmentation}
\label{sect:grav_frag}

Fig.~\ref{fig:rates} reports the collection of accretion rate histories onto the MYSOs that we obtained 
in this parameter study. The accretion rates (thin blue line, in $\rm M_{\odot}\, \rm yr^{-1}$) \textcolor{black}{are displayed}  
starting from the \textcolor{black}{early simulation time}, when the gravitational collapse 
is initiated, \textcolor{black}{up to the moment we stop the simulations, i.e. as soon as the protostellar 
mass has reached $M_{\star} = M_{\rm  c}/3$} for the runs with changing 
$M_{\rm c}$ and to $60\, \rm kyr$ for the runs with changing $\beta$, respectively. 
%
%
The rates \textcolor{black}{with different pre-stellar core mass} are in Fig.~\ref{fig:rates}a-h and the 
rates with changing $\beta$ are displayed in Fig.~\ref{fig:rates}i-t. 
In the rest of this paper, we will refer to these series of simulations as the "line of increasing 
$M_{\rm c}$" and the "line of increasing $\beta$", respectively, see also~\citet{vorobyov_apj_723_2010,vorobyov_apj_729_2011}. 
The last series of models with $M_{\rm c}=60\, \rm M_{\odot}$ and $\beta\le1\%$ explores 
the effect of lower initial spin \textcolor{black}{of the core onto} the formation of 
high-mass stars (Fig.~\ref{fig:rates}u-x).

After the very initial \textcolor{black}{rise of $\dot{M}$} during the free-fall collapse 
of the \textcolor{black}{parent molecular core, the protostar ceases accreting envelope material as the gas lands on a 
centrifugally-balanced disc, while it starts acquiring its mass by accretion of disc} material (Fig.~\ref{fig:rates}). 
The accretion rate shows variability once the disc has formed since it mirrors the anisotropies of 
the accretion flow~\citep{meyer_mnras_473_2018}. They are caused 
by the development of \textcolor{black}{dense spiral arms and clumps in the disc produced} by efficient gravitational fragmentation. 
These variations in the accretion rate continue after the disc formation and they are interspersed by violent 
accretion spikes of increasing occurrence as the disc growths~\citep{meyer_mnras_482_2019}. These strong bursts 
are \textcolor{black}{repeatedly produced by the quick inward-migration of dense fragments in the disc, themselves 
formed by gravitational fragmentation and} generating accretion-driven outbursts~\citep{meyer_mnras_464_2017}.

The Toomre parameter estimates the disc gravitational instability by \textcolor{black}{evaluating the respective effects 
of gas self-gravity versus that} of stabilizing disc thermal pressure and rotational shear induced by Keplerian 
rotation~\citep{toomre_apj_138_1963}. 
Hence, \textcolor{black}{the condition for Toomre--instability} is, 
\begin{equation}
    Q = \frac{ \kappa c_{\rm s} }{ \pi G \Sigma } \le Q_{\rm crit},
   \label{eq:toomre}    
\end{equation}
where $c_{\rm s}$ is the sound speed of the gas, $\Sigma$ the column mass density of the disc and 
$\kappa$ the local epicyclic frequency~\citep{durissen_prpl_607_2007}. Fragmentation of spiral 
arms into compact gaseous clumps may develop if $Q_{\rm crit}\le 1$, although recent studies derived 
$Q_{\rm crit} < 0.6$, see the study of~\citet{meyer_mnras_473_2018}. 
\textcolor{black}{
Q-unstable discs are} made of dense regions representing spiral arms, 
which are more prone to fragmentation~\citep{klassen_apj_823_2016}.

The exact nature of disc fragmentation is nevertheless a problem which complexity can not be reduced 
to the sole Toomre criterion. Let us review other criteria for the sake of completeness. 
\textcolor{black}{
The comparison between the local effects of disc thermodynamics regarding to the rotation-induced shear} 
is known as the co-called Gammie-criterion that reads \citep{meyer_mnras_473_2018} 
\textcolor{black}{
\begin{equation}
     \beta= t_{\rm cool} \Omega_{\rm K} \le \frac{2\pi}{\sqrt{1-Q^{2}}},
   \label{eq:gammie}    
\end{equation}
with the Keplerian frequency,
\begin{equation}
  \Omega_{\rm K} = \sqrt{ \frac{GM_{\star}}{r^{3}} },
  \label{eq:keplerian} 
\end{equation}
}
with $t_{\rm cool}$ the local cooling time-scale~\citep{2001ApJ...553..174G}. 
In the precedent papers of 
this series, we shown that this criterion,  
is satisfied in the \textcolor{black}{warm spiral arms location as well as in the 
blobs}, however it is not sufficient to characterise 
fragmentation as the interarm regions were  $\beta$-unstable~\citep{meyer_mnras_473_2018}. 
\textcolor{black}{Note that the Gammie criterion is approximate and exclusively 
applies to axisymmetric accretion discs.}
\textcolor{black}{
The last criterion based on the Hill radius measures the capability of spiral arm segments} 
to locally keep on gaining mass to eventually fragment by \textcolor{black}{confronting the consequences of self-gravity 
versus the stellar tidal forces engendering shears in the disc}~\citep{roger_mnras_423_2012}. 
A spiral arm of local cross-section $l$ is \textcolor{black}{therefore subject to instability} if,
\begin{equation}
  \frac{l}{2 R_{\rm Hill}} < 1, 
\end{equation}
with $R_{\rm Hill}$ the so-called Hill radius. Material lying more than 2 $R_{\rm Hill}$ of a given 
region fragment will not feel the gravity of the local dense region but will have its evolution 
governed by the overall disc dynamics. \textcolor{black}{For discs} around \textcolor{black}{high-mass} stars, \textcolor{black}{the  
Hill-radius-based criterion of~\citet{roger_mnras_423_2012}} has been shown to be more consistent 
with numerical simulations~\citep{meyer_mnras_473_2018}.

\subsection{Disc and masses of the MYSOs}
\label{sect:masses}

\textcolor{black}{
Top panels of Fig.~\ref{fig:masses} show the masses $M_{\rm d}$ of the disc} (in $\rm M_{\odot}$) 
\textcolor{black}{versus the protostellar} mass $M_{\star}$ (in $\rm M_{\odot}$) regarding to the 
line of increasing pre-stellar core mass ($\beta=4\%$, panel a) and for the line of increasing 
$\beta$-ratio ($M_{\rm c}=100\, \rm M_{\odot}$, panel b). 
For each model, the stellar mass is calculated \textcolor{black}{as being a proportion of the gas 
mass} leaving the computational domain per unit time through the inner region of the accretion 
disc, i.e., 
\begin{equation}
     M_{\star}(t) = \int_{0}^{t} \dot{M}(t^{'}) dt^{'}, 
\end{equation}
where $t$ denotes the time at which the protostellar mass $M_{\star}=M_{\star}(t)$ is calculated. 
Similarly, the disc mass $M_{\rm d}(t)$ is estimated for each output of the simulation, following 
the method used in~\citet{klassen_apj_823_2016}. 
The disc mass in the Figure is sampled starting from the end of the gravitational collapse and 
each data point is represented by a symbol and the color coding distinguishes the models with 
$M_{\rm c}=60$-$200\, \rm M_{\odot}$ (Fig.~\ref{fig:masses}a) and with $\beta=2$-$25\%$ 
(Fig.~\ref{fig:masses}b), respectively. Each coloured symbol therefore \textcolor{black}{corresponds 
to a MYSO produced} out of a distinct pre-stellar core, characterised by both a specific mass 
and spin. The overplotted solid green lines \textcolor{black}{are fits using a power law of} 
the model data, respectively.

\begin{figure*}
        \centering
        \begin{minipage}[b]{ 0.7\textwidth} 
                \includegraphics[width=1.0\textwidth]{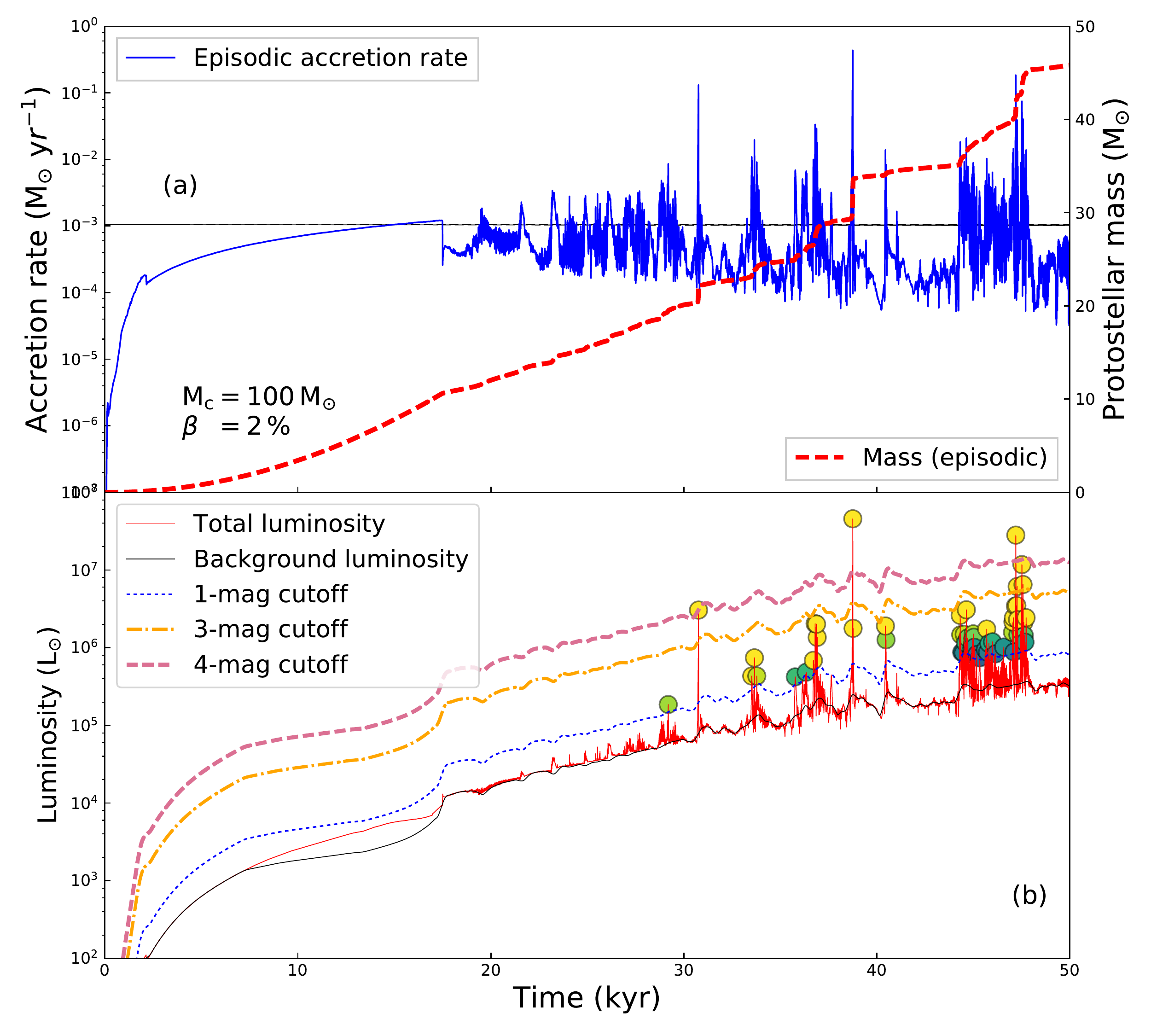}
        \end{minipage}     
        \begin{minipage}[b]{ 0.6\textwidth} 
                \includegraphics[width=1.0\textwidth]{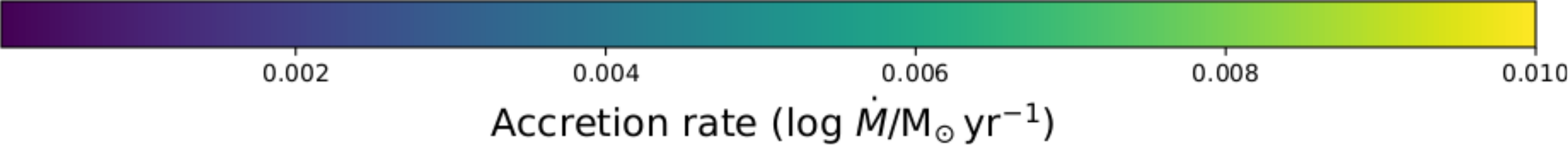}
        \end{minipage}         
        \caption{ 
        \textcolor{black}{
                 Top: Evolution of the accretion rate (in $\rm M_{\odot}\, \rm yr^{-1}$, thin solid line) 
                 and evolution of the mass of the protostar (in $\rm M_{\odot}$, thick dashed line)
                 in our simulation Run-100-2$\%$.
                 The thin horizontal black line marks a rate $\dot{M}=10^{-3}\, \rm M_{\odot}\, \rm yr^{-1}$ 
                 and the magenta dot marks when the protostar enters the high-mass regime 
                 ($M_{\star}=8\, \rm M_{\odot}$). 
        	     Bottom: Total luminosity (thin solid red line, in $\rm L_{\odot}$) of the same model (b), 
        	     background luminosity (thin solid black line), cut-off magnitudes for the 
        	     1-mag ($2.5$ times the background luminosity) to the 4-mag 
        	     ($2.5^{4}\approx39$ times the background luminosity) accretion bursts, respectively. 
        	     }
                 }      
        \label{fig:rate_lum}  
\end{figure*}

The \textcolor{black}{data distribution} in Fig.~\ref{fig:masses} suggests a correlation between $M_{\rm d}$ and 
$M_{\star}$. We perform the \textcolor{black}{least-square regressions} (solid green) and found the 
following relations, 
\textcolor{black}{
\begin{equation}
     \Big( M_\mathrm{d,\beta=4\%} \Big)  =  
     10^{ -0.46  \pm 0.042  }\Big( M_{\star,\beta=4\% }^{1.27 \pm 0.029} \Big),
   \label{eq:mass_fits_1}                  
\end{equation}
and,
\begin{equation}
     \Big( M_\mathrm{d,\frac{M_{\rm c}}{M_{\odot}}=100} \Big)  =  
     10^{ -0.29  \pm 0.063  }\Big( M_{\star,\frac{M_{\rm c}}{M_{\odot}}=100}^{1.17 \pm 0.049} \Big),
   \label{eq:mass_fits_2}                  
\end{equation}
}
where the subscripts $\beta$ and $M_{\rm c}$ stand for the lines of increasing spin pre-stellar core and 
spin, respectively.  
The time sampling of the disc mass history to construct Fig.~\ref{fig:masses}a may also influence the 
data distribution in the $M_{\rm d}$-$M_{\star}$ plane which, in its turn, make the finding of a best 
fit somehow uneasy. 
\textcolor{black}{
However, the power-law fits (solid green line) match fairly well except for $M_{\rm c}\gg 60\, \rm M_{\odot}$, 
meaning that the slope of $\approx 1.27$ is relatively good (see Eq.~\ref{eq:mass_fits_1}). 
}
%
%
In our example, the data for high-mass protostars are not as dispersed as those 
in~\citet{vorobyov_apj_729_2011} because all the considered models have the same 
gravitational-to-kinetic ratio $\beta=4\%$. One can immediately see that the models 
with the heavier pre-stellar cores $M_{\rm c}\ge 100\, \rm M_{\odot}$ are slightly off-set 
with respect to the fit, and that the slope of the overall fit might weaken if more models 
with higher pre-stellar cores $M_{\rm c}\gg 200\, \rm M_{\odot}$ will be considered. 
%

\begin{figure*}
        \centering
        \begin{minipage}[b]{ 0.95\textwidth} 
                \includegraphics[width=1.0\textwidth]{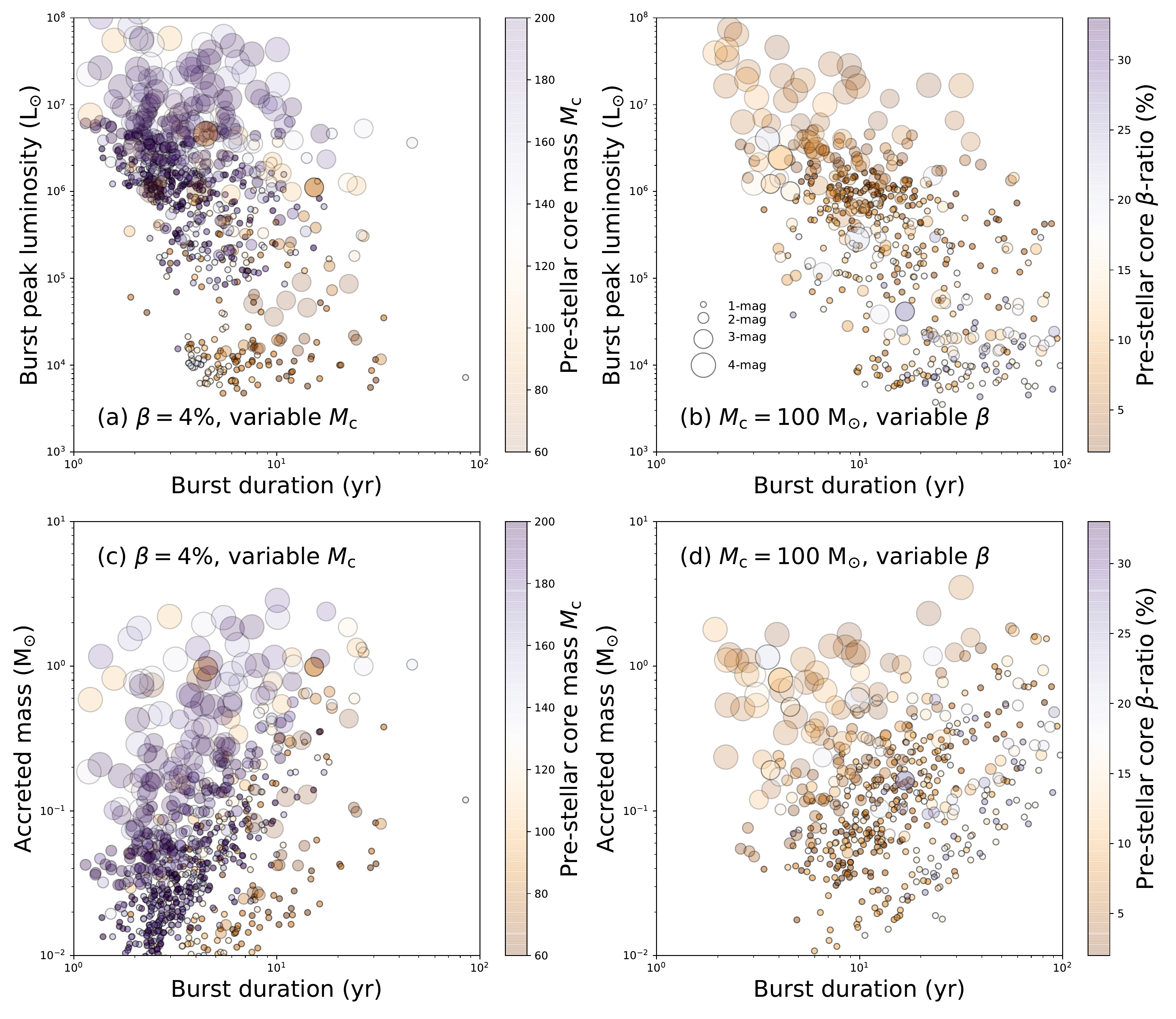}
        \end{minipage}           
        \caption{
                 \textcolor{black}{
                 Scatter plots representing the burst peak luminosity versus duration of the bursts (top panels) 
                 and scatter plots displaying the accretion rate peak versus the duration of the bursts (bottom panels). 
                 }
                 \textcolor{black}{\textcolor{black}{Colour scales distinguish} the models as a function of} the pre-stellar 
                 core mass $M_{\rm c}$ (top panels) and the pre-stellar core $\beta$-ratio (bottom panels).
                 The figures display the data for the line of increasing core mass $M_{\rm c}$ 
                 (left panels) and the line of increasing $\beta$ ratio (right panels), respectively.
                 }      
        \label{fig:burst_prop_spin}  
\end{figure*}

In Fig.~\ref{fig:masses}b we show the $M_{\rm d}$-$M_{\star}$ plane for the line of increasing $\beta$. 
It reveals more scattering of the data compare that for the line of increasing $M_{\rm c}$, meaning that the 
effect of the \textcolor{black}{core spin on the mass of the disc} is more important than that of the core mass. 
\textcolor{black}{
As found by~\citet{vorobyov_apj_729_2011}, \textcolor{black}{faster-rotating, lower-mass} cores 
tend to form heavier discs, which results in scattering in the disc mass distribution. In our case, this 
happens in the $10$--$30\, \rm M_{\odot}$ range. 
}
%
%
The fit might weaken if simulation models with smaller $\beta$-ratio are added. Models with 
initial rotational properties such that $\beta\gg 25\%$, populating the upper part of the 
Figure, are rather unrealistic, despite the fact that such models for massive star 
formation have been produced~\citep{klassen_apj_823_2016}. 
The models that produce high scattering above the fit are also the models in which the burst activity is weakened, 
indicating a smaller fragmentation probability of the accretion discs in them, see  Tables~\ref{tab:A}-\ref{tab:C}.

\subsection{Disc-to-star mass ratio}
\label{sect:mass_ratio}

\textcolor{black}{Bottom panels of Fig.~\ref{fig:masses} show the ratio of the disc-to-star masses}, defined as  
\begin{equation}
     \xi = \frac{ M_{\rm d}  }{  M_\ast }, 
\end{equation}
with $M_{\rm d}$ (in $\rm M_{\odot}$) the above discussed disc mass and $M_{\star}$ (in $\rm M_{\odot}$) 
the protostellar mass, respectively, for both the line of increasing pre-stellar core mass ($\beta=4\%$, 
panel c) and for the line of increasing $\beta$-ratio ($M_{\rm c}=100\, \rm M_{\odot}$, panel d). 
\textcolor{black}{
The disc mass evolution is sampled starting from the end of the gravitational collapse and 
each model is represented by a different symbol and color coding, which helps to distinguish the simulations 
with $M_{\rm c}=60$-$200\, \rm M_{\odot}$ (Fig.~\ref{fig:masses}c) and with $\beta=2$-$25\%$ 
(Fig.~\ref{fig:masses}d), respectively. 
}
Each coloured symbol therefore represents a single protostar which has formed out of a distinct pre-stellar 
core characterised with particular initial conditions \textcolor{black}{of $M_{\rm c}$ and $\beta$-ratios} 
that scan \textcolor{black}{our parameter space for MYSOs. The solid 
green line is a fit of the model data}.

The \textcolor{black}{data distribution} in Fig.~\ref{fig:masses} equivalently suggests a correlation between 
$M_{\rm d}$ and $M_{\star}$. We perform first least-square fits (solid green lines) 
and found that, 
\textcolor{black}{
\begin{equation}
     \Big( \xi_\mathrm{d,\beta=4\%} \Big)  =  
     10^{ -0.46  \pm 0.042  }\Big( \xi_{\star,\beta=4\% }^{0.27 \pm 0.029} \Big),
   \label{eq:mass_fits_3}                  
\end{equation}
and,
\begin{equation}
     \Big( M_\mathrm{d,\frac{M_{\rm c}}{M_{\odot}}=100} \Big)  =  
     10^{ -0.29  \pm 0.063  }\Big( M_{\star,\frac{M_{\rm c}}{M_{\odot}}=100}^{0.17 \pm 0.049} \Big),
   \label{eq:mass_fits_4}                  
\end{equation}
}
where the subscripts $\beta$ and $M_{\rm c}$ stand for the lines of increasing spin pre-stellar core 
and spin, respectively. 
Fig.~\ref{fig:masses}c plots the $\xi$-$M_{\star}$ correlation for the line of increasing $M_{\rm c}$. 
The power-law fits agree well except in the range of $M_{\star}\le15\, \rm M_{\star}$. 
The models with lower $M_{\rm c}$ \textcolor{black}{populate the figure's upper left part}, above the fits, while 
the models with higher $M_{\rm c}$ are located in the lower part of the figure, where more statistics 
might exist. 
No model seems to have $\xi<0.25$ 
and all of them \textcolor{black}{have $\xi>0.5$} as long as the protostellar mass has reached 
$M_{\star}\approx 15\, \rm M_{\odot}$. 
This can be explained by the substential mass gained by the discs around protostars which already 
entered the high-mass region, while the efficiency of mass transport via gravitational torques in 
their surrounding accretion discs is not strong enough to compete with the mass inflowing from the 
still collapsing molecular envelope. 
\textcolor{black}{
Young fragmenting discs with $\xi<0.25$ should therefore be very unusual along both the lines 
of increasing $M_{\rm c}$ and $\beta$ (Fig.~\ref{fig:masses}d).  
}

\begin{figure*}
        \centering
        \begin{minipage}[b]{ 0.7\textwidth} 
                \includegraphics[width=1.0\textwidth]{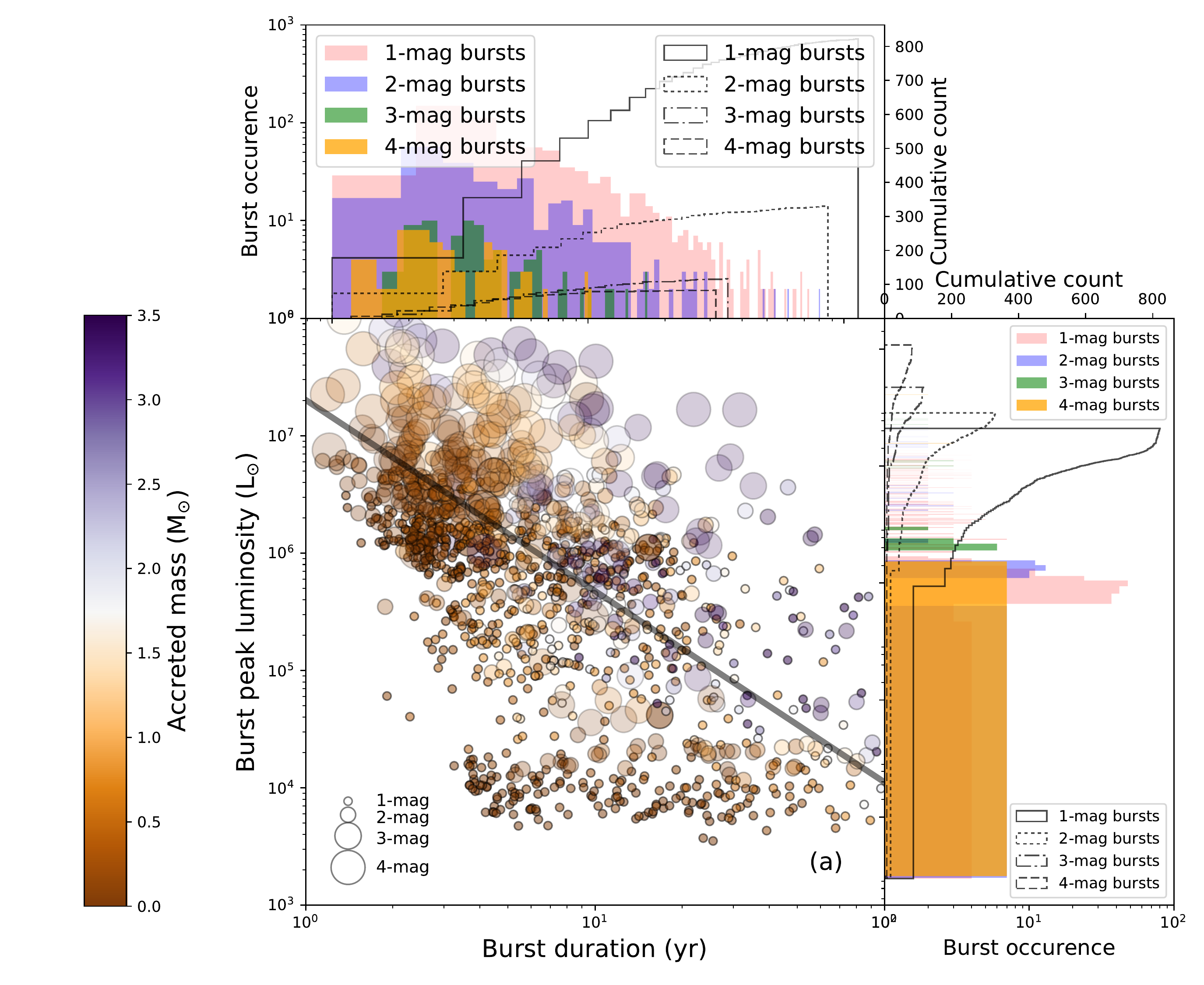}
        \end{minipage}   
        \begin{minipage}[b]{ 0.7\textwidth} 
                \includegraphics[width=1.0\textwidth]{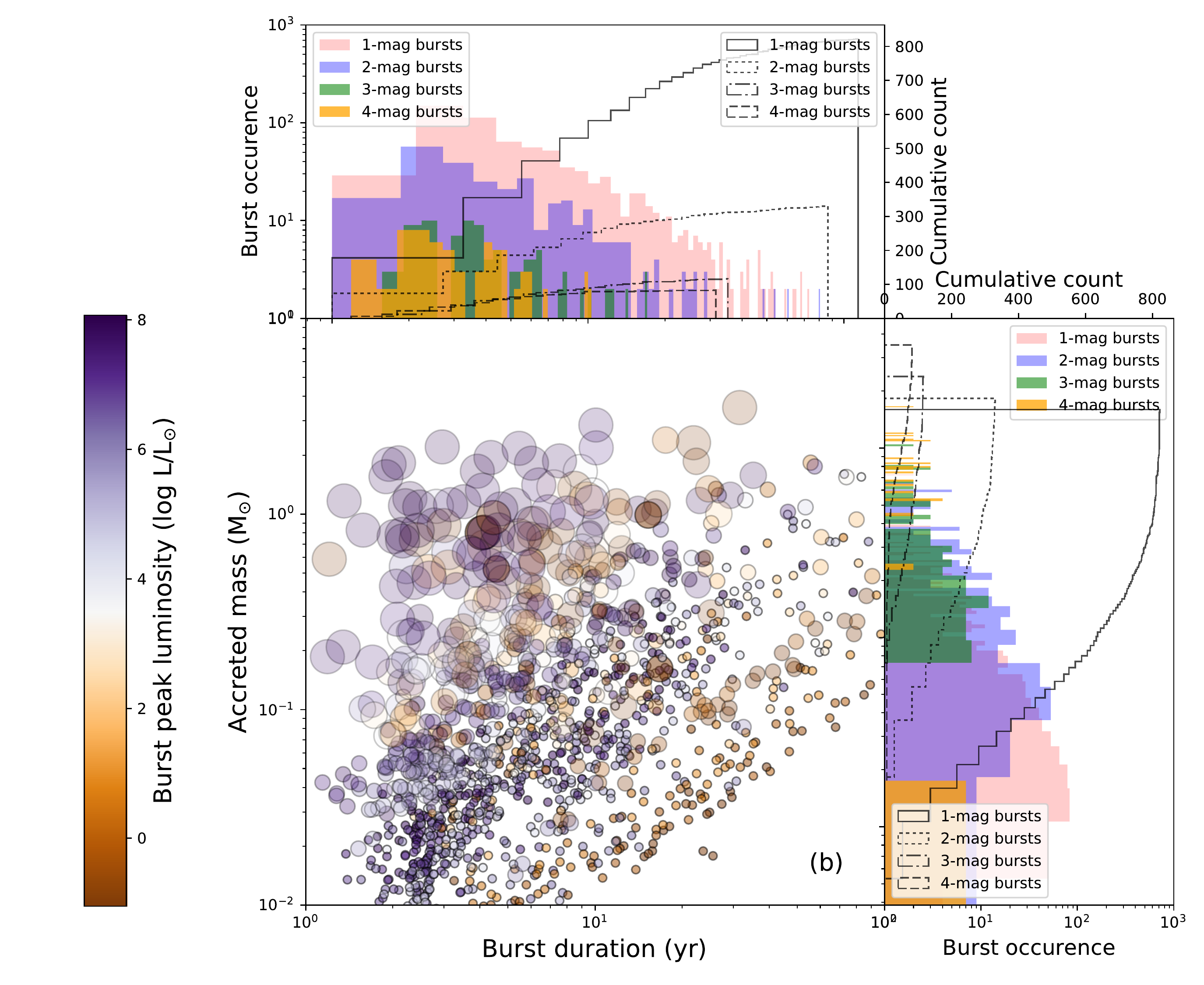}
        \end{minipage}         
        \caption{ 
                 \textcolor{black}{
                 Scatter plots representing the burst peak luminosity as a function of the duration of the burst (top panel) 
                 and scatter plots displaying the accretion rate peak as a function of duration of the bursts (bottom panel) 
                 for all  bursts in our parameter study. 
                 }
                 \textcolor{black}{
                 Colour scales distinguish the data as a function of the mass accreted 
                 by the protostar (top panel) and the bursts peak luminosity (bottom panel).
                 }
        		 \textcolor{black}{
        		 The marginal histograms show the burst occurrence versus the duration of the bursts, 
        		 peak luminosity and mass accreted by the protostar during each individual bursts. 
                 }
                 }
        \label{fig:burst_corr_2}  
\end{figure*}

The distribution of $\xi$ for the line of increasing $\beta$ is obviously more scattered 
than that of the line of increasing $M_{\rm c}$ as a consequence of the dispersion of the $M_{\rm c}$ 
distribution (see Fig.~\ref{fig:masses}) and the fit of the data deviates a lot for 
$M_{\star}\le30\, \rm M_{\odot}$. Slowly-spinning pre-stellar cores will produce 
lighter accretion discs and therefore populate the $\xi<0.5$ region of the figure, while fast-rotating 
core with high $\beta$-ratio will tend to populate the upper left region of the figure in which $\xi>1.0$, 
respectively. 
The increase of $\xi$ is the direct consequence of changes in the mass transport governing mechanism in 
circumstellar discs. When \textcolor{black}{the disc forms after the cloud collapse}, $\xi$ is 
rather low but it quickly increases as in the disc interior no physical process can yet cope with the infalling 
envelope. 
\textcolor{black}{
However, when the disc gains sufficient mass for gravitational instability to occur, the resulting torques 
stimulate protostellar accretion and the mass begins to grow, thus decelerating the initial 
increase in $\xi$. 
}
As soon as inward-migration of dense and heavy clumps is triggered, accompanied by accretion bursts, 
the disc mass is reduced by an equivalent amount of the clump mass gained by the protostar and 
$\xi$ decreases, typically in the $M_{\star}\ge30\, \rm M_{\odot}$ mass range. 
Again, this indicates that the variations of $\beta$-ratio in the initial conditions 
have a stronger effect on $\xi$ than the variations of the core mass $M_{\rm c}$.


\section{Bursts properties}
\label{sect:bursts}

We perform an analysis of the accretion-driven bursts contained in the lightcurves of \textcolor{black}{our MYSOs.} 
The burst properties are investigated \textcolor{black}{according to their parent core properties,} 
and we determine how stars gain their mass, either by quiescent accretion or by accretion-driven bursts.

\subsection{Protostellar luminosities}
\label{sect:lum}

We extract from each disc simulation the protostellar lightcurves and the properties of the corresponding 
accretion bursts. The total luminosity of the protostars,  
\begin{equation}
    L_{\rm tot} = L_{\star} + L_{\rm acc}, 
\end{equation}    
\textcolor{black}{
is calculated being the luminosity $L_{\star}$ of the protostellar photosphere taken 
from~\citet{hosokawa_apj_691_2009}, plus the accretion luminosity,  
}
\begin{equation}
    \textcolor{black}{
    L_{\rm acc} = f G \frac{ M_{\star} \dot{M} }{  R_{\star} }, 
    \label{lacc}
    }
\end{equation}    
\textcolor{black}{
where $M_{\star}$ is the mass of the MYSOs, $G$ is the universal gravitational constant, $\dot{M}$ denotes the 
protostellar mass accretion rate from the disc, and $R_{\star}$ is the protostellar radius. 
In Eq.~(\ref{lacc}) the coefficient} $f=3/4$ stands for the proportion of mass that is considered as being accreted 
by the star as compared to that going in a protostellar jet/outflow~\citep{meyer_mnras_482_2019}. 
\textcolor{black}{
Fig.~\ref{fig:rate_lum} illustrates how the mass transport from the accretion disc to the protostellar surface 
affects the variations of the lightcurve. 
}

\begin{figure*}
        \centering
        \begin{minipage}[b]{ 0.75\textwidth} 
                \includegraphics[width=1.0\textwidth]{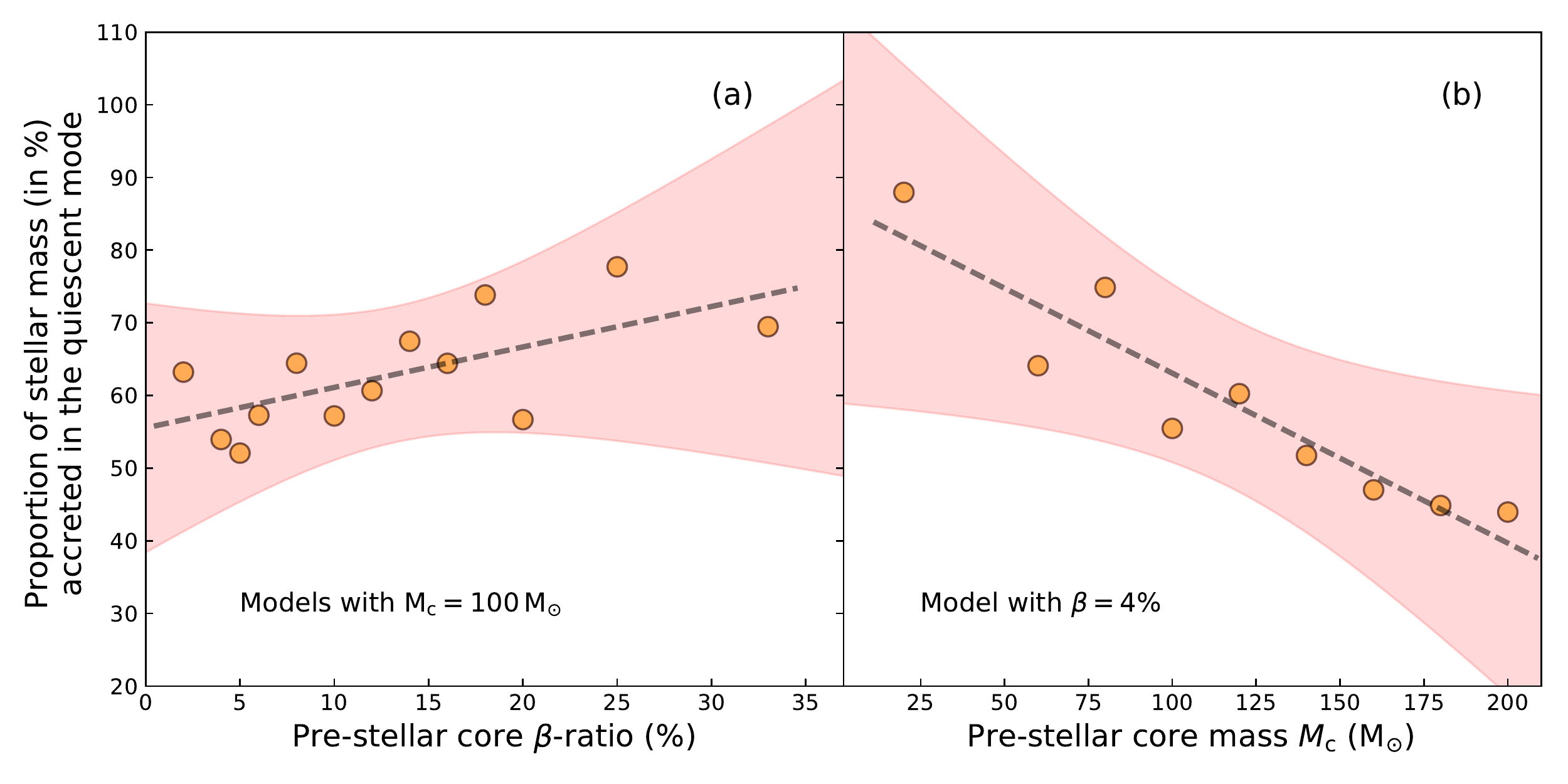}
        \end{minipage}           
        \caption{ 
                 Proportion of final protostellar mass gained during the quiescent phases 
                 of accretion for our models along the line of increasing core mass $M_{\rm c}$ 
                 (a, left panel) and for the line of increasing $\beta$ ratio (b, middle panel), respectively. 
                 Each orange dot represents a simulated protostar. 
                 The dashed black lines show the linear fits, 
                 respectively. The red zone represents the error interval to the fits. 
                 }      
        \label{fig:quiescent_plot}  
\end{figure*}

\textcolor{black}{
We analyse the accretion bursts together with their occurrence and characteristics throughout 
the modelled stellar lifetime.}
The method separates the background secular variability, which
accounts for spiral-arms-induced anisotropies \textcolor{black}{formed in the disc}, 
from the episodic accretion events caused by infalling dense gaseous clumps.
\textcolor{black}{
We first define $L_{\rm bg}$, the so-called background luminosity, which is calculated by filtering out all accretion bursts. It reads
}  
\begin{equation}
   L_\mathrm{bg}(t) = %
   \begin{cases}
     \Big \langle  L_{\star}(t) + L_{\rm acc}(t)           \Big \rangle  & \text{if $\dot{M} \le \dot{M}_{\rm crit} $  } \\
     \Big \langle  L_{\star}(t) + \delta L_{\rm acc}(t)  \Big \rangle    & \text{if $\dot{M} >   \dot{M}_{\rm crit} $  }, 
   \end{cases}
   \label{xxxx}
\end{equation} 
where,
\begin{equation}
   \delta = \frac{ \dot{M}_{\rm crit} }{ \dot{M} }, 
\end{equation} 
and with $\dot{M}_{\rm crit} = 5 \times 10^{-4}\, \rm M_{\odot}\, \rm yr^{-1}$,
\textcolor{black}{which replaces strong accretion bursts from $L_{\rm acc}$}. 
\textcolor{black}{
The time averaging in Eq.~(\ref{xxxx}) is $1000\, \rm yr$. 
}
\textcolor{black}{
We then derive the properties for the so-called $i$-mag bursts with $1 \le \mathrm{i} \le 4$, where an $i$-mag outburst is a burst with 
$L_{\rm acc}\ge2.5^{\mathrm{i}} L_{\rm bg}$~\citep{meyer_mnras_482_2019}. 
}
\textcolor{black}{
Our algorithm selecting the bursts makes sure that very mild luminosity variations smaller
than 1-mag, potentially originating from boundary effects, are not qualified as physical accretion 
bursts and that the duration of the bursts is sufficiently short that any secular variations of $L_{\rm tot}$ are not confused with an outburst.
}
All bursts and their properties are displayed as Appendix in our Tables~\ref{tab:A},~\ref{tab:B} 
and~\ref{tab:C}, respectively.

\subsection{Bursts properties}
\label{sect:burst_prop}

In Fig.~\ref{fig:burst_prop_spin} we display the correlation between the \textcolor{black}{maximum 
luminosity of the accretion bursts} (in $\rm L_{\odot}$) versus the burst duration (in $\rm yr$) (top panels) 
and the bursts peak accretion rate (in $\rm M_{\odot}\, \rm yr^{-1}$) versus the 
burst duration (in $\rm yr$) (bottom panels), \textcolor{black}{where the colour-coding} 
representing the pre-stellar core mass $M_{\rm c}$ (in $\rm M_{\odot}$) (top panels) and its 
corresponding $\beta$-ratio (in $\%$) (bottom panels), respectively. The panels display 
the data for the line of increasing core mass $M_{\rm c}$ (left panels) and the line of increasing $\beta$ 
ratio (right panels), respectively. The numbers and detailed properties of those burst are reported in the 
Tables in the Appendix.

The meaning of this figure is described in great details in~\citet{meyer_mnras_482_2019}. 
Fig.~\ref{fig:burst_prop_spin}a shows that along the line of increasing $M_{\rm c}$, 
the burst peak luminosity augments with $M_{\rm c}$. \textcolor{black}{The burst magnitude 
augments \textcolor{black}{as a function of} the burst luminosity}, except for the 4-mag 
bursts that are more dispersed in the figure (Table~\ref{tab:B}). 
Fig.~\ref{fig:burst_prop_spin}b illustrates that the most luminous flares are typically 
short-duration 3-mag and 4-mag bursts. These bright outbursts  
are generally shorter and more luminous in models with lower $\beta$-ratio than in models 
with higher $\beta$. 
The effect of the \textcolor{black}{increase} of pre-stellar core $M_{\rm c}$ 
results in a concentration of the bursts in the small duration-high luminosity part of the 
diagram, except for the 1-mag bursts which do not accrete much mass (Fig.~\ref{fig:burst_prop_spin}a). 
The effect of the \textcolor{black}{increase} of pre-stellar $\beta$-ratio results in 
the shift of the burst distribution to the region of longer bursts (Fig.~\ref{fig:burst_prop_spin}b). 
Fig.~\ref{fig:burst_prop_spin}c indicates that the 1-mag and 2-mag bursts accrete less mass 
by bursts than the 3-mag and 4-mag bursts. The most massive cores generate the shortest 
and least accreting bursts, while the lightest cores produce longest bursts. 
Fig.~\ref{fig:burst_prop_spin}d demonstrates that models with lower $\beta$ accrete 
more mass and generate more 3-mag and 4-mag burst than in the simulations with higher 
initial \textcolor{black}{core spin}.

\begin{table}
	\centering
	\caption{
	\textcolor{black}{
	Proportion of mass gained by the MYSOs in the quiescent phase of accretion (in $\%$). 
	The results are shown for the line of increasing $\beta$ and $\mathrm{M}_{\rm c}$, 
	respectively, as well as for all models together. 
	}
	}
	\begin{tabular}{lccr}
	\hline
	${\rm {Models}}$                                    & $\mathrm{min}$ ($\%$)   &  $\mathrm{mean}$ ($\%$)   &  $\mathrm{max}$ ($\%$)  \\ 
	\hline    
	{\rm Line of increasing $\beta$}                    & 52.07                   &  62.95                         &  77.71        \\  
	{\rm Line of increasing $\mathrm{M}_{\rm c}$}       & 43.96                   &  58.91                         &  87.95        \\  
	{\rm All models}                                    & 43.96                   &  61.30                         &  87.95      \\  	
        \hline
	\end{tabular}
\label{tab:models_prop}\\
\end{table}

\textcolor{black}{
In Fig.~\ref{fig:burst_corr_2} we display how the burst duration 
(in $\rm yr$) versus the peak luminosity of the bursts} (in $\rm L_{\odot}$) scatters 
(top panel) 
and the \textcolor{black}{duration of the bursts} (in $\rm yr$) versus their \textcolor{black}{maximum} 
accretion rate (in $\rm M_{\odot}\, \rm yr^{-1}$) 
(bottom panel) for each individual bursts without distinguishing the models with changing 
$M_{\rm c}$ and changing $\beta$. The colours \textcolor{black}{indicate the mass which has been transferred 
from the disc to the protostar through the} bursts (in $\rm M_{\odot}$) (top panel) 
and the peak luminosity reached during the bursts (in $\rm M_{\odot}$) (bottom panel), respectively. 
The bursts almost populate the entire Fig.~\ref{fig:burst_corr_2}a. 
The low-luminosity 1-mag bursts are typically in the $10^{3}$-$10^{6}\, \rm L_{\odot}$ 
region and they are characterized by a wide range of duration ($1$-$10^{2}\, \rm yr$). 
The bursts that accrete \textcolor{black}{less} mass are the 1-mag dimmer ones. 
The bursts accreting the larger amount of mass are mostly 3- and 4-mag bursts 
and they are distributed decreasing with the burst duration time in the upper 
region of the figure with respect to the fit of $L(t)$. 
A similar trend is visible in Fig.~\ref{fig:burst_corr_2}b, in which the more 
luminous bursts of 3- and 4-mag accrete more mass than the bursts producing 
dimmer 1- and 2-mag accretion-bursts. 
The luminous bursts are generally of shorter duration as compared to the fainter 
bursts which last longer.  
\textcolor{black}{ 
The bursts of similar magnitude and peak luminosity in Fig.~\ref{fig:burst_corr_2}b 
are distributed along diagonals (from bottom-left to upper-right), which reflects the fact that the accretion rate 
$\dot{M}$ varies slowly during a given burst. Therefore, if the burst duration 
augments, the accreted mass also increases linearly, and the bursts of similar 
luminosities appear as parallel diagonal lines in the mass-duration plane. 
}

\begin{figure*}
         \centering
         \begin{minipage}[b]{ 0.9\textwidth} 
                \includegraphics[width=1.0\textwidth]{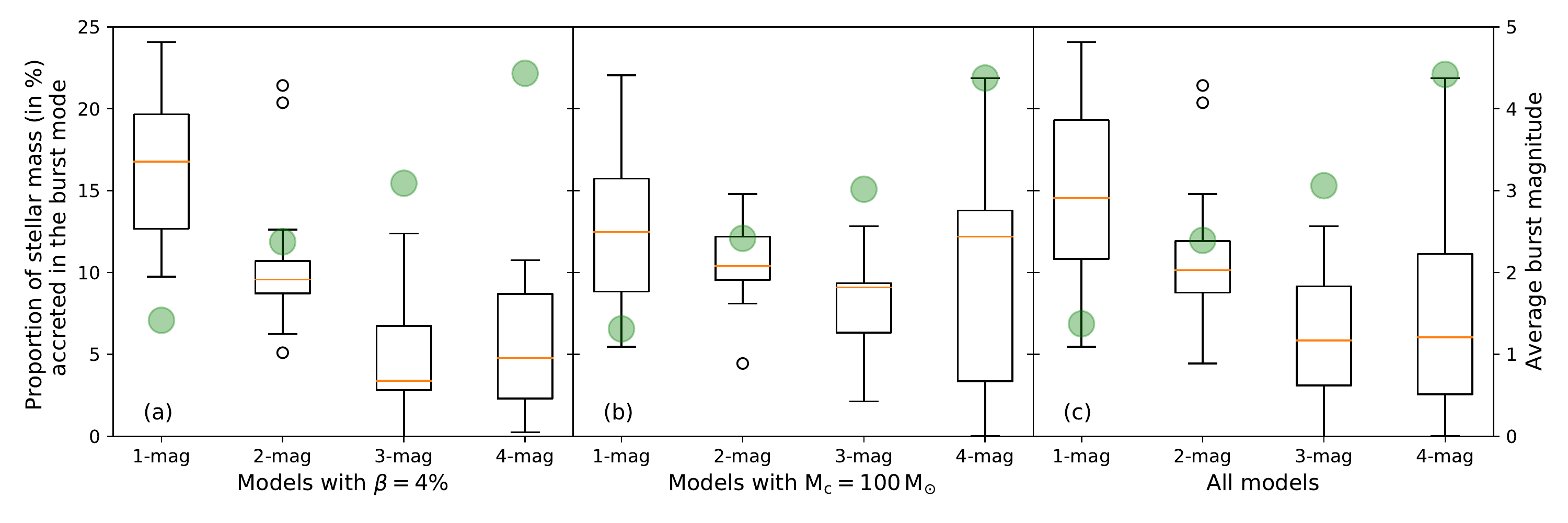}
         \end{minipage}  
         \caption{ 
         Box plot of the proportion of final protostellar mass accreted during the burst phases for our 
         grid of simulated protostars. The results are displayed as a function of the burst magnitude, which 
         can span from $2.5$ (1 magnitude bursts) to $2.5^{4}$ (4 magnitudes bursts) above the protostellar 
         background luminosity. 
         The orange line marks the mean value of a given series of bursts. 
		 The figures show the data for the line of increasing core mass $M_{\rm c}$ (a, left panel), 
		 the line of increasing $\beta$ ratio (b, middle panel) and for all data (c, right panel), respectively. 
		 \textcolor{black}{
		 The green dots indicate the average burst magnitude for each model. 
		 }
                 }      
         \label{fig:box_plot}  
\end{figure*}

\textcolor{black}{
One can note that the bursts with the largest accreted mass and shortest duration are also 
not necessarily the very most luminous ones (top left part of Fig.~\ref{fig:burst_corr_2}b). 
The total luminosity that we plot here reflects the variations of both the photospheric 
luminosity and the accretion luminosity, the latter being function of the accretion rate 
onto the protostar and of the inverse of the stellar radius~\citep{2019MNRAS.484.2482M}.
The protostars accreting the largest amount of mass consequently see their radius bloating 
while going to the red part of the Hertzsprung-Russell diagram. Consequently, even tough 
they accrete the largest mass and generate 4-mag bursts, they are fainter than some other bursts. 
}
\textcolor{black}{
The energy in the bloated atmosphere is then radiated away while the protostar returns to the 
quiescent phase of accretion and continues its pre-ZAMS evolution towards the main-sequence. 
It strongly impacts the nature of the ionizing flux released in the 
cavity that is normal to the disc plane (see also discussion Section~\ref{sect:discussion}). 
The proper timescale of this phenomenon is difficult to predict without self-consistent stellar evolution 
calculations, which time-dependently account for the physics of accretion, such as  
the {\sc genec}~\citep{haemmerle_phd_2014,haemmerle_585_aa_2016,haemmerle_aa_602_2017} or 
the {\sc stellar}~\citep{yorke_aa_54_1977,hosokawa_apj_691_2009,hosokawa_apj_721_2010} codes. 
Only then the structure and upper layer thermodynamics of the MYSOs can be calculated. 
Our stellar evolution calculations previously performed with Run-100-4$\%$ showed that when 
experiencing a 4-mag burst, MYSOs experience a sudden rise of their luminosity that is 
triggered by the brutal increase of the accretion rate at the moment of a disc clump 
accretion~\citep{2019MNRAS.484.2482M}. This induces the formation of an upper convective 
layer, provoking a luminosity wave propagating outwards~\citep{larson_mnras_157_1972}, and 
causes the swelling of the protostellar radius~\citep{hosokawa_apj_721_2010}. 
}

\begin{figure*}
         \centering
         \begin{minipage}[b]{ 0.9\textwidth} 
                \includegraphics[width=1.0\textwidth]{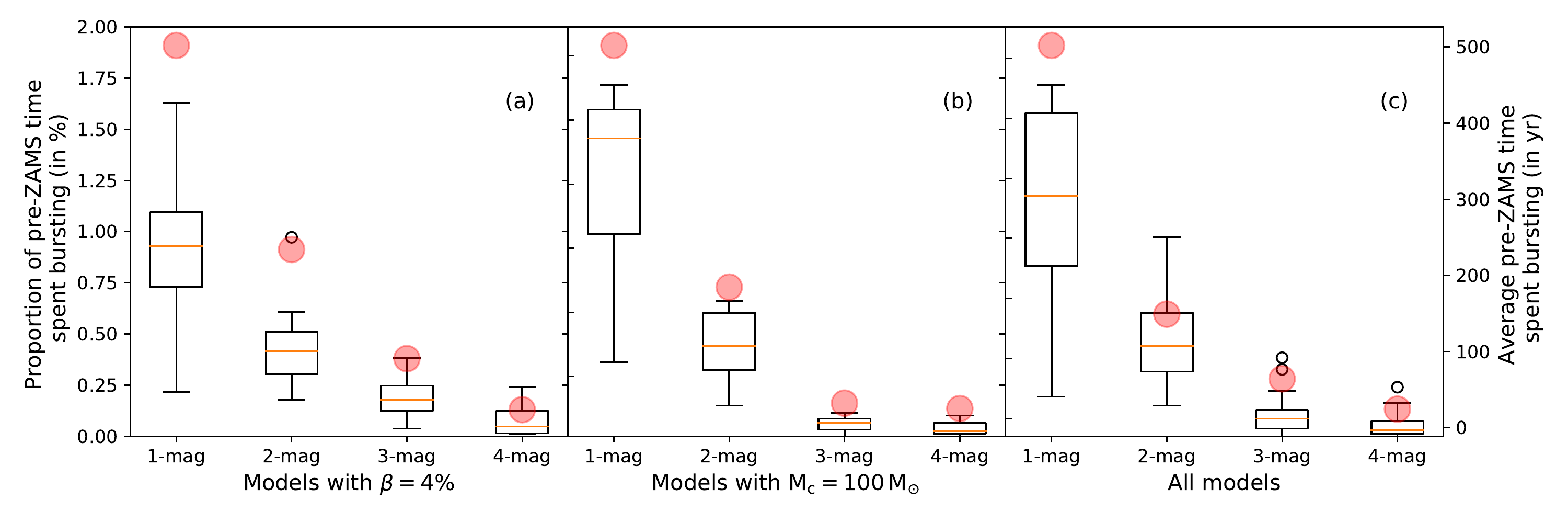}
         \end{minipage}  
         \caption{ 
		 \textcolor{black}{         
         Box plot of the proportion of pre-zero-age-main-sequence-time (pre-ZAMS) the protostellar 
         spend in the burst mode of accretion. 
         The results are displayed as a function of the burst magnitude, which 
         can span from $2.5$ (1 magnitude bursts) to $2.5^{4}$ (4 magnitudes bursts) above the protostellar 
         background luminosity. 
         The orange line marks the mean value of a given series of bursts. 
		 The figure shows the data for the line of increasing core mass $M_{\rm c}$ (a, left panel), 
		 the line of increasing $\beta$ ratio (b, middle panel) and for all data (c, right panel), respectively. 
		 The red dots indicate the average time the protostars spent in the burst phase (in $\rm yr$). 
		 }
                 }      
         \label{fig:box_plot_time}  
\end{figure*}

\textcolor{black}{
Our models for MYSOs show that this swelling lasts on the order of $~100$$-$$1000\, \rm yr$, depending on how 
much mass is accreted during the burst, and the protostellar flare may appear 
as a longer, lower amplitude burst. 
Furthermore, the situation is even more complex as there 
is no one-to-one correspondence between the accretion rate and total luminosity, as the 
star can act as a capacitor and release part of the accreted energy in a delayed manner. This induces a rise 
in the photospheric luminosity which might dominate the total luminosity in the late outburst stages 
when the MYSO returns to the quiescent phase. The protostellar flare may 
appear as a longer, lower amplitude burst. At least this can happen in the context of low-mass 
protostars, see~\citet{elbakyan_mnras_484_2019}. 
Recent observations show that some masers are very good tracers of the decrease of the radiation 
field, see section 3.3 of~\citet{chen_apj_890_2020} and Fig.~4 of~\citet{chen_natas_2020}, 
which can be interpreted as a clue of the burst duration. The flare of NGC 6334 I is going on 
still and that of S255 lasted for years~\citep{szymczak_aa_617_2018}, and stronger 
flares seem 
to be longer. 
}

This phenomenon is also probably influenced by the spatial resolution of the simulations, 
in the sense that higher resolution models will permit to better follow the collapse 
of the clump interiors, and by the size of the sink cell, inevitably introducing boundary 
effects. Indeed, our burst analysis of a higher-resolution disc model 
in~\citet{meyer_mnras_482_2019} shows that the 4-mag bursts are less frequent than in those 
with lower resolution, although this calculation had been integrated over a more reduced time. 
Further simulations with a much higher spatial resolution are is necessary to address this question 
in more detail.

\begin{figure}
        \centering
        \begin{minipage}[b]{ 0.45\textwidth} 
                \includegraphics[width=1.0\textwidth]{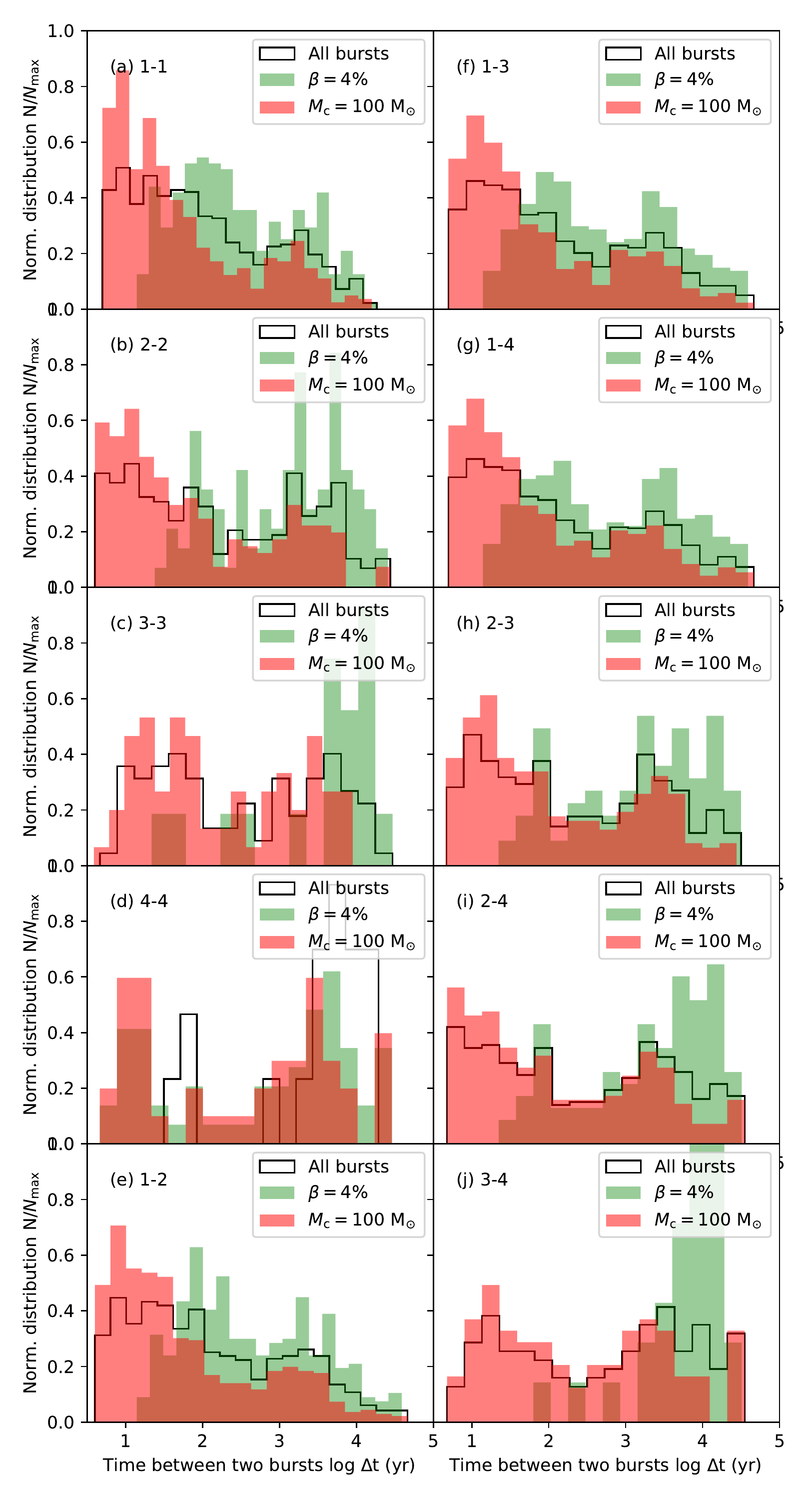}
        \end{minipage}           
        \caption{ 
        Normalised distribution $N/N_{\rm max}$ of the time intervals between two consecutive accretion-driven 
        bursts in our simulations. It is calculated for each possible combination allowed by our 
        1-, 2-, 3- and 4-mag bursts. We present the results for the 1- to 4-mag bursts exclusively 
        (panels a-d) while the panels e-g concerns the distribution of all consecutive bursts of magnitude 1 to 2 
        (1-2), 1 to 3 (1-3), and  1 to 4 (1-4), respectively. The other combinations are plotted in panels h-j. 
        All time intervals (in $\rm yr$) are plotted in the logarithmic scale. 
                 }      
        \label{fig:histograms}  
\end{figure}

The marginal histograms on the right and top sides of Fig.~\ref{fig:burst_corr_2}a,b 
concern the bursts occurrence of the whole set of bursts experienced by all our protostars. 
The data are plotted with different colours depending on the burst magnitude, while 
the black lines show their cumulative occurrence.  
The top histograms are the same of both panels (a) and (b) as they equally represent 
the burst duration. Our conclusion confirms that the maximum burst duration is 
below $100\, \rm yr$. 
We confirm that bursts with shorter duration induce stronger bursts, and
hence it will be more unlikely to monitor these events in the context of MYSOs, 
questioning the observability of 4-mag, FU-Orionis-like accretion bursts. 
The mass gained during a burst extends from about 10 Jupiter masses  to a solar mass, which is 
within the limits accreted during the outburst of, e.g., S255IR-NIRS 3~\citep{caratti_nature_2016}. 
As described in~\citet{meyer_mnras_482_2019}, the distribution of accreted masses mirrors the variety 
of disc fragments, i.e., the clumps and dense spiral arm segments generated by gravitational instability 
in the disc.

\textcolor{black}{
We note that strongest accretion bursts may happen alongside with the formation of 
low-mass binary companions to MYSOs. We demonstrated in ~\citet{meyer_mnras_473_2018} 
that it is possible to form simultaneously both close/spectroscopic objects around 
a MYSO, while it simultaneously undergoes an outburst.} 
This happens when migrating massive clumps \textcolor{black}{get rid of their envelope while}  
contracting into a dense nucleus, thus forming a secondary low-mass protostellar core. 
The burst luminosity distribution indicates that \textcolor{black}{1- and 2-mag bursts 
are more common that 3- and 4-mag bursts}. Their luminosity peak 
is at $\approx 10^{5}$–$10^{6}\, \rm L_{\odot}$, while the other, higher-luminosity 
bursts are much more uncommon. \textcolor{black}{Still, there are 3-mag bursts with luminosities 
$\ge 10^{6}\, \rm L_{\odot}$, and a few rare 4-mag} bursts peak 
at luminosities $\ge 10^{7}\, \rm L_{\odot}$, see 
also in~\citet{meyer_mnras_482_2019}.

\subsection{Quiescent versus burst phases of accretion}
\label{sect:accretion}

We calculate for each simulation model the proportion of final protostellar mass that is gained 
either in the quiescent or during the burst phases, respectively. The minimal, mean and maximal 
values for the quiescent phase are reported in our Table~\ref{tab:models_prop} for the lines of 
increasing $\beta$ and $\mathrm{M}_{\rm c}$, as well as for the other simulations' models. 
The models with different $\beta$-ratios indicate that the protostar acquires between $52.07\, \%$ 
and $77.71\, \%$ of their final mass during the quiescent phase, with a mean value of about 
$62.95\, \%$. The rest of the mass is therefore accreted during the time spent in the burst 
mode ($L_{\rm tot}\ge2.5L_{\rm bg}$). 
The simulations with constant $\beta$-ratio of $4\, \%$ but changing protostellar core mass $M_{\rm c}$ 
have a mean value of $58.91\, \%$ with extreme value of $43.96\, \%$ and $87.95\, \%$, respectively.

Fig.~\ref{fig:quiescent_plot} details the proportion of final protostellar mass gained during the 
quiescent phase of accretion, i.e. ignoring all burst phase, for all our models. 
One can see that it gradually increases with $\beta$ from $\approx 55\%$ for the model with 
$\beta=2\%$ to $\approx 70\%$ for the models with $\beta=33\%$ (Fig.~\ref{fig:quiescent_plot}a), 
meaning that less mass is gained in the burst mode in the case of highly-spinning cores. 
Inversely, the model with $M_{\rm c}=20\, \rm M_{\odot}$ spends $87.95\, \%$ of its protostellar 
lifetime in the quiescent phase and such quantity monotonically decreases to the model with 
$M_{\rm c}=200\, \rm M_{\odot}$ that spends half of its pre-main-sequence lifetime, namely 
$43.96\, \%$, in the quiescent mode (see Fig.~\ref{fig:quiescent_plot}b). 
It indicates that our results are more sensitive to \textcolor{black}{$M_{\rm c}$} than to 
its initial spin. The latter governs, for a give radius and \textcolor{black}{core's structure}, 
the \textcolor{black}{duration} of the free-fall gravitational collapse. Hence, the stars 
forming out of lightest pre-stellar cores are more prone to gain mass by quiescent disc accretion 
than by accretion-driven bursts, whereas the heaviest pre-stellar cores spend a larger fraction 
of their pre-zero-age-main-sequence in the burst phase.

In Fig.~\ref{fig:box_plot} we show the box plots of the fraction of the final protostellar mass accreted 
during the burst phase for the 1-mag ($L_{\rm tot}\ge2.5L_{\rm bg}$) to 
4-mag bursts ($L_{\rm tot}\ge2.5^{4}L_{\rm bg}$). The figures display the data for the line of increasing 
core mass $M_{\rm c}$ (a, left panel), the line of increasing $\beta$ ratio (b, middle panel) and for 
all data (c, right panel), respectively.
\textcolor{black}{
For each burst samples, i.e. the lines of increasing $M_{\rm c}$ (a), increasing $\beta$ (b), 
or both (c), we draw a box extending from the lower/first quartile $Q_{\rm L}$ (i.e. \textcolor{black}{the data 
lower half's median}) to upper quartile $Q_{\rm U}$ (i.e. \textcolor{black}{the data upper half's median}) 
of the considered sample, with an orange line at the median of all the data. 
With $\mathrm{IQR}=Q_{\rm U}-Q_{\rm L}$ being the interquartile range, the box whiskers extend from the box 
to $1.5 \times Q_{\rm U}$ and to $1.5 \times Q_{\rm L}$, respectively. Flying points marked as white 
circles are those past the range \textcolor{black}{[$Q_{\rm L} - 3\mathrm{IQR}/2$ , $Q_{\rm U} + 3\mathrm{IQR}/2$]}. 
Hence, the extend of the whiskers marks the dispersion of most bursts, except marginal ones represented 
as circles and laying outside of the whiskers. 
\textcolor{black}{
The green dots in the figure indicate the average magnitude of the bursts for all models. 
The burst magnitude is the exponent $i$ defined as, 
\begin{equation}
   \frac{ L_{\rm tot} }{ L_{\rm bg} } = 2.5^{i}, 
\end{equation} 
which corresponds to
\begin{equation}
   i  =   \frac{1}{\log(  2.5 )}   \log \Big( \frac{ L_{\rm tot} }{ L_{\rm bg} } \Big), 
\end{equation} 
with $L_{\rm tot}$ and $L_{\rm bg}$ the total luminosity and the background luminosity, respectively. 
The arithmetic average is then performed for both the lines of the increasing $\beta$ (Fig.~\ref{fig:box_plot}a), 
and $\rm M_{\rm c}$ (Fig.~\ref{fig:box_plot}b) and for all models (Fig.~\ref{fig:box_plot}c). 
Note that the average 1-mag burst can only be in the (1-2)-mag limit, the 2-mag burst can only 
be in the (2-3)-mag limit, and so forth. 
Interestingly, 
the data exhibit a significant homogeneity (Fig.~\ref{fig:box_plot}a,b) 
meaning that, on the average, the mean burst magnitude is independent of the pre-stellar core properties. 
Our approach is modelling bursts can therefore be compared to observations, see Section~5.4. 
}
}

Concerning the line of increasing $M_{\rm c}$ (Fig.~\ref{fig:box_plot}a), most mass accreted during 
the \textcolor{black}{burst phase} is gained as 1-mag bursts, with a median amount of mass $\approx 17\%$ of 
the final protostellar mass. The amount of material accreted during the 2- and 3-mag bursts 
decreases with median values of $10\%$ and $4\%$, respectively. Finally, for models at constant 
$\beta$-ratio, the mass accumulated during the 4-mag FU-Orionis-like bursts is slightly higher 
than that of the 3-mag burst, however with a larger dispersion than, e.g. the 2-mag bursts. 
The situation is globally similar for the line of increasing $\beta$ (Fig.~\ref{fig:box_plot}b) as the 
median mass accreted by the protostar decreases from $13\%$ and $10\%$ for the 1-, 2- and 3-mag bursts, 
nevertheless the 4-mag bursts behave differently with a mean mass similar to that of the 1-mag bursts, 
but with a huge dispersion spanning from $<5\%$ to $>20\%$. This indicates that the $\beta$-ratio of the 
pre-stellar core affects much more the manner stars gain their mass than the initial core mass. 
Regarding to the whole data set (Fig.~\ref{fig:box_plot}c), a decreasing trend of the accreted 
mass during \textcolor{black}{the accretion phases showing bursts versus the burst} magnitude is found with 
$14\%$, $10\%$ and $6\%$ for the 1-, 2- and 3-mag bursts, respectively, and another $6\%$ for 
the 4-mag burst, the latter being however attached to a huge dispersion of the values produced 
by differences in the models with changing $\beta$.

\textcolor{black}{
In Fig.~\ref{fig:box_plot_time}, we display the statistics for the proportion of 
pre-ZAMS time the MYSOs spend in the burst mode of accretion (in $\%$), together with the 
average time protostars experience accretion phases that are characterised by either 
1-mag, 2-mag, 3-mag or 4-mag bursts, respectively (in $\rm yr$). 
The models with $\beta=4\%$ have a rather large dispersion of the proportion of time they spend in
1-mag bursts, which spread between $0.25\%$ and $1.6\%$ of the calculated time, with a mean 
value around $\approx 1\%$. These values gradually diminish as the burst magnitude augments 
and we find that our MYSOs spend very little ($\le 0.1\%$) of their time experiencing 4-mag 
bursts (Fig.~\ref{fig:box_plot_time}a). 
The same is true for the models with $M_{\rm c}=100\, \rm M_{\odot}$, although the values are 
slightly larger for the 1-mag and the 2-mag bursts (Fig.~\ref{fig:box_plot_time}b), because 
our models with high initial $\beta$-ratio of their molecular pre-stellar core spend more time 
in the burst mode  than those with lower $\beta$-ratio (Table~\ref{tab:B}). 
The statistics for all models (Fig.~\ref{fig:box_plot_time}c) therefore indicates that  
MYSOs spend about $2\%$ of their pre-ZAMS time in the burst mode of accretion. The rare 
events are the fast 4-mag bursts responsible for the excursions in the cold regions of 
the Hertzsprung-Russell diagram~\citep{2019MNRAS.484.2482M}. 
The findings in our parameter study therefore confirm the previously obtained results on 
the basis of a much smaller sample of massive protostars, and which stated that MYOs spend 
about $1.7\%$ of their early formation phase in the burst mode of 
accretion~\citep{meyer_mnras_482_2019}. 
}


\section{Discussion}
\label{sect:discussion}

\textcolor{black}{
This section presents different caveats in our method, further discuss the results in the light 
of known 
} 
young high-mass stars which experienced an outburst, and compare 
\textcolor{black}{
them their low-mass counterparts. 
} 
Finally, we consider our outcomes by discussing them in the context of the temporal variabilities of 
massive protostellar jets.

\subsection{Limitation of the model}
\label{sect:model}

Our parameter study is based on numerical models underlying assumptions regarding to the numerical methods. 
The simplifications have already been thoroughly discussed in our pilot paper~\citet{meyer_mnras_464_2017}.
\textcolor{black}{
Particularly, we demonstrate therein that the discs in our simulations are adequately resolved, 
by comparing the Truelove criterion, i.e. the minimal inverse Jeans number, in the mid-plane 
of the accretion disc as a function 
of radius for three different grid resolutions. Note that the model Run-100-4$\%$ in our study 
is the Run-1 of~\citet{meyer_mnras_473_2018}, see their Figs.~11 and~12. 
}
They limitations principally concern the spatial resolution of the computational grid and the consideration 
of additional physical processes such as magnetic fields and associated non-ideal effects in the numerical 
simulations. 
\textcolor{black}{
Photoionization is neglected in our scheme because we concentrate on studying the 
accretion disc, not the bipolar \textcolor{black}{lobes filled with ionising radiation, which 
develop} perpendicular to it~\citep{yorke_aa_108_1982,rosen_mnras_463_2016}. 
}
That is why our computational mesh has a cosine-like grid along the polar direction, degrading  
the resolution of the protostellar cavity. 
Consequently, omitting photoionization in the scheme does not \textcolor{black}{drastically} 
change the \textcolor{black}{outer disc physics ($\sim 100-1000\, \rm au$)} that we concentrate 
on, while resulting in a substantial speed-up of the code.

\begin{figure}
        \centering
        \begin{minipage}[b]{ 0.45\textwidth} 
                \includegraphics[width=1.0\textwidth]{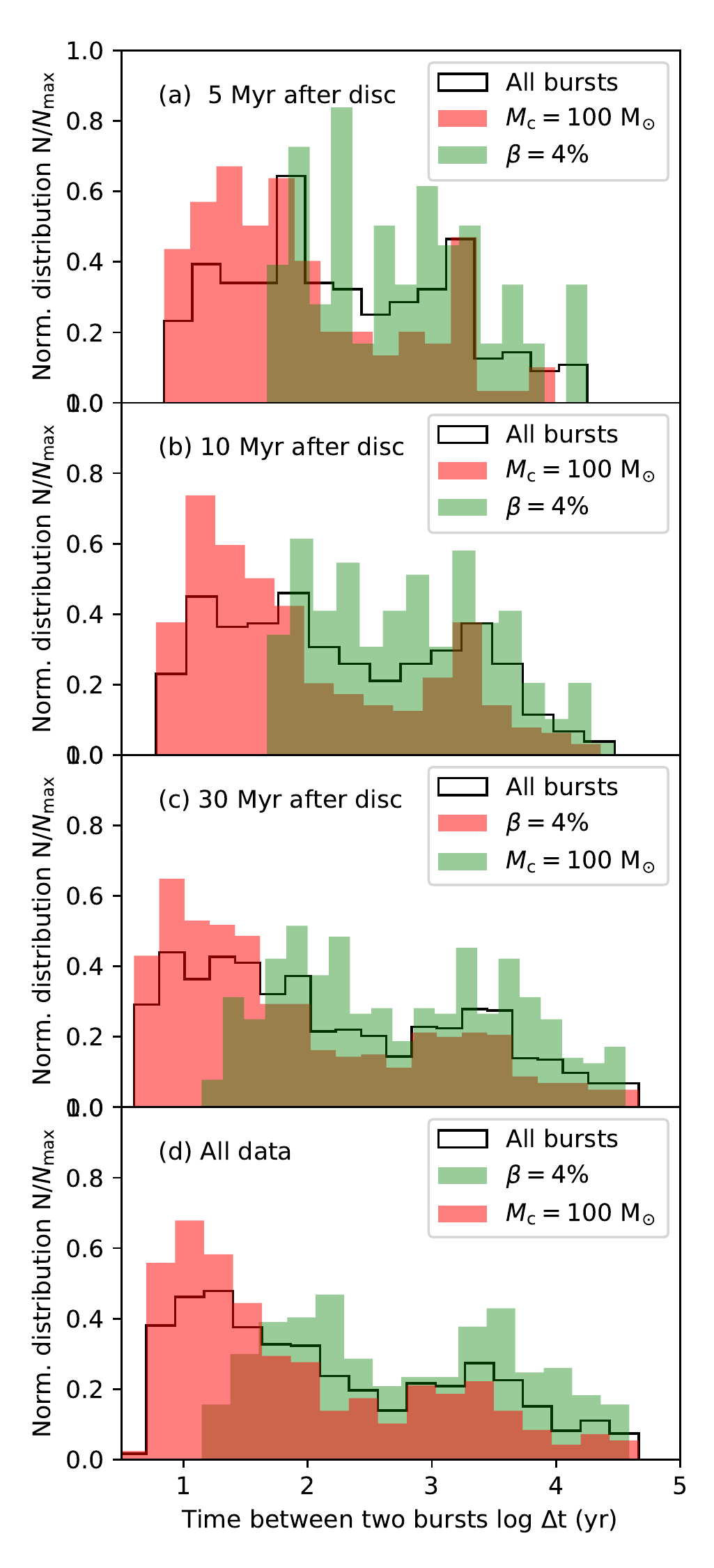}
        \end{minipage}           
        \caption{
        \textcolor{black}{
        Time evolution of the 
        normalised distribution $N/N_{\rm max}$ of the time intervals between two consecutive accretion-driven 
        bursts in our simulations. It is calculated for each possible combination allowed by our 
        1-, 2-, 3- and 4-mag bursts. The distribution is shown at times $5\, \rm kyr$ (a), $10\, \rm kyr$ (d), 
        and $30\, \rm kyr$ (c) \textcolor{black}{once the disc has formed} and for all bursts of all models (d). 
        All time intervals (in $\rm yr$) are plotted in the logarithmic scale. 
        }
        }      
        \label{fig:histograms_evolution}  
\end{figure}

\textcolor{black}{
Nevertheless, this physical mechanism not only governs the ionizing flux evacuated in the outflow lobes, 
but also impacts the structure of accretion discs by photoevaporation~\citep{hollenbach_apj_428_1994,mckee_apj_681_2008}. 
The stellar feedback (Lyman continuum, X-ray and ultra-violet photons) irradiating the circumstellar 
medium can cause the ionization of the gas at the disc surface, thus leading to its evaporation as a 
steady flow into the ISM. Without that, the accretion flow at the disc truncation radius stops 
shielding the neutral disc material. 
It launches so-called irradiated disc winds which host complex chemical reactions between 
enriched species and dust particles present in the disc. 
This particularly happens in the late phases of disc evolution, e.g. at the T-Tauri phases or even 
late, when giant planets have formed and orbit inside of it, 
see~\citet{ercolano_mnras_460_2016,weber_mnras_496_2020,franz_aa_635_2020} and references therein. 
One should not expect photionisation to destroy the entire discs or even to affect the development of 
gravitational instability in the discs~\citep{yorke_aa_315_1996,richling_aa_327_1997,richling_aa_340_1998,richling_apj_539_2000}, and 
consequently it should not be a determinant factor in the burst mode of accretion in massive star formation. 
The flux of ionizing stellar radiation is a direct function of the protostellar properties, 
themselves depending on the accretion history. As stated above, high accretion rates induce 
bloating of the stellar radius and a decrease of its effective temperature and ionizing 
luminosity, released either in the polar lobes or towards the equatorial plane where the disc lies. 
Consequently, the \hii region generated by the protostar becomes intermittent, with variations 
reflecting the episodic disc accretion history onto the protostellar surface~\citep{hosokawa_2015}. 
}

With the photon flux being switch-off towards the colder part of the Herztsprung-Russell diagram during 
the excursions of these stars undergoing a burst, one should expect 
the \hii regions to disappear when $\dot{M}$ reaches its maximum peak. 
The ionized lobed region then gradually reappears as the star recovers 
pre-ZAMS surface properties corresponding to its quiescent 
phase of accretion, after radiating away the clump entropy during a phase of lower-amplitude burst.  
Such a process has been revealed in the context of primordial, supermassive 
stars~\citet{Hosokawa_2011Sci,hosokawa_apj_760_2012,hosokawa_apj__778_2013}
We postulated that this mechanism of blinking \hii regions should also be at work 
in massive star formation and constitutes a major difference between present-day young low-mass 
and high-mass stars~\citep{2019MNRAS.484.2482M}. 

\begin{figure*}
         \centering
         \begin{minipage}[b]{ 0.9\textwidth} 
                \includegraphics[width=1.0\textwidth]{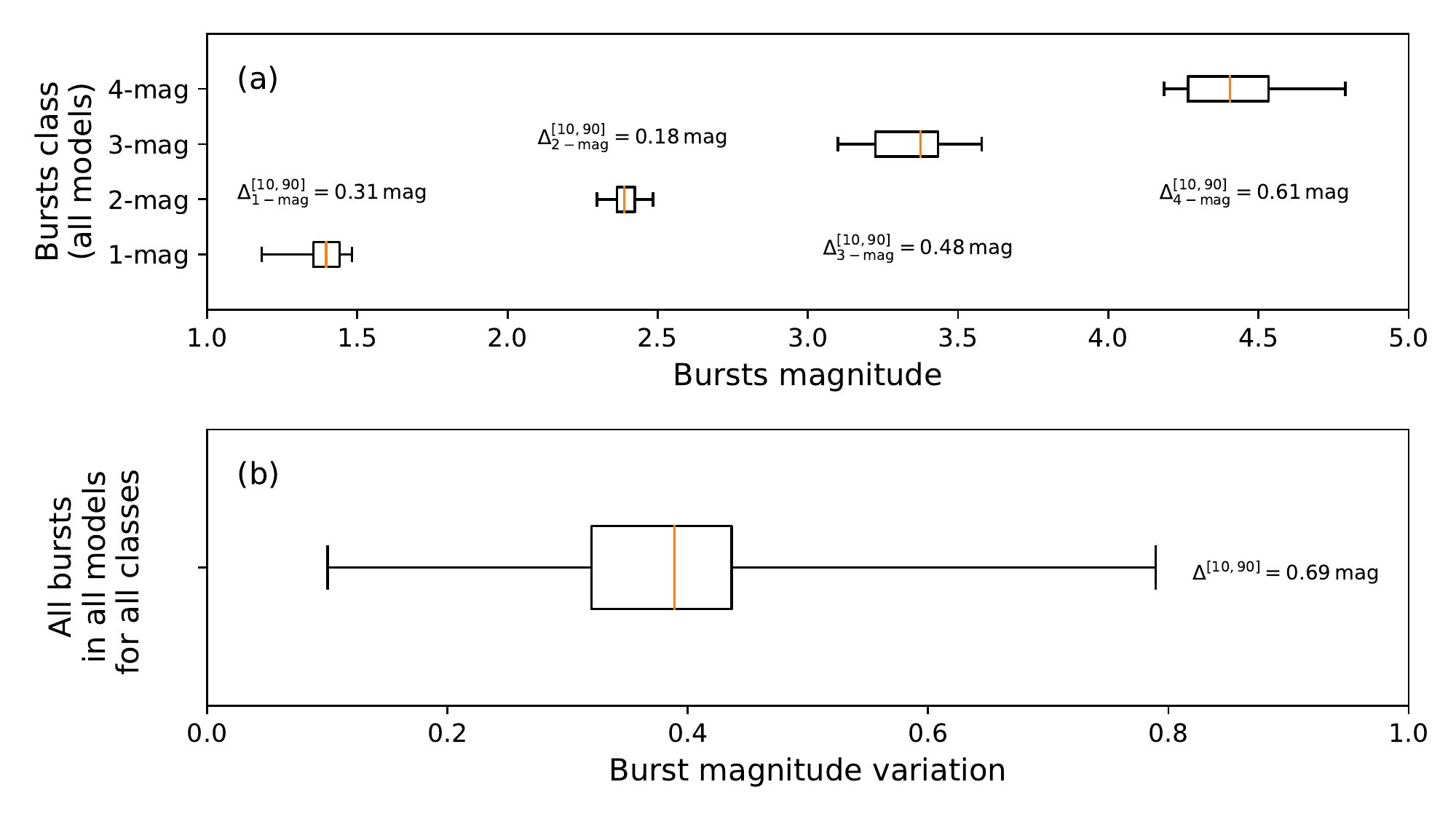}
         \end{minipage}  
         \caption{ 
         Box plot of the burst magnitude distribution as a function of the burst class, for all 
         models in this study. The bursts statistics are displayed per magnitude class (a) and for all 
         bursts of all classes in all our simulations (b). The figure indicates the magnitude difference 
         between the $10$-th and $90$-th percentiles of each burst class. 
         }  
         \label{fig:box_plot_mag}  
\end{figure*}

Future improvements might principally consist of changing the initial conditions in terms of internal structures 
of the pre-stellar core to make it more realistic and of increasing the spatial resolution of the grid simulation, 
so that we can further resolve disc fragmentation \textcolor{black}{when circumstellar 
clumps migrate} in the vicinity of the protostar. 
Indeed, the filamentary nature of the parent pre-stellar cores in which young massive star form should definitely 
affect the manner star gain their mass, however, this will not support the midplane-symmetry which we impose in our 
simulations to divide the computational costs by a factor of $2$. Similarly, a higher spatial resolution 
will permit to investigate trajectories of migrating clumps to the inner disc region. 
Nevertheless, circumventing the caveats of our current models would be at the cost of unaffordable computational
resources which will not permit a scan of the huge star formation parameter space. 
Longer simulations and a smaller sink cell radius $r_{\rm in}$ would also permit us to better 
simulate the fall, evolution, probable distortion and/or segmentation of the clumps before they are accreted 
onto the star. Again, this would in its turn strongly modify the time-step controlling the 
time-marching algorithm of the calculations and therefore the overall cost of the calculations.

\subsection{Time interval between bursts}
\label{sect:jets}

Fig.~\ref{fig:histograms} displays the normalised distribution $N/N_{\rm max}$ of the time intervals 
between two accretion-driven bursts in our simulations, 
\begin{equation}
    \Delta = t_{j+1}^{\{i-\mathrm{mag}\}} - t_{j}^{\{i-\mathrm{mag}\}}, 
    \label{time_deltat}
\end{equation}    
where the subscripts $j$ and $j+1$ designate two consecutive bursts  selected on the basis of their 
magnitude ($i-\mathrm{mag}$, with $1 \le i \le 4$) with respect to $L_{\rm bg}$.
The distribution is calculated for each possible combination allowed by our 1-, 2-, 3- and 4-mag bursts. We present the 
results for the 1- to 4-mag bursts in Fig.~\ref{fig:histograms}a-d, the panels in 
Fig.~\ref{fig:histograms}e-g show the cumulative distribution of all bursts of magnitude 1 to 2 (1-2),
1 to 3 (1-3), and  1 to 4 (1-4), respectively. The other combinations are plotted in Fig.~\ref{fig:histograms}h-j. 
All time intervals are plotted in the logarithmic scale in $\rm yr$.  
In each panel we distinguish the results obtained for the models in the line of increasing 
$M_{\rm c}$ (green colour, $\beta=4\%$) and for the line of increasing $\beta=4\%$ 
(red colour, $M_{\rm c}=100\, \rm M_{\odot}$). The distribution including all bursts are shown 
with a thin black line in each panel. The number of bursts taken into account in the histograms 
decreases from panel (a) to panel (d) as a natural 
consequence of the occurrence of 1- to 4-mag bursts (Tables~\ref{tab:A}-\ref{tab:C}). Panel (g) 
is the  plot in which all bursts in this study are considered. 

Clearly, the inter-burst intervals span a wide range from several years to tens of thousands of years.
When considering bursts of all durations, the short inter-burst intervals prevail.
Bursts of higher amplitude (3- and 4-mag) have a bimodal distribution for the duration of quiescent phases between the bursts
(Fig.~\ref{fig:histograms}c-d). The inclusion of 1- and 2-mag bursts diminishes 
the bimodality in favour of the shorter inter-burst time intervals (Fig.~\ref{fig:histograms}a-b). 
\textcolor{black}{
The differences between panels~\ref{fig:histograms}a,f and panels~\ref{fig:histograms}h,j highlight 
the fact that the bimodality is produced by the inter-burst time intervals between the lower-magnitude 
bursts (1,2-mag bursts) on the one hand, and the higher-magnitude bursts (3,4 mag bursts), on the other hand. 
The same is true for panel~\ref{fig:histograms}i, while the disappearance of the bimodality is 
obvious in panels~\ref{fig:histograms}e and \ref{fig:histograms}f. 
}
This information may be used in future studies to compare the inter-burst time intervals with the jet spacings 
such as in~\citet{vorobyov_aa_613_2018}.

\textcolor{black}{
Fig.~\ref{fig:histograms_evolution} presents all the 
time intervals between the bursts calculated in our simulations during a period of 
$5\, \rm kyr$ (a), $10\, \rm kyr$ (b) and $30\, \rm kyr$ (c) \textcolor{black}{once the disc has formed}. 
The first three panels illustrate the development of the inter-burst time intervals  as the disc evolves. 
The last panel (d) shows \textcolor{black}{the histogram displaying the distribution of the time intervals 
between the bursts}, for all bursts of all models and 
is the same distribution as in Fig.~\ref{fig:histograms}g. 
The distribution is initially rather dispersed, especially along the line of increasing 
$\rm M_{\rm c}$ (green bins), see Fig.~\ref{fig:histograms_evolution}a. 
At this time, the disc begins to fragment and form gaseous clumps and 
the bursts are still mild. The models with higher $\beta$-ratios fragment 
faster and therefore the corresponding inter-burst intervals are shorter 
(green bins) than along the line of increasing $\rm M_{\rm c}$ (red bins), 
see Fig.~\ref{fig:histograms_evolution}b-c. 
At later times, both series of model reach an equilibrium distribution 
that is made of two types of bursts separated by 
$\Delta t\sim\, 10^{2}\, \rm yr$ and $\Delta t\ge\, 10^{3}\, \rm yr$, 
respectively. This bimodality is further illustrated for all bursts 
(black line in Fig.~\ref{fig:histograms_evolution}d). 
FU-Orionis-like bursts (and therefore close binary companions) should be observed 
in older, massive MYSOs, surrounded by rather extended and fragmented discs. 
}

\subsection{Protostellar jets as indicators of the burst history? }
\label{sect:indicator}

\textcolor{black}{
Protostellar outflows and jets are part of the accretion--ejection mechanism that
carries angular momentum of the accreted matter away, and thereby prevents
the accreting protostar from spinning up to a break-up velocity. 
There are observational indications that the angles of the outflows from the 
high-mass young stellar object are wider for more evolved and luminous 
stars~\citep{2007prpl.conf..245A}. 
}
\textcolor{black}{
\textcolor{black}{In~\citet{meyer_mnras_482_2019}} we already mentioned that tracing 
of the outflows allows to show that there we about four bursts in the luminous S255 NIRS3 
\textcolor{black}{during a time interval of $\approx7000\, \rm yr$ before 
present-day observations}~\citep{wang_aa_527_2011,zinchenko_apj_810_2015,burns_mnras_460_2016}  
and that the burst in NGC 6334I-MM1 was not a single event~\citep{brogan_apj_866_2018}. 
}

\textcolor{black}{
Well pronounced jets are observed in the number of the younger massive stars in the infrared 
and radio ranges (see, e.g. infrared survey by~\citet{caratti_aa_573_2015} or 
radio surveys by~\citet{purser_mnras_475_2018} and~\citet{obonyo_486_MNRAS_2019}. 
The jets manifest themselves as elongated structures in the 
\textcolor{black}{close surroundings of the source} and further knots sometimes organized in chains. 
So, they have potential to provide information on the history of eruptions. Protostellar 
jets are observed over a large source mass range (Frank et al. 2014), and recent studies show that 
the jets from the massive stars show similarity in physical parameters and origin with the jets from 
the low-mass stars~\citep{caratti_aa_573_2015,fedriani_natco_2019}. 
}

%
%
%

\textcolor{black}{
Well-accepted is the fact that the outflowing matter may not be constant in mass
and velocity. 
}
Measurements of the shock velocities in the jets from the massive stars vary 
from hundreds to thousands $\rm km\, \rm s^{-1}$~\citep{mcloed_nature_554_2018,purser_mnras_475_2018}. 
Therefore, faster material that is ejected at later times will
catch up and run into slower material ahead of it, creating a new mini-bow
shock. The knotty jets then are chains of these small bow-type structures. The proper motion
measurements show that the dynamical times between the ejection of
knots in the chains are on the order of a few 
decades~\citep{1992A&A...263..292E,1998AJ....115.1554E,1997AJ....114.2095D}, whereas the
times for the larger bow-type structure at their ends are on the order of centuries,
and for the largest structures in the parsec-scale jets they are even on the order
of millennia~\citep{1997AJ....114..280E,1997AJ....114.2708R}.
\textcolor{black}{
Moreover, it should be noted that the knot's brightness} in the jets from 
the \textcolor{black}{MYSOs} is subject to time variability~\citep{obonyo_486_MNRAS_2019}.

The interesting question arises if these jets then are a frozen record of the
accretion history of the source, and \textcolor{black}{these} jet knots could be used for a direct
comparison with accretion events of the sources in model calculations, like
the ones presented in this paper (see also \citet{vorobyov_aa_613_2018}). 
As described above the chains of knots are not 
a one-to-one image of the source's accretion history, 
and this principally because of possible merging of the shocks with different velocities 
and because the brightness of the shock knots sometimes varies with time.
Little is known, however, also from a modeling point of view, whether all bursts 
lead to ejections of matter, how bursts can change the outflow speed, and if 
indeed stronger bursts are leading to faster outflowing material as well.

\textcolor{black}{
Keeping all the above mentioned caveats in mind, we note that in most
jets with regular chains of knots the measured proper motions for each knot
are not hugely different, so that one can assume that they are an indicator
to a certain kind of similar burst events. At the mentioned time intervals
of decades, these would then correspond to the first peak in the bi-modal
burst distribution. 
}
\textcolor{black}{
The second peak in the bimodal burst distribution,
at $10^3$ to $10^4$ years can correspond to the dynamical age of 
the bright knots of the jet observed in the distant source in the Large 
Magellanic Cloud - 8 knots are detected in the 11 pc jet with the lifetime 
about $28$$-$$37\, \rm kyr$~\citep{mcloed_nature_554_2018}. These knots probably represent 
the giant bow shocks seen at the end of the jets, or even multiple times in
some parsec-scale jets.   
}

\subsection{What distinguishes massive star formation from its low-mass counterpart ?}
\label{sect:lsf}

\subsubsection{Massive stars principally gain their mass through bursts}
\label{sect:comparison_lms_hms_1}

A series of differences between the formation processes in lower-mass 
and higher-mass star regimes arise from our study. First of all, our accretion histories 
systematically exhibit accretion variability and accretion-driven outbursts along both the lines of 
increasing $M_{\rm c}$ and $\beta$, but also for the models with the lowest $M_{\rm c}$ (Fig.~\ref{fig:rates}). 
Although our results may be affected by physical \textcolor{black}{mechanisms that are so far neglected, like} 
the magnetisation of the pre-stellar core or other non-ideal magneto-hydrodynamical effects, 
accretion bursts  seem to be a systematic feature in the formation of massive protostars. 
\textcolor{black}{When lower-mass stellar objects form}, on the contrary, 
accretion bursts caused by clump infall seem to 
exist only for cloud cores of sufficiently high mass and angular momentum 
\citep[see fig.~A1 in][]{elbakyan_mnras_484_2019}. A lower limit on the 
cloud core mass seems to exists also for accretion bursts triggered by the 
magnetorotational instability in the innermost parts of low-mass 
disks (Kadam et al. 2020, submitted).
Our study, based on a large sample of models, confirms the conclusions 
of~\citet{meyer_mnras_482_2019} stating that the MYSOs gain an \textcolor{black}{important 
part of their final mass} during the burst phase of accretion, sometimes 
amounting to 50\% or even more. On the contrary, the low-mass stars accrete 
on average about 5\% of their final mass with a peak value of 33\%~\citep{dunham_apj_747_2012}.
The efficiency of gravitational instability in discs is consequently always 
at work in massive discs, which 
is consistent with the work of~\citet{kratter_mnras_373_2006,rafikov_apj_662_2007,rafikov_apj_704_2009}, 
reporting that massive discs around high-mass protostars inevitably lead to fragmentation.  
%

\subsubsection{Protostars in FU-Orionis-like burst phases 
evolve towards the red part of the Herzsprung-Russell diagram}
\label{sect:comparison_lms_hms_2}

A series of similarities should also be underlined between the different mass regimes of star formation. 
Indeed, this picture of centrifugally balanced discs onto which inflowing material lands and competes 
with the disc thermodynamics and rotational shear equivalently applies to both regimes. Once disc 
fragmentation is triggered, the gaseous clumps  migrate inwards, producing bursts once they are tidally destroyed near the star. Concurrently, the star migrates in the Herzsprung-Russell diagram~\citep{elbakyan_mnras_484_2019}, 
irradiating the discs which should be noticed in infrared. 
Nevertheless, the low- and high-mass protostars show different types of excursions. The low-mass 
stars \textcolor{black}{become bluer} in the Herzsprung-Russell diagram, while the high-mass 
stars do it to redder, upper right part. 
Similar in both mass regimes is also the nearly-linear relationship between the disc and protostellar masses, the respective 
\textcolor{black}{effects of the initial $M_{\rm c}$ and $\beta$-ratio of the pre-stellar cores} onto the 
disc properties, implying a comparable global evolution if accretion discs at all scales and masses are 
ruled by analogous physical mechanisms. 
The observational study on \textcolor{black}{young low-mass protostars} by~\citet{contreras_mnras_486_2019} interestingly 
reports that "Surprisingly many objects in this group show high-amplitude irregular variability over 
timescales shorter than $10$ years, in contrast with the view that high-amplitude objects always
have long outbursts". This is consistent with our findings in the sense that our 3- and 4-mag 
bursts are characterized by a wide range of burst duration (see Fig.~\ref{fig:burst_corr_2}). 
All this correspondences strongly motivate further works on the detailed features of disc fragmentation in 
the context of massive protostars.

\subsubsection{The scatter in burst magnitudes is wider in massive star formation}
\label{sect:comparison_lms_hms_3}

\textcolor{black}{
We show in Fig.~\ref{fig:box_plot_mag} the scattering of the burst magnitude for all bursts in our 
data as a function of their burst class (from the 1-mag bursts to the 4-mag ones). 
The box plots present the data using whiskers extending from the $10$-th to the $90$-th percentile, which 
allows us to visualize the extent of the variation in magnitudes for all bursts encompassed within one burst 
class (e.g., between 1- and 2-mag, 2- and 3-mag, etc). 
The variation reads
\begin{equation}
   \Delta_\mathrm{i-mag}^{[10,90]}  =  | W_\mathrm{i-mag}^{\mathrm{90-th}} - W_\mathrm{i-mag}^{\mathrm{10-th}} |, 
\end{equation} 
where $W_\mathrm{i-mag}^{\mathrm{90-th}}$ and $W_\mathrm{i-mag}^{\mathrm{10-th}}$ are the burst magnitude at the 
extent of the whiskers, respectively, and $i$ denotes a considered burst class ($1\le i \le4$). 
\textcolor{black}{
The upper panel displays the bursts variation statistics for all burst in our parameter study as a function of 
the burst class (a) and for all the collection of bursts in our parameter study (b). 
}
We find variations of $\Delta_\mathrm{1-mag}^{[10,90]}=0.31\, \mathrm{mag}$, 
$\Delta_\mathrm{2-mag}^{[10,90]}=0.18\, \mathrm{mag}$, 
$\Delta_\mathrm{3-mag}^{[10,90]}=0.48\, \mathrm{mag}$ 
and $\Delta_\mathrm{4-mag}^{[10,90]}=0.61\, \mathrm{mag}$, respectively. 
\textcolor{black}{
The extend of the burst variation regardless of their magnitude class (Fig.~\ref{fig:box_plot_mag}b) 
gives $\Delta_\mathrm{1-mag}^{[10,90]}=0.69\, \mathrm{mag}$. 
}
Particularly, one can compare the obtained burst variations with the value of $0.22$ found for low-mass 
 sources at $3.6\, \mu m$ in the context of low-mass stellar objects in the Serpens South 
star formation region~\citep{2018AJ....155...99W}. 
\textcolor{black}{
According to our study, the average luminosity variation in massive star formation is 
larger than that in low-mass star formation, which constitutes a remarkable difference 
between these two regimes.
}
}


\section{Conclusion}
\label{sect:conclusion}

\textcolor{black}{
This work explores the effects
} 
of both the initial mass ($M_{\rm c}=60$--$200\, \rm M_{\odot}$) 
and rotational-to-gravitational energy ratio ($\beta=0.5$--$33\%$) of a representative sample 
of molecular pre-stellar cores by means of three-dimensional gravito-radiation-hydrodynamics 
simulations. We \textcolor{black}{utilise} the method previously detailed in~\citet{meyer_mnras_482_2019}. 
\textcolor{black}{
Our simulations model the evolution of molecular cores and how the collapsing material
} 
lands onto centrifugally balanced accretion discs surrounding  young massive protostars. 
\textcolor{black}{
The efficient gravitational instability in the disc results in the aggregation of disc material in 
clumps within spiral structures. These blobs of gas can gravitationally fall 
towards the protostar and generate luminous accretion-driven outbursts~\citep{meyer_mnras_464_2017}, 
affecting both the properties of the disc and its central MYSOs. 
} 
We calculate in each model the accretion rate histories and lightcurves of the evolving 
massive protostars. As soon as a simulated protostar leaves the quiescent regime of accretion and enters 
the burst mode, we analyse the properties of the corresponding flare, such as its duration, peak 
luminosity, accreted mass, and intensity. These quantities are statistically analysed for the large sample 
of bursts that we extract from our grid of hydrodynamical simulations.

\textcolor{black}{
Under an assumption of negligible magnetic fields, which may have a major effect on accretion disc 
physics~\citep{2011ApJ...735..122F} and star formation processes~\citep{2020MNRAS.tmp.2057M},
}
we found that \textcolor{black}{cores with higher mass $M_{\rm c}$ and/or $\beta$-ratio 
tends to produce circumstellar discs more susceptible} to experience accretion bursts. 
All massive protostars in our sample have accretion bursts, even those with pre-stellar 
cores of low $\beta$-ratio $\le1\%$. This constitutes, under our assumptions, a major 
\textcolor{black}{
difference between the  mechanisms happening in the low-mass and massive regimes of star formation.
}
%
%
All our disc masses scale as a power-law with \textcolor{black}{the mass of the protostars} and disc-to-mass 
ratios $M_{\rm d}/M_{\star}>1$ are obtained in models with higher $\beta$ or small $M_{\rm c}$, 
as at equal age more massive discs are obtained from cores of greater $M_{\rm c}$ but 
larger $\beta$. 
\textcolor{black}{
Our results confirm that massive protostars accrete about $40$-$60\%$ of 
their mass in the burst mode and stronger bursts appear in the later phase 
of the disc evolution.  
}

\textcolor{black}{
Our numerical experiments keep on indicating that present-day massive formation is a 
scaled-up version of low-mass star formation, both being ruled by the burst mode 
of accretion. 
As for their low-mass counterparts, young massive stars experience a strong and sudden increase 
of their accretion rate, e.g. when a disc fragment falls onto the star. 
This results in large amplitude fluctuations of its total luminosity, a swelling 
of the stellar radius and a decrease of the flux released in the protostar's 
associated \hii region. 
Under our assumptions, we calculate within the $10$$-$th and $90$$-$th percentile of 
the collection of bursts in our simulations of forming massive stars, the extend of their 
luminosity variations is $\approx 0.69$, which is much larger than that observed for low-mass 
protostars~\citep{2018AJ....155...99W}. This constitutes a major difference between 
the high- and low-mass regimes of star formation, to be verified by means of future 
observations. 
}
Last, we discuss the structure of massive protostellar jets as potential 
indicators of their driving star's burst history. 
We propose that the high-frequency component of the burst bimodal distribution would 
correspond to the regular chain of knots along the overall jet morphology, while 
the second, low-frequency component peaking at $10^3$-$10^4\, \rm yr$ would 
be associated \textcolor{black}{with the giant bow shock at the top of these} jets. 
Our results motivate further investigations of the burst mode of accretion in 
forming higher-mass stars and its connection with the morphology of massive 
protostellar jets.


\section*{Acknowledgements}

\textcolor{black}{
The authors thank the anonymous referee for comments which improved the quality of the paper.
}
\textcolor{black}{
D.~M.-A.~Meyer thanks W.~Kley for priceless advice on disc physics. 
}
The authors acknowledge the North-German Supercomputing Alliance (HLRN) for providing HPC 
resources that have contributed to the research results reported in this paper. 
The computational results presented have been partly achieved using the Vienna Scientific Cluster (VSC).
E. I. Vorobyov, V. G. Elbakyan and A. M. Sobolev acknowledge
support by 
the Ministry of Science and Higher Education of the Russian Federation under the grant 075-15-2020-780.

\section*{Data availability}

\textcolor{black}{
This research made use of the {\sc pluto} code developed at the University of Torino  
by A.~Mignone (http://plutocode.ph.unito.it/).
The figures have been produced using the Matplotlib plotting 
library for the Python programming language (https://matplotlib.org/). 
The data underlying this article will be shared on reasonable request to the corresponding author.
}


\bibliographystyle{mn2e}

\footnotesize{
\bibliography{grid}

\begin{thebibliography}{}

\bibitem[\protect\citeauthoryear{{A{\~n}ez-L{\'o}pez}, {Osorio}, {Busquet},
  {Girart}, {Mac{\'\i}as}, {Carrasco-Gonz{\'a}lez}, {Curiel}, {Estalella},
  {Fern{\'a}ndez-L{\'o}pez}, {Galv{\'a}n-Madrid}, {Kwon} \&
  {Torrelles}}{{A{\~n}ez-L{\'o}pez} et~al.}{2020}]{anezlopez_apj_888_2020}
{A{\~n}ez-L{\'o}pez} N.,  {Osorio} M.,  {Busquet} G.,  {Girart} J.~M.,
  {Mac{\'\i}as} E.,  {Carrasco-Gonz{\'a}lez} C.,  {Curiel} S.,  {Estalella} R.,
   {Fern{\'a}ndez-L{\'o}pez} M.,  {Galv{\'a}n-Madrid} R.,  {Kwon} J.,
  {Torrelles} J.~M.,  2020, \apj, 888, 41

\bibitem[\protect\citeauthoryear{{Ahmadi}, {Beuther}, {Mottram}, {Bosco},
  {Linz}, {Henning}, {Winters}, {Kuiper}, {Pudritz}, {S{\'a}nchez-Monge},
  {Keto}, {Beltran}, {Bontemps}, {Cesaroni}, {Csengeri}, {Feng} \&
  {Galvan-Madrid}}{{Ahmadi} et~al.}{2018}]{ahmadi_aa_618_2018}
{Ahmadi} A.,  {Beuther} H.,  {Mottram} J.~C.,  {Bosco} F.,  {Linz} H.,
  {Henning} T.,  {Winters} J.~M.,  {Kuiper} R.,  {Pudritz} R.,
  {S{\'a}nchez-Monge} {\'A}.,  {Keto} E.,  {Beltran} M.,  {Bontemps} S.,
  {Cesaroni} R.,  {Csengeri} T.,  {Feng} S.,    {Galvan-Madrid} R. e.~a.,
  2018, \aap, 618, A46

\bibitem[\protect\citeauthoryear{{Ahmadi}, {Kuiper} \& {Beuther}}{{Ahmadi}
  et~al.}{2019}]{ahmadi_aa_632_2019}
{Ahmadi} A.,  {Kuiper} R.,    {Beuther} H.,  2019, \aap, 632, A50

\bibitem[\protect\citeauthoryear{{Andr{\'e} Oliva} \& {Kuiper}}{{Andr{\'e}
  Oliva} \& {Kuiper}}{2020}]{2020arXiv200813653A}
{Andr{\'e} Oliva} G.,  {Kuiper} R.,  2020, arXiv e-prints, p. arXiv:2008.13653

\bibitem[\protect\citeauthoryear{{Arce}, {Shepherd}, {Gueth}, {Lee},
  {Bachiller}, {Rosen} \& {Beuther}}{{Arce} et~al.}{2007}]{2007prpl.conf..245A}
{Arce} H.~G.,  {Shepherd} D.,  {Gueth} F.,  {Lee} C.~F.,  {Bachiller} R.,
  {Rosen} A.,    {Beuther} H.,  2007, in {Reipurth} B.,  {Jewitt} D.,   {Keil}
  K.,  eds, Protostars and Planets V {Molecular Outflows in Low- and High-Mass
  Star-forming Regions}.
p.~245

\bibitem[\protect\citeauthoryear{{Beuther}, {Ahmadi}, {Mottram}, {Linz},
  {Maud}, {Henning}, {Kuiper}, {Walsh}, {Johnston} \& {Longmore}}{{Beuther}
  et~al.}{2019}]{2018arXiv181110245B}
{Beuther} H.,  {Ahmadi} A.,  {Mottram} J.~C.,  {Linz} H.,  {Maud} L.~T.,
  {Henning} T.,  {Kuiper} R.,  {Walsh} A.~J.,  {Johnston} K.~G.,    {Longmore}
  S.~N.,  2019, \aap, 621, A122

\bibitem[\protect\citeauthoryear{{Boley}, {Linz}, {Dmitrienko}, {Georgiev},
  {Rabien}, {Busoni}, {G{\"a}ssler}, {Bonaglia} \& {Orban de Xivry}}{{Boley}
  et~al.}{2019}]{2019arXiv191208510B}
{Boley} P.~A.,  {Linz} H.,  {Dmitrienko} N.,  {Georgiev} I.~Y.,  {Rabien} S.,
  {Busoni} L.,  {G{\"a}ssler} W.,  {Bonaglia} M.,    {Orban de Xivry} G.,
  2019, arXiv e-prints, p. arXiv:1912.08510

\bibitem[\protect\citeauthoryear{{Bonnell}, {Bate} \& {Zinnecker}}{{Bonnell}
  et~al.}{1998}]{1998MNRAS.298...93B}
{Bonnell} I.~A.,  {Bate} M.~R.,    {Zinnecker} H.,  1998, \mnras, 298, 93

\bibitem[\protect\citeauthoryear{{Bosco}, {Beuther}, {Ahmadi}, {Mottram},
  {Kuiper}, {Linz} \& {Maud}}{{Bosco} et~al.}{2019}]{bosco_aa_629_2019}
{Bosco} F.,  {Beuther} H.,  {Ahmadi} A.,  {Mottram} J.~C.,  {Kuiper} R.,
  {Linz} H.,    {Maud} e.~a.,  2019, \aap, 629, A10

\bibitem[\protect\citeauthoryear{{Breen}, {Sobolev}, {Kaczmarek}, {Ellingsen},
  {McCarthy} \& {Voronkov}}{{Breen} et~al.}{2019}]{breen_apj_876_2019}
{Breen} S.~L.,  {Sobolev} A.~M.,  {Kaczmarek} J.~F.,  {Ellingsen} S.~P.,
  {McCarthy} T.~P.,    {Voronkov} M.~A.,  2019, \apjl, 876, L25

\bibitem[\protect\citeauthoryear{{Brogan}, {Hunter}, {Cyganowski}, {Chibueze},
  {Friesen}, {Hirota}, {MacLeod}, {McGuire} \& {Sobolev}}{{Brogan}
  et~al.}{2018}]{brogan_apj_866_2018}
{Brogan} C.~L.,  {Hunter} T.~R.,  {Cyganowski} C.~J.,  {Chibueze} J.~O.,
  {Friesen} R.~K.,  {Hirota} T.,  {MacLeod} G.~C.,  {McGuire} B.~A.,
  {Sobolev} A.~M.,  2018, \apj, 866, 87

\bibitem[\protect\citeauthoryear{{Burns}}{{Burns}}{2018}]{burns_2018IAUS}
{Burns} R.~A.,  2018, in {Tarchi} A.,  {Reid} M.~J.,   {Castangia} P.,  eds,
  Astrophysical Masers: Unlocking the Mysteries of the Universe Vol.~336 of IAU
  Symposium, {Water masers in bowshocks: Addressing the radiation pressure
  problem of massive star formation}.
pp 263--266

\bibitem[\protect\citeauthoryear{{Burns}, {Handa}, {Imai}, {Nagayama},
  {Omodaka}, {Hirota}, {Motogi}, {van Langevelde} \& {Baan}}{{Burns}
  et~al.}{2017}]{burns_mnras_467_2017}
{Burns} R.~A.,  {Handa} T.,  {Imai} H.,  {Nagayama} T.,  {Omodaka} T.,
  {Hirota} T.,  {Motogi} K.,  {van Langevelde} H.~J.,    {Baan} W.~A.,  2017,
  \mnras, 467, 2367

\bibitem[\protect\citeauthoryear{{Burns}, {Handa}, {Nagayama}, {Sunada} \&
  {Omodaka}}{{Burns} et~al.}{2016}]{burns_mnras_460_2016}
{Burns} R.~A.,  {Handa} T.,  {Nagayama} T.,  {Sunada} K.,    {Omodaka} T.,
  2016, \mnras, 460, 283

\bibitem[\protect\citeauthoryear{{Burns}, {Sugiyama}, {Hirota}, {Kim},
  {Sobolev}, {Stecklum}, {MacLeod} \& {Yonekura}}{{Burns}
  et~al.}{2020}]{burns_natas_2020}
{Burns} R.~A.,  {Sugiyama} K.,  {Hirota} T.,  {Kim} K.-T.,  {Sobolev} A.~M.,
  {Stecklum} B.,  {MacLeod} G.~C.,    {Yonekura} Y. e.~a.,  2020, Nature
  Astronomy, 4, 506

\bibitem[\protect\citeauthoryear{{Caratti o Garatti}, {Stecklum}, {Garcia
  Lopez}, {Eisloffel}, {Ray}, {Sanna}, {Cesaroni}, {Walmsley}, {Oudmaijer}, {de
  Wit}, {Moscadelli}, {Greiner}, {Krabbe}, {Fischer}, {Klein} \&
  {Ibanez}}{{Caratti o Garatti} et~al.}{2017}]{caratti_nature_2016}
{Caratti o Garatti} A.,  {Stecklum} B.,  {Garcia Lopez} R.,  {Eisloffel} J.,
  {Ray} T.~P.,  {Sanna} A.,  {Cesaroni} R.,  {Walmsley} C.~M.,  {Oudmaijer}
  R.~D.,  {de Wit} W.~J.,  {Moscadelli} L.,  {Greiner} J.,  {Krabbe} A.,
  {Fischer} C.,  {Klein} R.,    {Ibanez} J.~M.,  2017, Nature Physics, 13, 276

\bibitem[\protect\citeauthoryear{{Caratti o Garatti}, {Stecklum}, {Linz},
  {Garcia Lopez} \& {Sanna}}{{Caratti o Garatti}
  et~al.}{2015}]{caratti_aa_573_2015}
{Caratti o Garatti} A.,  {Stecklum} B.,  {Linz} H.,  {Garcia Lopez} R.,
  {Sanna} A.,  2015, \aap, 573, A82

\bibitem[\protect\citeauthoryear{{Carrasco-Gonz{\'a}lez}, {Rodr{\'\i}guez},
  {Anglada}, {Mart{\'\i}}, {Torrelles} \& {Osorio}}{{Carrasco-Gonz{\'a}lez}
  et~al.}{2010}]{carrasco_sci_330_2010}
{Carrasco-Gonz{\'a}lez} C.,  {Rodr{\'\i}guez} L.~F.,  {Anglada} G.,
  {Mart{\'\i}} J.,  {Torrelles} J.~M.,    {Osorio} M.,  2010, Science, 330,
  1209

\bibitem[\protect\citeauthoryear{{Cesaroni}, {Galli}, {Lodato}, {Walmsley} \&
  {Zhang}}{{Cesaroni} et~al.}{2006}]{cesaroni_natur_444_2006}
{Cesaroni} R.,  {Galli} D.,  {Lodato} G.,  {Walmsley} M.,    {Zhang} Q.,  2006,
  \nat, 444, 703

\bibitem[\protect\citeauthoryear{{Cesaroni}, {Hofner}, {Araya} \&
  {Kurtz}}{{Cesaroni} et~al.}{2010}]{cesaroni_aa_509_2010}
{Cesaroni} R.,  {Hofner} P.,  {Araya} E.,    {Kurtz} S.,  2010, \aap, 509, A50

\bibitem[\protect\citeauthoryear{{Chen}, {Ren}, {Zhang}, {Shen} \&
  {Qiu}}{{Chen} et~al.}{2017}]{chen_apj_835_2017}
{Chen} X.,  {Ren} Z.,  {Zhang} Q.,  {Shen} Z.,    {Qiu} K.,  2017, \apj, 835,
  227

\bibitem[\protect\citeauthoryear{{Chen}, {Sobolev}, {Breen}, {Shen},
  {Ellingsen} \& {MacLeod}}{{Chen} et~al.}{2020}]{chen_apj_890_2020}
{Chen} X.,  {Sobolev} A.~M.,  {Breen} S.~L.,  {Shen} Z.-Q.,  {Ellingsen} S.~P.,
     {MacLeod} G. C. e.~a.,  2020, \apjl, 890, L22

\bibitem[\protect\citeauthoryear{{Chen}, {Sobolev}, {Ren}, {Parfenov}, {Breen},
  {Ellingsen}, {Shen}, {Li}, {MacLeod}, {Baan}, {Brogan}, {Hirota}, {Hunter},
  {Linz}, {Menten}, {Sugiyama}, {Stecklum}, {Gong} \& {Zheng}}{{Chen}
  et~al.}{2020}]{chen_natas_2020}
{Chen} X.,  {Sobolev} A.~M.,  {Ren} Z.-Y.,  {Parfenov} S.,  {Breen} S.~L.,
  {Ellingsen} S.~P.,  {Shen} Z.-Q.,  {Li} B.,  {MacLeod} G.~C.,  {Baan} W.,
  {Brogan} C.,  {Hirota} T.,  {Hunter} T.~R.,  {Linz} H.,  {Menten} K.,
  {Sugiyama} K.,  {Stecklum} B.,  {Gong} Y.,    {Zheng} X.,  2020, Nature
  Astronomy

\bibitem[\protect\citeauthoryear{{Chini}, {Hoffmeister}, {Nasseri}, {Stahl} \&
  {Zinnecker}}{{Chini} et~al.}{2012}]{chini_424_mnras_2012}
{Chini} R.,  {Hoffmeister} V.~H.,  {Nasseri} A.,  {Stahl} O.,    {Zinnecker}
  H.,  2012, \mnras, 424, 1925

\bibitem[\protect\citeauthoryear{{Contreras Pe{\~n}a}, {Naylor} \&
  {Morrell}}{{Contreras Pe{\~n}a} et~al.}{2019}]{contreras_mnras_486_2019}
{Contreras Pe{\~n}a} C.,  {Naylor} T.,    {Morrell} S.,  2019, \mnras, 486,
  4590

\bibitem[\protect\citeauthoryear{{Cunningham}, {Moeckel} \&
  {Bally}}{{Cunningham} et~al.}{2009}]{Cunningham_apj_692_2009}
{Cunningham} N.~J.,  {Moeckel} N.,    {Bally} J.,  2009, \apj, 692, 943

\bibitem[\protect\citeauthoryear{{Devine}, {Bally}, {Reipurth} \&
  {Heathcote}}{{Devine} et~al.}{1997}]{1997AJ....114.2095D}
{Devine} D.,  {Bally} J.,  {Reipurth} B.,    {Heathcote} S.,  1997, \aj, 114,
  2095

\bibitem[\protect\citeauthoryear{{Dunham} \& {Vorobyov}}{{Dunham} \&
  {Vorobyov}}{2012}]{dunham_apj_747_2012}
{Dunham} M.~M.,  {Vorobyov} E.~I.,  2012, \apj, 747, 52

\bibitem[\protect\citeauthoryear{{Durisen}, {Boss}, {Mayer}, {Nelson}, {Quinn}
  \& {Rice}}{{Durisen} et~al.}{2007}]{durissen_prpl_607_2007}
{Durisen} R.~H.,  {Boss} A.~P.,  {Mayer} L.,  {Nelson} A.~F.,  {Quinn} T.,
  {Rice} W.~K.~M.,  2007, Protostars and Planets V, pp 607--622

\bibitem[\protect\citeauthoryear{{Eisl{\"o}ffel} \& {Mundt}}{{Eisl{\"o}ffel} \&
  {Mundt}}{1992}]{1992A&A...263..292E}
{Eisl{\"o}ffel} J.,  {Mundt} R.,  1992, \aap, 263, 292

\bibitem[\protect\citeauthoryear{{Eisl{\"o}ffel} \& {Mundt}}{{Eisl{\"o}ffel} \&
  {Mundt}}{1997}]{1997AJ....114..280E}
{Eisl{\"o}ffel} J.,  {Mundt} R.,  1997, \aj, 114, 280

\bibitem[\protect\citeauthoryear{{Eisl{\"o}ffel} \& {Mundt}}{{Eisl{\"o}ffel} \&
  {Mundt}}{1998}]{1998AJ....115.1554E}
{Eisl{\"o}ffel} J.,  {Mundt} R.,  1998, \aj, 115, 1554

\bibitem[\protect\citeauthoryear{{Elbakyan}, {Vorobyov}, {Rab}, {Meyer},
  {G{\"u}del}, {Hosokawa} \& {Yorke}}{{Elbakyan}
  et~al.}{2019}]{elbakyan_mnras_484_2019}
{Elbakyan} V.~G.,  {Vorobyov} E.~I.,  {Rab} C.,  {Meyer} D.~M.-A.,  {G{\"u}del}
  M.,  {Hosokawa} T.,    {Yorke} H.,  2019, \mnras, 484, 146

\bibitem[\protect\citeauthoryear{{Ercolano} \& {Owen}}{{Ercolano} \&
  {Owen}}{2016}]{ercolano_mnras_460_2016}
{Ercolano} B.,  {Owen} J.~E.,  2016, \mnras, 460, 3472

\bibitem[\protect\citeauthoryear{{Fedriani}, {Caratti o Garatti}, {Purser},
  {Sanna}, {Tan}, {Garcia-Lopez}, {Ray}, {Coffey}, {Stecklum} \&
  {Hoare}}{{Fedriani} et~al.}{2019}]{fedriani_natco_2019}
{Fedriani} R.,  {Caratti o Garatti} A.,  {Purser} S.~J.~D.,  {Sanna} A.,  {Tan}
  J.~C.,  {Garcia-Lopez} R.,  {Ray} T.~P.,  {Coffey} D.,  {Stecklum} B.,
  {Hoare} M.,  2019, Nature Communications, 10, 3630

\bibitem[\protect\citeauthoryear{{Flock}, {Dzyurkevich}, {Klahr}, {Turner} \&
  {Henning}}{{Flock} et~al.}{2011}]{2011ApJ...735..122F}
{Flock} M.,  {Dzyurkevich} N.,  {Klahr} H.,  {Turner} N.~J.,    {Henning} T.,
  2011, \apj, 735, 122

\bibitem[\protect\citeauthoryear{{Forgan}, {Ilee}, {Cyganowski}, {Brogan} \&
  {Hunter}}{{Forgan} et~al.}{2016}]{forgan_mnras_463_2016}
{Forgan} D.~H.,  {Ilee} J.~D.,  {Cyganowski} C.~J.,  {Brogan} C.~L.,
  {Hunter} T.~R.,  2016, \mnras, 463, 957

\bibitem[\protect\citeauthoryear{{Franz}, {Picogna}, {Ercolano} \&
  {Birnstiel}}{{Franz} et~al.}{2020}]{franz_aa_635_2020}
{Franz} R.,  {Picogna} G.,  {Ercolano} B.,    {Birnstiel} T.,  2020, \aap, 635,
  A53

\bibitem[\protect\citeauthoryear{{Fuente}, {Neri}, {Mart{\'{\i}}n-Pintado},
  {Bachiller}, {Rodr{\'{\i}}guez-Franco} \& {Palla}}{{Fuente}
  et~al.}{2001}]{fuente_aa_366_2001}
{Fuente} A.,  {Neri} R.,  {Mart{\'{\i}}n-Pintado} J.,  {Bachiller} R.,
  {Rodr{\'{\i}}guez-Franco} A.,    {Palla} F.,  2001, \aap, 366, 873

\bibitem[\protect\citeauthoryear{{Gammie}}{{Gammie}}{2001}]{2001ApJ...553..174G}
{Gammie} C.~F.,  2001, \apj, 553, 174

\bibitem[\protect\citeauthoryear{{Ginsburg}, {Bally}, {Goddi}, {Plambeck} \&
  {Wright}}{{Ginsburg} et~al.}{2018}]{2018arXiv180410622G}
{Ginsburg} A.,  {Bally} J.,  {Goddi} C.,  {Plambeck} R.,    {Wright} M.,  2018,
  \apj, 860, 119

\bibitem[\protect\citeauthoryear{{Haemmerl{\'e}}}{{Haemmerl{\'e}}}{2014}]{haemmerle_phd_2014}
{Haemmerl{\'e}} L.,  2014, PhD thesis, University of Geneva

\bibitem[\protect\citeauthoryear{{Haemmerl{\'e}}, {Eggenberger}, {Meynet},
  {Maeder} \& {Charbonnel}}{{Haemmerl{\'e}}
  et~al.}{2016}]{haemmerle_585_aa_2016}
{Haemmerl{\'e}} L.,  {Eggenberger} P.,  {Meynet} G.,  {Maeder} A.,
  {Charbonnel} C.,  2016, \aap, 585, A65

\bibitem[\protect\citeauthoryear{{Haemmerl{\'e}}, {Eggenberger}, {Meynet},
  {Maeder}, {Charbonnel} \& {Klessen}}{{Haemmerl{\'e}}
  et~al.}{2017}]{haemmerle_aa_602_2017}
{Haemmerl{\'e}} L.,  {Eggenberger} P.,  {Meynet} G.,  {Maeder} A.,
  {Charbonnel} C.,    {Klessen} R.~S.,  2017, \aap, 602, A17

\bibitem[\protect\citeauthoryear{{Harries}}{{Harries}}{2015}]{harries_mnras_448_2015}
{Harries} T.~J.,  2015, \mnras, 448, 3156

\bibitem[\protect\citeauthoryear{{Harries}, {Douglas} \& {Ali}}{{Harries}
  et~al.}{2017}]{harries_2017}
{Harries} T.~J.,  {Douglas} T.~A.,    {Ali} A.,  2017, \mnras, 471, 4111

\bibitem[\protect\citeauthoryear{{Hollenbach}, {Johnstone}, {Lizano} \&
  {Shu}}{{Hollenbach} et~al.}{1994}]{hollenbach_apj_428_1994}
{Hollenbach} D.,  {Johnstone} D.,  {Lizano} S.,    {Shu} F.,  1994, \apj, 428,
  654

\bibitem[\protect\citeauthoryear{{Hosokawa}, {Hirano}, {Kuiper}, {Yorke},
  {Omukai} \& {Yoshida}}{{Hosokawa} et~al.}{2016}]{hosokawa_2015}
{Hosokawa} T.,  {Hirano} S.,  {Kuiper} R.,  {Yorke} H.~W.,  {Omukai} K.,
  {Yoshida} N.,  2016, \apj, 824, 119

\bibitem[\protect\citeauthoryear{{Hosokawa} \& {Omukai}}{{Hosokawa} \&
  {Omukai}}{2009}]{hosokawa_apj_691_2009}
{Hosokawa} T.,  {Omukai} K.,  2009, \apj, 691, 823

\bibitem[\protect\citeauthoryear{{Hosokawa}, {Omukai}, {Yoshida} \&
  {Yorke}}{{Hosokawa} et~al.}{2011}]{Hosokawa_2011Sci}
{Hosokawa} T.,  {Omukai} K.,  {Yoshida} N.,    {Yorke} H.~W.,  2011, Science,
  334, 1250

\bibitem[\protect\citeauthoryear{{Hosokawa}, {Yorke}, {Inayoshi}, {Omukai} \&
  {Yoshida}}{{Hosokawa} et~al.}{2013}]{hosokawa_apj__778_2013}
{Hosokawa} T.,  {Yorke} H.~W.,  {Inayoshi} K.,  {Omukai} K.,    {Yoshida} N.,
  2013, \apj, 778, 178

\bibitem[\protect\citeauthoryear{{Hosokawa}, {Yorke} \& {Omukai}}{{Hosokawa}
  et~al.}{2010}]{hosokawa_apj_721_2010}
{Hosokawa} T.,  {Yorke} H.~W.,    {Omukai} K.,  2010, \apj, 721, 478

\bibitem[\protect\citeauthoryear{{Hosokawa}, {Yoshida}, {Omukai} \&
  {Yorke}}{{Hosokawa} et~al.}{2012}]{hosokawa_apj_760_2012}
{Hosokawa} T.,  {Yoshida} N.,  {Omukai} K.,    {Yorke} H.~W.,  2012, \apjl,
  760, L37

\bibitem[\protect\citeauthoryear{{Hunter}}{{Hunter}}{2019}]{2019asrc_confE_91H}
{Hunter} T.,  2019, in ALMA2019: Science Results and Cross-Facility Synergies
  {Resolving the source of the massive protostellar accretion outburst in
  NGC6334I-MM1B}.
p.~91

\bibitem[\protect\citeauthoryear{{Ilee}, {Cyganowski}, {Brogan}, {Hunter},
  {Forgan}, {Haworth}, {Clarke} \& {Harries}}{{Ilee}
  et~al.}{2018}]{2018ApJ...869L..24I}
{Ilee} J.~D.,  {Cyganowski} C.~J.,  {Brogan} C.~L.,  {Hunter} T.~R.,  {Forgan}
  D.~H.,  {Haworth} T.~J.,  {Clarke} C.~J.,    {Harries} T.~J.,  2018, \apjl,
  869, L24

\bibitem[\protect\citeauthoryear{{Ilee}, {Cyganowski}, {Nazari}, {Hunter},
  {Brogan}, {Forgan} \& {Zhang}}{{Ilee} et~al.}{2016}]{ilee_mnras_462_2016}
{Ilee} J.~D.,  {Cyganowski} C.~J.,  {Nazari} P.,  {Hunter} T.~R.,  {Brogan}
  C.~L.,  {Forgan} D.~H.,    {Zhang} Q.,  2016, \mnras, 462, 4386

\bibitem[\protect\citeauthoryear{{Jankovic}, {Haworth}, {Ilee}, {Forgan},
  {Cyganowski}, {Walsh}, {Brogan}, {Hunter} \& {Mohanty}}{{Jankovic}
  et~al.}{2019}]{jankovic_mnras_482_2019}
{Jankovic} M.~R.,  {Haworth} T.~J.,  {Ilee} J.~D.,  {Forgan} D.~H.,
  {Cyganowski} C.~J.,  {Walsh} C.,  {Brogan} C.~L.,  {Hunter} T.~R.,
  {Mohanty} S.,  2019, \mnras, 482, 4673

\bibitem[\protect\citeauthoryear{{Johnston}, {Hoare}, {Beuther}, {Kuiper},
  {Kee}, {Linz}, {Boley}, {Maud}, {Ilee}, {Ahmadi} \& {Robitaille}}{{Johnston}
  et~al.}{2019}]{2019arXiv191109692J}
{Johnston} K.~G.,  {Hoare} M.~G.,  {Beuther} H.,  {Kuiper} R.,  {Kee} N.~D.,
  {Linz} H.,  {Boley} P.,  {Maud} L.~T.,  {Ilee} J.~D.,  {Ahmadi} A.,
  {Robitaille} T.~P.,  2019, arXiv e-prints, p. arXiv:1911.09692

\bibitem[\protect\citeauthoryear{{Johnston}, {Robitaille}, {Beuther}, {Linz},
  {Boley}, {Kuiper}, {Keto}, {Hoare} \& {van Boekel}}{{Johnston}
  et~al.}{2015}]{johnston_apj_813_2015}
{Johnston} K.~G.,  {Robitaille} T.~P.,  {Beuther} H.,  {Linz} H.,  {Boley} P.,
  {Kuiper} R.,  {Keto} E.,  {Hoare} M.~G.,    {van Boekel} R.,  2015, \apjl,
  813, L19

\bibitem[\protect\citeauthoryear{{Keto} \& {Wood}}{{Keto} \&
  {Wood}}{2006}]{keto_apj_637_2006}
{Keto} E.,  {Wood} K.,  2006, \apj, 637, 850

\bibitem[\protect\citeauthoryear{{Klassen}, {Pudritz}, {Kuiper}, {Peters} \&
  {Banerjee}}{{Klassen} et~al.}{2016}]{klassen_apj_823_2016}
{Klassen} M.,  {Pudritz} R.~E.,  {Kuiper} R.,  {Peters} T.,    {Banerjee} R.,
  2016, \apj, 823, 28

\bibitem[\protect\citeauthoryear{{Kobulnicky}, {Kiminki}, {Lundquist}, {Burke},
  {Chapman}, {Keller}, {Lester}, {Rolen}, {Topel}, {Bhattacharjee}, {Smullen},
  {Vargas {\'A}lvarez}, {Runnoe}, {Dale} \& {Brotherton}}{{Kobulnicky}
  et~al.}{2014}]{2014ApJS..213...34K}
{Kobulnicky} H.~A.,  {Kiminki} D.~C.,  {Lundquist} M.~J.,  {Burke} J.,
  {Chapman} J.,  {Keller} E.,  {Lester} K.,  {Rolen} E.~K.,  {Topel} E.,
  {Bhattacharjee} A.,  {Smullen} R.~A.,  {Vargas {\'A}lvarez} C.~A.,  {Runnoe}
  J.~C.,  {Dale} D.~A.,    {Brotherton} M.~M.,  2014, \apjs, 213, 34

\bibitem[\protect\citeauthoryear{{Kratter} \& {Matzner}}{{Kratter} \&
  {Matzner}}{2006}]{kratter_mnras_373_2006}
{Kratter} K.~M.,  {Matzner} C.~D.,  2006, \mnras, 373, 1563

\bibitem[\protect\citeauthoryear{{Kraus}, {Kluska}, {Kreplin}, {Bate},
  {Harries}, {Hofmann}, {Hone}, {Monnier}, {Weigelt}, {Anugu}, {de Wit} \&
  {Wittkowski}}{{Kraus} et~al.}{2017}]{kraus_apj_835_2017}
{Kraus} S.,  {Kluska} J.,  {Kreplin} A.,  {Bate} M.,  {Harries} T.~J.,
  {Hofmann} K.-H.,  {Hone} E.,  {Monnier} J.~D.,  {Weigelt} G.,  {Anugu} A.,
  {de Wit} W.~J.,    {Wittkowski} M.,  2017, \apjl, 835, L5

\bibitem[\protect\citeauthoryear{{Krumholz}, {Klein} \& {McKee}}{{Krumholz}
  et~al.}{2007}]{krumholz_apj_656_2007}
{Krumholz} M.~R.,  {Klein} R.~I.,    {McKee} C.~F.,  2007, \apj, 656, 959

\bibitem[\protect\citeauthoryear{{Laor} \& {Draine}}{{Laor} \&
  {Draine}}{1993}]{laor_apj_402_1993}
{Laor} A.,  {Draine} B.~T.,  1993, \apj, 402, 441

\bibitem[\protect\citeauthoryear{{Larson}}{{Larson}}{1969}]{larson_mnras_145_1969}
{Larson} R.~B.,  1969, \mnras, 145, 271

\bibitem[\protect\citeauthoryear{{Larson}}{{Larson}}{1972}]{larson_mnras_157_1972}
{Larson} R.~B.,  1972, \mnras, 157, 121

\bibitem[\protect\citeauthoryear{{Liu}, {Su}, {Zinchenko}, {Wang}, {Meyer},
  {Wang} \& {Hsieh}}{{Liu} et~al.}{2020}]{2020arXiv201009199L}
{Liu} S.-Y.,  {Su} Y.-N.,  {Zinchenko} I.,  {Wang} K.-S.,  {Meyer} D. M.~A.,
  {Wang} Y.,    {Hsieh} I.-T.,  2020, arXiv e-prints, p. arXiv:2010.09199

\bibitem[\protect\citeauthoryear{{MacLeod}, {Smits}, {Goedhart}, {Hunter},
  {Brogan}, {Chibueze}, {van den Heever}, {Thesner}, {Banda} \&
  {Paulsen}}{{MacLeod} et~al.}{2018}]{macleod_mnras_478_2018}
{MacLeod} G.~C.,  {Smits} D.~P.,  {Goedhart} S.,  {Hunter} T.~R.,  {Brogan}
  C.~L.,  {Chibueze} J.~O.,  {van den Heever} S.~P.,  {Thesner} C.~J.,  {Banda}
  P.~J.,    {Paulsen} J.~D.,  2018, \mnras, 478, 1077

\bibitem[\protect\citeauthoryear{{MacLeod}, {Sugiyama}, {Hunter}, {Quick},
  {Baan}, {Breen}, {Brogan}, {Burns} \& {Caratti o Garatti}}{{MacLeod}
  et~al.}{2019}]{macLeod_mnras_489_2019}
{MacLeod} G.~C.,  {Sugiyama} K.,  {Hunter} T.~R.,  {Quick} J.,  {Baan} W.,
  {Breen} S.~L.,  {Brogan} C.~L.,  {Burns} R.~A.,    {Caratti o Garatti} A.
  e.~a.,  2019, \mnras, 489, 3981

\bibitem[\protect\citeauthoryear{{Mahy}, {Rauw}, {De Becker}, {Eenens} \&
  {Flores}}{{Mahy} et~al.}{2013}]{2013A&A...550A..27M}
{Mahy} L.,  {Rauw} G.,  {De Becker} M.,  {Eenens} P.,    {Flores} C.~A.,  2013,
  \aap, 550, A27

\bibitem[\protect\citeauthoryear{{Maud}, {Cesaroni}, {Kumar}, {Rivilla},
  {Ginsburg}, {Klaassen} \& {Harsono}}{{Maud} et~al.}{2019}]{maud_aa_627_2019}
{Maud} L.~T.,  {Cesaroni} R.,  {Kumar} M.~S.~N.,  {Rivilla} V.~M.,  {Ginsburg}
  A.,  {Klaassen} P.~D.,    {Harsono} D. e.~a.,  2019, \aap, 627, L6

\bibitem[\protect\citeauthoryear{{Maud}, {Cesaroni}, {Kumar}, {van der Tak},
  {Allen}, {Hoare}, {Klaassen}, {Harsono} \& {Hogerheijde}}{{Maud}
  et~al.}{2018}]{maud_aa_620_2018}
{Maud} L.~T.,  {Cesaroni} R.,  {Kumar} M.~S.~N.,  {van der Tak} F.~F.~S.,
  {Allen} V.,  {Hoare} M.~G.,  {Klaassen} P.~D.,  {Harsono} D.,
  {Hogerheijde} M.~R. e.~a.,  2018, \aap, 620, A31

\bibitem[\protect\citeauthoryear{{Maud}, {Hoare}, {Galv{\'a}n-Madrid}, {Zhang},
  {de Wit}, {Keto}, {Johnston} \& {Pineda}}{{Maud}
  et~al.}{2017}]{maud_467_mnras_2017}
{Maud} L.~T.,  {Hoare} M.~G.,  {Galv{\'a}n-Madrid} R.,  {Zhang} Q.,  {de Wit}
  W.~J.,  {Keto} E.,  {Johnston} K.~G.,    {Pineda} J.~E.,  2017, \mnras, 467,
  L120

\bibitem[\protect\citeauthoryear{{McKee}, {Stacy} \& {Li}}{{McKee}
  et~al.}{2020}]{2020MNRAS.tmp.2057M}
{McKee} C.~F.,  {Stacy} A.,    {Li} P.~S.,  2020, \mnras

\bibitem[\protect\citeauthoryear{{McKee} \& {Tan}}{{McKee} \&
  {Tan}}{2008}]{mckee_apj_681_2008}
{McKee} C.~F.,  {Tan} J.~C.,  2008, \apj, 681, 771

\bibitem[\protect\citeauthoryear{{McLeod}, {Reiter}, {Kuiper}, {Klaassen} \&
  {Evans}}{{McLeod} et~al.}{2018}]{mcloed_nature_554_2018}
{McLeod} A.~F.,  {Reiter} M.,  {Kuiper} R.,  {Klaassen} P.~D.,    {Evans}
  C.~J.,  2018, \nat, 554, 334

\bibitem[\protect\citeauthoryear{{Meyer}, {Haemmerl{\'e}} \&
  {Vorobyov}}{{Meyer} et~al.}{2019}]{2019MNRAS.484.2482M}
{Meyer} D.~M.~A.,  {Haemmerl{\'e}} L.,    {Vorobyov} E.~I.,  2019, \mnras, 484,
  2482

\bibitem[\protect\citeauthoryear{{Meyer}, {Kreplin}, {Kraus}, {Vorobyov},
  {Haemmerle} \& {Eisl{\"o}ffel}}{{Meyer} et~al.}{2019}]{meyer_487_MNRAS_2019}
{Meyer} D.~M.~A.,  {Kreplin} A.,  {Kraus} S.,  {Vorobyov} E.~I.,  {Haemmerle}
  L.,    {Eisl{\"o}ffel} J.,  2019, \mnras, 487, 4473

\bibitem[\protect\citeauthoryear{{Meyer}, {Kuiper}, {Kley}, {Johnston} \&
  {Vorobyov}}{{Meyer} et~al.}{2018}]{meyer_mnras_473_2018}
{Meyer} D.~M.-A.,  {Kuiper} R.,  {Kley} W.,  {Johnston} K.~G.,    {Vorobyov}
  E.,  2018, \mnras, 473, 3615

\bibitem[\protect\citeauthoryear{{Meyer}, {Vorobyov}, {Elbakyan}, {Stecklum},
  {Eisl{\"o}ffel} \& {Sobolev}}{{Meyer} et~al.}{2019}]{meyer_mnras_482_2019}
{Meyer} D.~M.-A.,  {Vorobyov} E.~I.,  {Elbakyan} V.~G.,  {Stecklum} B.,
  {Eisl{\"o}ffel} J.,    {Sobolev} A.~M.,  2019, \mnras, 482, 5459

\bibitem[\protect\citeauthoryear{{Meyer}, {Vorobyov}, {Kuiper} \&
  {Kley}}{{Meyer} et~al.}{2017}]{meyer_mnras_464_2017}
{Meyer} D.~M.-A.,  {Vorobyov} E.~I.,  {Kuiper} R.,    {Kley} W.,  2017, \mnras,
  464, L90

\bibitem[\protect\citeauthoryear{{Mignone}, {Bodo}, {Massaglia}, {Matsakos},
  {Tesileanu}, {Zanni} \& {Ferrari}}{{Mignone}
  et~al.}{2007}]{mignone_apj_170_2007}
{Mignone} A.,  {Bodo} G.,  {Massaglia} S.,  {Matsakos} T.,  {Tesileanu} O.,
  {Zanni} C.,    {Ferrari} A.,  2007, \apjs, 170, 228

\bibitem[\protect\citeauthoryear{{Mignone}, {Zanni}, {Tzeferacos}, {van
  Straalen}, {Colella} \& {Bodo}}{{Mignone}
  et~al.}{2012}]{migmone_apjs_198_2012}
{Mignone} A.,  {Zanni} C.,  {Tzeferacos} P.,  {van Straalen} B.,  {Colella} P.,
     {Bodo} G.,  2012, \apjs, 198, 7

\bibitem[\protect\citeauthoryear{{Motogi}, {Hirota}, {Machida}, {Yonekura},
  {Honma}, {Takakuwa} \& {Matsushita}}{{Motogi}
  et~al.}{2019}]{motogi_apj_877_2019}
{Motogi} K.,  {Hirota} T.,  {Machida} M.~N.,  {Yonekura} Y.,  {Honma} M.,
  {Takakuwa} S.,    {Matsushita} S.,  2019, \apjl, 877, L25

\bibitem[\protect\citeauthoryear{{Obonyo}, {Lumsden}, {Hoare}, {Purser},
  {Kurtz} \& {Johnston}}{{Obonyo} et~al.}{2019}]{obonyo_486_MNRAS_2019}
{Obonyo} W.~O.,  {Lumsden} S.~L.,  {Hoare} M.~G.,  {Purser} S.~J.~D.,  {Kurtz}
  S.~E.,    {Johnston} K.~G.,  2019, \mnras, 486, 3664

\bibitem[\protect\citeauthoryear{{Offner} \& {McKee}}{{Offner} \&
  {McKee}}{2011}]{2011ApJ...736...53O}
{Offner} S. S.~R.,  {McKee} C.~F.,  2011, \apj, 736, 53

\bibitem[\protect\citeauthoryear{{Padoan}, {Haugb{\o}lle} \&
  {Nordlund}}{{Padoan} et~al.}{2014}]{2014ApJ...797...32P}
{Padoan} P.,  {Haugb{\o}lle} T.,    {Nordlund} {\r{A}}.,  2014, \apj, 797, 32

\bibitem[\protect\citeauthoryear{{Peters}, {Banerjee}, {Klessen}, {Mac Low},
  {Galv{\'a}n-Madrid} \& {Keto}}{{Peters} et~al.}{2010}]{peters_apj_711_2010}
{Peters} T.,  {Banerjee} R.,  {Klessen} R.~S.,  {Mac Low} M.-M.,
  {Galv{\'a}n-Madrid} R.,    {Keto} E.~R.,  2010, \apj, 711, 1017

\bibitem[\protect\citeauthoryear{{Purser}, {Lumsden}, {Hoare} \&
  {Cunningham}}{{Purser} et~al.}{2018}]{purser_mnras_475_2018}
{Purser} S.~J.~D.,  {Lumsden} S.~L.,  {Hoare} M.~G.,    {Cunningham} N.,  2018,
  \mnras, 475, 2

\bibitem[\protect\citeauthoryear{{Purser}, {Lumsden}, {Hoare}, {Urquhart},
  {Cunningham}, {Purcell}, {Brooks}, {Garay}, {G{\'u}zman} \&
  {Voronkov}}{{Purser} et~al.}{2016}]{purser_mnras_460_2016}
{Purser} S.~J.~D.,  {Lumsden} S.~L.,  {Hoare} M.~G.,  {Urquhart} J.~S.,
  {Cunningham} N.,  {Purcell} C.~R.,  {Brooks} K.~J.,  {Garay} G.,
  {G{\'u}zman} A.~E.,    {Voronkov} M.~A.,  2016, \mnras, 460, 1039

\bibitem[\protect\citeauthoryear{{Rafikov}}{{Rafikov}}{2007}]{rafikov_apj_662_2007}
{Rafikov} R.~R.,  2007, \apj, 662, 642

\bibitem[\protect\citeauthoryear{{Rafikov}}{{Rafikov}}{2009}]{rafikov_apj_704_2009}
{Rafikov} R.~R.,  2009, \apj, 704, 281

\bibitem[\protect\citeauthoryear{{Reipurth}, {Bally} \& {Devine}}{{Reipurth}
  et~al.}{1997}]{1997AJ....114.2708R}
{Reipurth} B.,  {Bally} J.,    {Devine} D.,  1997, \aj, 114, 2708

\bibitem[\protect\citeauthoryear{{Reiter}, {Kiminki}, {Smith} \&
  {Bally}}{{Reiter} et~al.}{2017}]{reiter_mnras_470_2017}
{Reiter} M.,  {Kiminki} M.~M.,  {Smith} N.,    {Bally} J.,  2017, \mnras, 470,
  4671

\bibitem[\protect\citeauthoryear{{Richling} \& {Yorke}}{{Richling} \&
  {Yorke}}{1997}]{richling_aa_327_1997}
{Richling} S.,  {Yorke} H.~W.,  1997, \aap, 327, 317

\bibitem[\protect\citeauthoryear{{Richling} \& {Yorke}}{{Richling} \&
  {Yorke}}{1998}]{richling_aa_340_1998}
{Richling} S.,  {Yorke} H.~W.,  1998, \aap, 340, 508

\bibitem[\protect\citeauthoryear{{Richling} \& {Yorke}}{{Richling} \&
  {Yorke}}{2000}]{richling_apj_539_2000}
{Richling} S.,  {Yorke} H.~W.,  2000, \apj, 539, 258

\bibitem[\protect\citeauthoryear{{Rogers} \& {Wadsley}}{{Rogers} \&
  {Wadsley}}{2012}]{roger_mnras_423_2012}
{Rogers} P.~D.,  {Wadsley} J.,  2012, \mnras, 423, 1896

\bibitem[\protect\citeauthoryear{{Rosen}, {Krumholz}, {McKee} \&
  {Klein}}{{Rosen} et~al.}{2016}]{rosen_mnras_463_2016}
{Rosen} A.~L.,  {Krumholz} M.~R.,  {McKee} C.~F.,    {Klein} R.~I.,  2016,
  \mnras, 463, 2553

\bibitem[\protect\citeauthoryear{{Rosen}, {Li}, {Zhang} \& {Burkhart}}{{Rosen}
  et~al.}{2019}]{rosen_apj_887_2019}
{Rosen} A.~L.,  {Li} P.~S.,  {Zhang} Q.,    {Burkhart} B.,  2019, \apj, 887,
  108

\bibitem[\protect\citeauthoryear{{Samal}, {Chen}, {Takami}, {Jose} \&
  {Froebrich}}{{Samal} et~al.}{2018}]{samal_mnras_477_2018}
{Samal} M.~R.,  {Chen} W.~P.,  {Takami} M.,  {Jose} J.,    {Froebrich} D.,
  2018, \mnras, 477, 4577

\bibitem[\protect\citeauthoryear{{Sanna}, {K{\"o}lligan}, {Moscadelli},
  {Kuiper}, {Cesaroni}, {Pillai}, {Menten}, {Zhang}, {Caratti o Garatti},
  {Goddi}, {Leurini} \& {Carrasco-Gonz{\'a}lez}}{{Sanna}
  et~al.}{2019}]{sanna_aa_623_2019}
{Sanna} A.,  {K{\"o}lligan} A.,  {Moscadelli} L.,  {Kuiper} R.,  {Cesaroni} R.,
   {Pillai} T.,  {Menten} K.~M.,  {Zhang} Q.,  {Caratti o Garatti} A.,  {Goddi}
  C.,  {Leurini} S.,    {Carrasco-Gonz{\'a}lez} C.,  2019, \aap, 623, A77

\bibitem[\protect\citeauthoryear{{Sanna}, {Moscadelli}, {Surcis}, {van
  Langevelde}, {Torstensson} \& {Sobolev}}{{Sanna}
  et~al.}{2017}]{sanna_aa_603_2017}
{Sanna} A.,  {Moscadelli} L.,  {Surcis} G.,  {van Langevelde} H.~J.,
  {Torstensson} K.~J.~E.,    {Sobolev} A.~M.,  2017, \aap, 603, A94

\bibitem[\protect\citeauthoryear{{Seifried}, {Banerjee}, {Klessen}, {Duffin} \&
  {Pudritz}}{{Seifried} et~al.}{2011}]{seifried_mnras_417_2011}
{Seifried} D.,  {Banerjee} R.,  {Klessen} R.~S.,  {Duffin} D.,    {Pudritz}
  R.~E.,  2011, \mnras, 417, 1054

\bibitem[\protect\citeauthoryear{{Shchekinov} \& {Sobolev}}{{Shchekinov} \&
  {Sobolev}}{2004}]{Shchekinov_aa_418_2004}
{Shchekinov} Y.~A.,  {Sobolev} A.~M.,  2004, \aap, 418, 1045

\bibitem[\protect\citeauthoryear{{Shu}}{{Shu}}{1977}]{shu_apj_214_1977}
{Shu} F.~H.,  1977, \apj, 214, 488

\bibitem[\protect\citeauthoryear{{Szymczak}, {Olech}, {Wolak}, {G{\'e}rard} \&
  {Bartkiewicz}}{{Szymczak} et~al.}{2018}]{szymczak_aa_617_2018}
{Szymczak} M.,  {Olech} M.,  {Wolak} P.,  {G{\'e}rard} E.,    {Bartkiewicz} A.,
   2018, \aap, 617, A80

\bibitem[\protect\citeauthoryear{{Testi}}{{Testi}}{2003}]{testi_2003}
{Testi} L.,  2003, in {De Buizer} J.~M.,  {van der Bliek} N.~S.,  eds, Galactic
  Star Formation Across the Stellar Mass Spectrum Vol.~287 of Astronomical
  Society of the Pacific Conference Series, {Intermediate Mass Stars (Invited
  Review)}.
pp 163--173

\bibitem[\protect\citeauthoryear{{Toomre}}{{Toomre}}{1963}]{toomre_apj_138_1963}
{Toomre} A.,  1963, \apj, 138, 385

\bibitem[\protect\citeauthoryear{{Vorobyov}}{{Vorobyov}}{2009}]{vorobyov_apj_704_2009}
{Vorobyov} E.~I.,  2009, \apj, 704, 715

\bibitem[\protect\citeauthoryear{{Vorobyov}}{{Vorobyov}}{2010}]{vorobyov_apj_723_2010}
{Vorobyov} E.~I.,  2010, \apj, 723, 1294

\bibitem[\protect\citeauthoryear{{Vorobyov}}{{Vorobyov}}{2011a}]{vorobyov_apj_728_2011}
{Vorobyov} E.~I.,  2011a, \apjl, 728, L45

\bibitem[\protect\citeauthoryear{{Vorobyov}}{{Vorobyov}}{2011b}]{vorobyov_apj_729_2011}
{Vorobyov} E.~I.,  2011b, \apj, 729, 146

\bibitem[\protect\citeauthoryear{{Vorobyov} \& {Basu}}{{Vorobyov} \&
  {Basu}}{2006}]{voroboyov_apj_650_2006}
{Vorobyov} E.~I.,  {Basu} S.,  2006, \apj, 650, 956

\bibitem[\protect\citeauthoryear{{Vorobyov} \& {Basu}}{{Vorobyov} \&
  {Basu}}{2010}]{vorobyov_apj_719_2010}
{Vorobyov} E.~I.,  {Basu} S.,  2010, \apj, 719, 1896

\bibitem[\protect\citeauthoryear{{Vorobyov} \& {Basu}}{{Vorobyov} \&
  {Basu}}{2015}]{vorobyov_apj_805_2015}
{Vorobyov} E.~I.,  {Basu} S.,  2015, \apj, 805, 115

\bibitem[\protect\citeauthoryear{{Vorobyov}, {Elbakyan}, {Plunkett}, {Dunham},
  {Audard}, {Guedel} \& {Dionatos}}{{Vorobyov}
  et~al.}{2018}]{vorobyov_aa_613_2018}
{Vorobyov} E.~I.,  {Elbakyan} V.~G.,  {Plunkett} A.~L.,  {Dunham} M.~M.,
  {Audard} M.,  {Guedel} M.,    {Dionatos} O.,  2018, \aap, 613, A18

\bibitem[\protect\citeauthoryear{{Wang}, {Beuther}, {Bik}, {Vasyunina},
  {Jiang}, {Puga}, {Linz}, {Rod{\'o}n}, {Henning} \& {Tamura}}{{Wang}
  et~al.}{2011}]{wang_aa_527_2011}
{Wang} Y.,  {Beuther} H.,  {Bik} A.,  {Vasyunina} T.,  {Jiang} Z.,  {Puga} E.,
  {Linz} H.,  {Rod{\'o}n} J.~A.,  {Henning} T.,    {Tamura} M.,  2011, \aap,
  527, A32

\bibitem[\protect\citeauthoryear{{Weber}, {Ercolano}, {Picogna}, {Hartmann} \&
  {Rodenkirch}}{{Weber} et~al.}{2020}]{weber_mnras_496_2020}
{Weber} M.~L.,  {Ercolano} B.,  {Picogna} G.,  {Hartmann} L.,    {Rodenkirch}
  P.~J.,  2020, \mnras, 496, 223

\bibitem[\protect\citeauthoryear{{Wolk}, {G{\"u}nther}, {Poppenhaeger},
  {Winston}, {Rebull}, {Stauffer}, {Gutermuth}, {Cody}, {Hillenbrand},
  {Plavchan}, {Covey} \& {Song}}{{Wolk} et~al.}{2018}]{2018AJ....155...99W}
{Wolk} S.~J.,  {G{\"u}nther} H.~M.,  {Poppenhaeger} K.,  {Winston} E.,
  {Rebull} L.~M.,  {Stauffer} J.~R.,  {Gutermuth} R.~A.,  {Cody} A.~M.,
  {Hillenbrand} L.~A.,  {Plavchan} P.,  {Covey} K.~R.,    {Song} I.,  2018,
  \aj, 155, 99

\bibitem[\protect\citeauthoryear{{Wurster} \& {Bate}}{{Wurster} \&
  {Bate}}{2019a}]{wurster_mnras_486_2019}
{Wurster} J.,  {Bate} M.~R.,  2019a, \mnras, 486, 2587

\bibitem[\protect\citeauthoryear{{Wurster} \& {Bate}}{{Wurster} \&
  {Bate}}{2019b}]{2019arXiv190612276W}
{Wurster} J.,  {Bate} M.~R.,  2019b, arXiv e-prints, p. arXiv:1906.12276

\bibitem[\protect\citeauthoryear{{Yorke}, {Bodenheimer} \&
  {Tenorio-Tagle}}{{Yorke} et~al.}{1982}]{yorke_aa_108_1982}
{Yorke} H.~W.,  {Bodenheimer} P.,    {Tenorio-Tagle} G.,  1982, \aap, 108, 25

\bibitem[\protect\citeauthoryear{{Yorke} \& {Kruegel}}{{Yorke} \&
  {Kruegel}}{1977}]{yorke_aa_54_1977}
{Yorke} H.~W.,  {Kruegel} E.,  1977, \aap, 54, 183

\bibitem[\protect\citeauthoryear{{Yorke} \& {Sonnhalter}}{{Yorke} \&
  {Sonnhalter}}{2002}]{2002ApJ...569..846Y}
{Yorke} H.~W.,  {Sonnhalter} C.,  2002, \apj, 569, 846

\bibitem[\protect\citeauthoryear{{Yorke} \& {Welz}}{{Yorke} \&
  {Welz}}{1996}]{yorke_aa_315_1996}
{Yorke} H.~W.,  {Welz} A.,  1996, \aap, 315, 555

\bibitem[\protect\citeauthoryear{{Zhao}, {Caselli}, {Li} \&
  {Krasnopolsky}}{{Zhao} et~al.}{2018}]{zhao_mnras_473_2018}
{Zhao} B.,  {Caselli} P.,  {Li} Z.-Y.,    {Krasnopolsky} R.,  2018, \mnras,
  473, 4868

\bibitem[\protect\citeauthoryear{{Zinchenko}, {Liu}, {Su}, {Salii}, {Sobolev},
  {Zemlyanukha}, {Beuther}, {Ojha}, {Samal} \& {Wang}}{{Zinchenko}
  et~al.}{2015}]{zinchenko_apj_810_2015}
{Zinchenko} I.,  {Liu} S.-Y.,  {Su} Y.-N.,  {Salii} S.~V.,  {Sobolev} A.~M.,
  {Zemlyanukha} P.,  {Beuther} H.,  {Ojha} D.~K.,  {Samal} M.~R.,    {Wang} Y.,
   2015, \apj, 810, 10

\bibitem[\protect\citeauthoryear{{Zinchenko}, {Liu}, {Su}, {Wang} \&
  {Wang}}{{Zinchenko} et~al.}{2019}]{2019arXiv191111447Z}
{Zinchenko} I.~I.,  {Liu} S.-Y.,  {Su} Y.-N.,  {Wang} K.-S.,    {Wang} Y.,
  2019, arXiv e-prints, p. arXiv:1911.11447

\end{thebibliography}
}



\begin{table*}
	\centering
	\caption{
	Summary of burst characteristics along the line of increasing $\beta$. 
	$N_{\mathrm{bst}}$ is the number of bursts at a given magnitude cut-off.  
	$L_{\mathrm{max}}/L_{\mathrm{min}}/L_{\mathrm{mean}}$ are the maximum, minimum and mean burst luminosities, respectively. 
	Similarly, $\dot{M}_{\mathrm{max}}/\dot{M}_{\mathrm{min}}/\dot{M}_{\mathrm{mean}}$ are the maximum, minimum and mean 
	accretion rates through the central sink cell and 
	$t_{\mathrm{bst}}^{\mathrm{max}}$/$t_{\mathrm{bst}}^{\mathrm{min}}$/$t_{\mathrm{bst}}^{\mathrm{mean}}$
	are the maximum, minimum and mean bursts duration, while 
	$t_{\mathrm{bst}}^{\mathrm{tot}}$ is the integrated bursts duration time throughout the star's live. 
	}
        \begin{tabular}{lcccccr}
        \hline  
        Model &  & $N_{\mathrm{bst}}$  & $L_{\mathrm{max}}/L_{\mathrm{min}}/L_{\mathrm{mean}}$   ($10^{5}\, \rm L_{\odot}$) & $\dot{M}_{\mathrm{max}}/\dot{M}_{\mathrm{min}}/\dot{M}_{\mathrm{mean}}$ ($\rm M_{\odot}\, \rm yr^{-1}$) & $t_{\mathrm{bst}}^{\mathrm{max}}$/$t_{\mathrm{bst}}^{\mathrm{min}}$/$t_{\mathrm{bst}}^{\mathrm{mean}}$ $\rm (yr)$  & $t_{\mathrm{bst}}^{\mathrm{tot}}$ $\rm (yr)$\tabularnewline
        \hline
        \multicolumn{7}{c}{\textbf{1-mag cutoff}}\tabularnewline
${\rm Run-100-2}$$\%$  &    & 38 & 21.12 / 1.855 / 10.75 & 0.0194 / 0.0033 / 0.0076 & 81  / 5  / 16 & 626\tabularnewline
${\rm Run-100-4}$$\%$  &    & 33 & 21.43 / 0.558 / 7.91  & 0.0180 / 0.0031 / 0.0068 & 39  / 5  / 13 & 444\tabularnewline
${\rm Run-100-5}$$\%$  &    & 21 & 15.09 / 0.067 / 4.44  & 0.0190 / 0.0014 / 0.0083 & 88  / 9  / 26 & 553\tabularnewline
${\rm Run-100-6}$$\%$  &    & 36 & 18.69 / 0.149 / 5.77  & 0.0222 / 0.0023 / 0.0094 & 73  / 3  / 22 & 807\tabularnewline
${\rm Run-100-8}$$\%$  &    & 29 & 9.60  / 0.057 / 3.28  & 0.0301 / 0.0012 / 0.0082 & 53  / 5  / 15 & 449\tabularnewline
${\rm Run-100-10}$$\%$ &    & 44 & 13.07 / 0.063 / 3.83  & 0.0228 / 0.0013 / 0.0060 & 94  / 5  / 16 & 705\tabularnewline
${\rm Run-100-12}$$\%$ &    & 30 & 6.19  / 0.057 / 1.84  & 0.0198 / 0.0012 / 0.0094 & 110 / 4  / 26 & 789\tabularnewline
${\rm Run-100-14}$$\%$ &    & 30 & 10.73 / 0.052 / 3.04  & 0.0216 / 0.0012 / 0.0081 & 114 / 6  / 27 & 806\tabularnewline
${\rm Run-100-16}$$\%$ &    & 30 & 4.99  / 0.035 / 1.16  & 0.0209 / 0.0007 / 0.0073 & 76  / 7  / 28 & 829\tabularnewline
${\rm Run-100-18}$$\%$ &    & 26 & 2.54  / 0.037 / 0.64  & 0.0160 / 0.0008 / 0.0054 & 97  / 6  / 31 & 794\tabularnewline
${\rm Run-100-20}$$\%$ &    & 7  & 3.02  / 0.064 / 1.54  & 0.0241 / 0.0016 / 0.0087 & 78  / 5  / 27 & 188\tabularnewline
${\rm Run-100-25}$$\%$ &    & 21 & 0.83  / 0.043 / 0.21  & 0.0148 / 0.0012 / 0.0048 & 89  / 5  / 41 & 855\tabularnewline
${\rm Run-100-33}$$\%$ &    & 6  & 0.34  / 0.036 / 0.12  & 0.0116 / 0.0012 / 0.0035 & 79  / 12 / 39 & 234\tabularnewline
        \textbf{Total all models}  &    & \textbf{27}{} & \textbf{21.43 / 0.035 / 3.42} & \textbf{0.0301 / 0.0007 / 0.0072} & \textbf{114 / 3 / 25} & \textbf{622{}}\tabularnewline
	\multicolumn{7}{c}{\textbf{2-mag cutoff}}\tabularnewline
${\rm Run-100-2}$$\%$  &    & 13 & 50.44 / 17.752 / 26.67 & 0.0337 / 0.0127 / 0.0197 & 40 / 3  / 11 & 139\tabularnewline
${\rm Run-100-4}$$\%$  &    & 22 & 45.60 / 10.495 / 26.62 & 0.0356 / 0.0117 / 0.0226 & 56 / 3  / 13 & 277\tabularnewline
${\rm Run-100-5}$$\%$  &    &  6 & 35.66 / 6.385  / 15.34 & 0.0488 / 0.0104 / 0.0265 & 17 / 6  / 11 & 69\tabularnewline
${\rm Run-100-6}$$\%$  &    &  8 & 35.15 / 4.124  / 23.22 & 0.0519 / 0.0150 / 0.0307 & 14 / 6  / 9  & 73\tabularnewline
${\rm Run-100-8}$$\%$  &    & 11 & 22.62 / 0.221  / 6.81  & 0.0603 / 0.0044 / 0.0322 & 44 / 4  / 12 & 127\tabularnewline
${\rm Run-100-10}$$\%$ &    & 11 & 27.96 / 0.201  / 8.32  & 0.0494 / 0.0040 / 0.0236 & 74 / 6  / 20 & 217\tabularnewline
${\rm Run-100-12}$$\%$ &    &  8 & 23.31 / 0.146  / 5.18  & 0.0422 / 0.0031 / 0.0155 & 78 / 8  / 35 & 281\tabularnewline
${\rm Run-100-14}$$\%$ &    & 10 & 12.68 / 0.144  / 3.48  & 0.0478 / 0.0033 / 0.0191 & 80 / 5  / 30 & 305\tabularnewline
${\rm Run-100-16}$$\%$ &    &  7 & 4.25  / 0.211  / 1.10  & 0.0330 / 0.0042 / 0.0145 & 60 / 4  / 31 & 217\tabularnewline
${\rm Run-100-18}$$\%$ &    &  9 & 3.38  / 0.188  / 1.24  & 0.0669 / 0.0037 / 0.0196 & 87 / 6  / 36 & 320\tabularnewline
${\rm Run-100-20}$$\%$ &    &  7 & 7.63  / 0.129  / 1.75  & 0.0482 / 0.0035 / 0.0166 & 81 / 10 / 37 & 262\tabularnewline
${\rm Run-100-25}$$\%$ &    &  6 & 2.99  / 0.150  / 0.68  & 0.0250 / 0.0038 / 0.0082 & 91 / 16 / 36 & 216\tabularnewline
${\rm Run-100-33}$$\%$ &    & 10 & 1.84  / 0.116  / 0.46  & 0.0566 / 0.0031 / 0.0117 & 68 / 9  / 35 & 351\tabularnewline
\textbf{Total all models}  &   & \textbf{4}{} & \textbf{50.44 / 0.116 / 9.30} & \textbf{0.0669 / 0.0031 / 0.0200} & \textbf{91 / 3 / 24} & \textbf{220{}}\tabularnewline
        \multicolumn{7}{c}{\textbf{3-mag cutoff}}\tabularnewline
${\rm Run-100-2}$$\%$  &   & 4 & 117.38 / 30.481 / 68.54 & 0.1303 / 0.0386 / 0.0711 & 14 / 5 / 9  & 37\tabularnewline
${\rm Run-100-4}$$\%$  &   & 4 & 65.74  / 13.512 / 50.64 & 0.0537 / 0.0385 / 0.0466 & 29 / 4 / 13 & 52\tabularnewline
${\rm Run-100-5}$$\%$  &   & 2 & 37.41  / 4.992  / 21.20 & 0.0616 / 0.0448 / 0.0532 & 35 / 8 / 22 & 43\tabularnewline
${\rm Run-100-6}$$\%$  &   & 3 & 44.86  / 11.743 / 26.04 & 0.0663 / 0.0636 / 0.0651 & 16 / 6 / 11 & 32\tabularnewline
${\rm Run-100-8}$$\%$  &   & 3 & 70.36  / 40.645 / 52.45 & 0.0683 / 0.0371 / 0.0558 & 8  / 3 / 6  & 18\tabularnewline
${\rm Run-100-10}$$\%$ &   & 3 & 38.50  / 9.240  / 23.18 & 0.0708 / 0.0375 / 0.0559 & 8  / 3 / 6  & 17\tabularnewline
${\rm Run-100-12}$$\%$ &   & 2 & 6.63   / 0.575  / 3.60  & 0.0902 / 0.0124 / 0.0513 & 25 / 10/ 18 & 35\tabularnewline
${\rm Run-100-14}$$\%$ &   & 1 & 12.23  / 12.234 / 12.23 & 0.0525 / 0.0525 / 0.0525 & 4  / 4 / 4  & 4\tabularnewline
${\rm Run-100-16}$$\%$ &   & 1 & 10.07  / 10.068 / 10.07 & 0.1148 / 0.1148 / 0.1148 & 5  / 5 / 5  & 5\tabularnewline
${\rm Run-100-18}$$\%$ &   & - & -                       & -                        & -           & -\tabularnewline
${\rm Run-100-20}$$\%$ &   & 4 & 15.17  / 0.384  / 7.72  & 0.0511 / 0.0083 / 0.0345 & 23 / 4 / 12 & 46\tabularnewline
${\rm Run-100-25}$$\%$ &   & 1 & 0.42   / 0.415  / 0.42  & 0.0097 / 0.0097 / 0.0097 & 17 /17 / 17 & 17\tabularnewline
${\rm Run-100-33}$$\%$ &   & 5 & 9.83   / 0.620  / 4.09  & 0.1124 / 0.0132 / 0.0738 & 21 / 3 / 7  & 35\tabularnewline
\textbf{Total all models} &   & \textbf{3}{} & \textbf{117.38 / 0.384 / 24.18} & \textbf{0.1303 / 0.0083 / 0.0570} & \textbf{35 / 3 / 11} & \textbf{28{}}\tabularnewline
        \multicolumn{7}{c}{\textbf{4-mag cutoff}}\tabularnewline
${\rm Run-100-2}$$\%$  &   & 4 & 456.50 / 167.523 / 277.06 & 0.4224 / 0.1057 / 0.2104 & 22 / 4 / 11 & 44\tabularnewline
${\rm Run-100-4}$$\%$  &   & 5 & 745.03 / 140.465 / 307.95 & 0.5235 / 0.0929 / 0.2260 & 9  / 2 / 5 & 27\tabularnewline
${\rm Run-100-5}$$\%$  &   & 8 & 644.26 / 37.481  / 221.32 & 0.4384 / 0.0804 / 0.2146 & 10 / 2 / 4 & 32\tabularnewline
${\rm Run-100-6}$$\%$  &   & 2 & 190.73 / 166.192 / 178.46 & 0.2117 / 0.1108 / 0.1612 & 32 / 2 / 18 & 37\tabularnewline
${\rm Run-100-8}$$\%$  &   & 2 & 195.60 / 30.315  / 112.96 & 0.3164 / 0.1457 / 0.2311 & 6  / 3 / 4 & 9\tabularnewline
${\rm Run-100-10}$$\%$ &   & 4 & 432.37 / 100.012 / 261.88 & 0.9294 / 0.1007 / 0.4241 & 7  / 2 / 3 & 14\tabularnewline
${\rm Run-100-12}$$\%$ &   & 1 & 24.54  / 24.540  / 24.54  & 0.1966 / 0.1966 / 0.1966 & 4  / 4 / 4 & 4\tabularnewline
${\rm Run-100-14}$$\%$ &   & - & - & - & - & -\tabularnewline
${\rm Run-100-16}$$\%$ &   & 3 & 42.30  / 12.521  / 28.73  & 0.2903 / 0.1527 / 0.2155 & 5  / 3 / 4 & 11\tabularnewline
${\rm Run-100-18}$$\%$ &   & - & - & - & - & -\tabularnewline
${\rm Run-100-20}$$\%$ &   & 2 & 40.11  / 2.822   / 21.47  & 0.3285 / 0.0600 / 0.1942 & 10 / 4 / 7 & 13\tabularnewline
${\rm Run-100-25}$$\%$ &   & - & - & - & - & -\tabularnewline
${\rm Run-100-33}$$\%$ &   & 1 & 12.92  / 12.920  / 12.92  & 0.3145 / 0.3145 / 0.3145 & 2  / 2 / 2 & 2\tabularnewline
\textbf{Total all models} &   & \textbf{3}{} & \textbf{745.03 / 2.822 / 144.74} & \textbf{0.9294 / 0.0600 / 0.2388} & \textbf{32 / 2 / 6} & \textbf{19{}}\tabularnewline
        \hline    
        \end{tabular}
\label{tab:A}
\end{table*}

\begin{table*}
	\centering
	\caption{
	Same as Tab.~\ref{tab:A} for the line of increasing $\rm M_{\rm c}$. 
	}
        \begin{tabular}{lcccccr}
        \hline  
        Model &  & $N_{\mathrm{bst}}$  & $L_{\mathrm{max}}/L_{\mathrm{min}}/L_{\mathrm{mean}}$   ($10^{5}\, \rm L_{\odot}$) & $\dot{M}_{\mathrm{max}}/\dot{M}_{\mathrm{min}}/\dot{M}_{\mathrm{mean}}$ ($\rm M_{\odot}\, \rm yr^{-1}$) & $t_{\mathrm{bst}}^{\mathrm{max}}$/$t_{\mathrm{bst}}^{\mathrm{min}}$/$t_{\mathrm{bst}}^{\mathrm{mean}}$ $\rm (yr)$  & $t_{\mathrm{bst}}^{\mathrm{tot}}$ $\rm (yr)$\tabularnewline
        \hline
        \multicolumn{7}{c}{\textbf{1-mag cutoff}}\tabularnewline
${\rm Run-60-4}$$\%$  &    & 30  & 3.28  / 0.05 / 0.74  & 0.022 / 0.001 / 0.008 & 30.8 / 2.3  / 10.5 & 315\tabularnewline
${\rm Run-80-4}$$\%$  &    & 49  & 4.43  / 0.05 / 0.58  & 0.022 / 0.001 / 0.006 & 33.6 / 1.9  / 8.7  & 428\tabularnewline
${\rm Run-120-4}$$\%$ &    & 58  & 17.24 / 0.06 / 4.45  & 0.019 / 0.001 / 0.007 & 18.1 / 2.6  / 5.1  & 296\tabularnewline
${\rm Run-140-4}$$\%$ &    & 44  & 18.1  / 0.07 / 5.44  & 0.020 / 0.001 / 0.008 & 85   / 1.9  / 6.2  & 274\tabularnewline
${\rm Run-160-4}$$\%$ &    & 79  & 22.68 / 0.79 / 10.6  & 0.018 / 0.003 / 0.008 & 17.1 / 1.6  / 3.9  & 309\tabularnewline
${\rm Run-180-4}$$\%$ &    & 94  & 27.57 / 0.15 / 11.1  & 0.020 / 0.004 / 0.008 & 12.6 / 1.8  / 3.7  & 352\tabularnewline
${\rm Run-200-4}$$\%$ &    & 103 & 44.07 / 0.7  / 13.56 & 0.022 / 0.004 / 0.010 & 16.2 / 1.4  / 4.0  & 407\tabularnewline
        \textbf{Total all models} &   & \textbf{59} & \textbf{44.07 / 0.05 / 5.97} & \textbf{0.022 / 0.001 / 0.007} & \textbf{85 / 1.4 / 6.0} & \textbf{309}\tabularnewline
	\multicolumn{7}{c}{\textbf{2-mag cutoff}}\tabularnewline
${\rm Run-60-4}$$\%$  &    & 12  & 4.71  / 0.15  / 1.17  & 0.043 / 0.003 / 0.13  & 24.6 / 5.3 / 11.8 & 142\tabularnewline
${\rm Run-80-4}$$\%$  &    & 11  & 5.20  / 0.12  / 1.32  & 0.048 / 0.003 / 0.018 & 32.6 / 5.3 / 13.2 & 145\tabularnewline
${\rm Run-120-4}$$\%$ &    & 14  & 37.46 / 18.08 / 20.25 & 0.060 / 0.121 / 0.029 & 24.1 / 2.9 / 7.7  & 108\tabularnewline
${\rm Run-140-4}$$\%$ &    & 16  & 46.55 / 2.8   / 20.28 & 0.058 / 0.013 / 0.03  & 46.4 / 1.5 / 9.8  & 156\tabularnewline
${\rm Run-160-4}$$\%$ &    & 33  & 46.29 / 3.45  / 25.78 & 0.054 / 0.01  / 0.024 & 13.9 / 1.4 / 4.4  & 144\tabularnewline
${\rm Run-180-4}$$\%$ &    & 45  & 75.39 / 1.84  / 28.56 & 0.047 / 0.013 / 0.026 & 12.0 / 1.4 / 4.0  & 182\tabularnewline
${\rm Run-200-4}$$\%$ &    & 70  & 81.13 / 3.77  / 39.84 & 0.057 / 0.014 / 0.027 & 10.5 / 1.1 / 3.5  & 243\tabularnewline
        \textbf{Total all models} &  & \textbf{53} & \textbf{81.13 / 0.12 / 18.17} & \textbf{0.060 / 0.003 / 0.067} & \textbf{ 46.4 / 1.1 / 7.6 } & \textbf{149}\tabularnewline
        \multicolumn{7}{c}{\textbf{3-mag cutoff}}\tabularnewline
${\rm Run-60-4}$$\%$  &    & 6  & 0.90  / 0.36  / 0.61  & 0.02  / 0.008 / 0.014 & 22.6 / 7.7 / 13.1 & 78\tabularnewline
${\rm Run-80-4}$$\%$  &    & 1  & 11.15                 & 0.065                 & 15.3              & 15\tabularnewline
${\rm Run-120-4}$$\%$ &    & 8  & 105   / 12.65 / 51.07 & 0.088 / 0.051 / 0.07  & 22.4 / 2.0 / 6.5  & 52\tabularnewline
${\rm Run-140-4}$$\%$ &    & 9  & 109.4 / 35.5  / 65.9  & 0.11  / 0.04  / 0.075 & 26.8 / 1.7 / 8.1  & 73\tabularnewline
${\rm Run-160-4}$$\%$ &    & 9  & 106.4 / 36.4  / 65.7  & 0.119 / 0.035 / 0.059 & 6.5  / 1.9 / 3.6  & 33\tabularnewline
${\rm Run-180-4}$$\%$ &    & 20 & 221.8 / 16.49 / 93.17 & 0.136 / 0.038 / 0.075 & 17.5 / 1.6 / 4.9  & 98\tabularnewline
${\rm Run-200-4}$$\%$ &    & 22 & 217.6 / 46.35 / 92.89 & 0.119 / 0.036 / 0.061 & 16.4 / 1.7 / 4.3  & 96\tabularnewline
        \textbf{Total all models} &  & \textbf{10} & \textbf{221.8 / 0.36 / 50.5} & \textbf{0.136 / 0.008 / 0.06} & \textbf{26.8 / 1.6 / 8.3} & \textbf{65}\tabularnewline
        \multicolumn{7}{c}{\textbf{4-mag cutoff}}\tabularnewline
${\rm Run-60-4}$$\%$  &    & 2  & 10.46 / 10.22 / 10.34  &  0.31  / 0.20  / 0.25  & 4.1 / 2.4  / 3.3 &  6.5\tabularnewline
${\rm Run-80-4}$$\%$  &    & 1  & 46.16                  &  0.215                 &    4.5           &  4.5\tabularnewline
${\rm Run-120-4}$$\%$ &    & 4  & 420.6 / 121.5  / 199.1 &  0.43  / 0.086 / 0.246 & 6.7  / 2.0 / 4.3 &  17\tabularnewline
${\rm Run-140-4}$$\%$ &    & 7  & 582.2 / 224.8  / 407.1 &  0.446 / 0.157 / 0.318 & 4.4  / 1.2 / 2.5 &  18\tabularnewline
${\rm Run-160-4}$$\%$ &    & 10 & 1171  / 55.49  / 394.3 &  0.87  / 0.13  / 0.33  & 10.1 / 1.9 / 4.9 &  49\tabularnewline
${\rm Run-180-4}$$\%$ &    & 6  & 1025  / 89.7   / 381.6 &  0.86  / 0.086 / 0.307 & 10.1 / 1.4 / 5.6 &  33\tabularnewline
${\rm Run-200-4}$$\%$ &    & 14 & 575   / 117.2  / 276.4 &  0.337 / 0.091 / 0.176 & 8    / 1.4 / 4.3 &  60\tabularnewline
        \textbf{Total all models} &  & \textbf{6} & \textbf{1025/10.22/264.7} & \textbf{0.87/0.13/0.30} & \textbf{10.1/1.2/3.9} & \textbf{24}\tabularnewline
        \hline    
        \end{tabular}
\label{tab:B}
\end{table*}

\begin{table*}
	\centering
	\caption{
	Same as Tab.~\ref{tab:A} for the models with $M_{\rm c}=60\, \rm M_{\odot}$ and $\beta\le1$.  
	}
        \begin{tabular}{lcccccr}
        \hline  
        Model &  & $N_{\mathrm{bst}}$  & $L_{\mathrm{max}}/L_{\mathrm{min}}/L_{\mathrm{mean}}$   ($10^{5}\, \rm L_{\odot}$) & $\dot{M}_{\mathrm{max}}/\dot{M}_{\mathrm{min}}/\dot{M}_{\mathrm{mean}}$ ($\rm M_{\odot}\, \rm yr^{-1}$) & $t_{\mathrm{bst}}^{\mathrm{max}}$/$t_{\mathrm{bst}}^{\mathrm{min}}$/$t_{\mathrm{bst}}^{\mathrm{mean}}$ $\rm (yr)$  & $t_{\mathrm{bst}}^{\mathrm{tot}}$ $\rm (yr)$\tabularnewline
        \hline
        \multicolumn{7}{c}{\textbf{1-mag cutoff}}\tabularnewline
${\rm Run-60-0.1}$$\%$ &   & -  &  -                   &                        - &            - & -\tabularnewline
${\rm Run-60-0.5}$$\%$ &   & 4  &  11.88 / 5.32 / 7.83 & 0.0121 / 0.0063 / 0.0100 & 45  / 7 / 19 & 77 \tabularnewline
${\rm Run-60-0.8}$$\%$ &   & 10 &   0.70 / 0.03 / 0.15 & 0.0265 / 0.0011 / 0.0048 & 109 / 6 / 59 & 594\tabularnewline
${\rm Run-60-1}$$\%$   &   & 8  &  15.55 / 2.11 / 8.18 & 0.0173 / 0.0065 / 0.0100 & 56  / 9 / 19 & 152\tabularnewline
\textbf{All models}   &   & 7{} & \textbf{15.55 / 0.03 / 5.39} & \textbf{0.0265 / 0.0011 / 0.0083} & \textbf{109 / 6 / 32} & \textbf{274}\tabularnewline
	\multicolumn{7}{c}{\textbf{2-mag cutoff}}\tabularnewline
${\rm Run-60-0.1}$$\%$ &   & -  & -                     & -                        & -            & -\tabularnewline
${\rm Run-60-0.5}$$\%$ &   & 2  & 23.41 / 11.02 / 17.21 & 0.0290 / 0.0255 / 0.0272 & 20 / 12 / 16 & 32\tabularnewline
${\rm Run-60-0.8}$$\%$ &   & 10 & 2.32  / 0.10  / 0.56  & 0.0495 / 0.0027 / 0.0154 & 79 / 4  / 34 & 336\tabularnewline
${\rm Run-60-1}$$\%$   &   & 4  & 25.52 / 5.80  / 14.74 & 0.0406 / 0.0222 / 0.0311 & 49 / 9  / 21 & 86\tabularnewline
\textbf{All models}   &   & \textbf{4}{} & \textbf{25.52 / 0.10 / 10.84} & \textbf{0.0495 / 0.0027 / 0.0246} & \textbf{79 / 4 / 24} & \textbf{151}\tabularnewline
        \multicolumn{7}{c}{\textbf{3-mag cutoff}}\tabularnewline
${\rm Run-60-0.1}$$\%$ &   & - & - & - & - & -\tabularnewline
${\rm Run-60-0.5}$$\%$ &   & - & - & - & - & -\tabularnewline
${\rm Run-60-0.8}$$\%$ &   & 4 & 0.88 / 0.33 / 0.52 & 0.0187 / 0.0074 / 0.0111 & 25 / 6 / 17 & 70\tabularnewline
${\rm Run-60-1}$$\%$   &   & - & - & - & - & -\tabularnewline
\textbf{All models}   &   & \textbf{4{}} & \textbf{0.88 / 0.33 / 0.52} & \textbf{0.0187 / 0.0074 / 0.0111} & \textbf{25 / 6 / 17} & \textbf{70}\tabularnewline
        \hline    
        \end{tabular}
\label{tab:C}
\end{table*}



\end{document}